\magnification = \magstep0
\hsize=15.0truecm  \hoffset=0.0truecm  \vsize=20.1truecm  \voffset=1.5truecm
\font\sc=cmr8   
\font\sit=cmti8
\font\sbf=cmbx8

\def\scbaselines{\baselineskip=8pt    \lineskip=0pt   \lineskiplimit=0pt}
\def\smedbaselines{\baselineskip=9pt    \lineskip=0pt   \lineskiplimit=0pt}

\def\refbaselines{\baselineskip=11pt    \lineskip=0pt   \lineskiplimit=0pt}

\def\vsl{\vskip\baselineskip}   \def\vs{\vskip 6pt} \def\vsss{\vskip 3pt}
\parindent=10pt \nopagenumbers

\def\omit#1{\empty}

\input colordvi
%
%  VERSION DOCTORED BY JK FOR TABLES ONLY
%
%----------%
%  TABLES  %
%----------%
\def\endtable{\endgroup}
\def\tableheight{\vrule width 0pt height 8.5pt depth 3.5pt}
{\catcode`|=\active \catcode`&=\active 
    \gdef\tabledelim{\catcode`|=\active \let|=\vbar
                     \catcode`&=\active \let&=\nobar} }
\def\table{\begingroup
    \def\twidth{\hsize}
    \def\tablewidth##1{\def\twidth{##1}}
    \def\defaultheight{\vrule width 0pt height 8.5pt depth 3.5pt}
    \def\heightdepth##1{\dimen0=##1
        \ifdim\dimen0>5pt 
            \divide\dimen0 by 2 \advance\dimen0 by 2.5pt
            \dimen1=\dimen0 \advance\dimen1 by -5pt
            \vrule width 0pt height \the\dimen0  depth \the\dimen1
        \else  \divide\dimen0 by 2
            \vrule width 0pt height \the\dimen0  depth \the\dimen0 \fi}
    \def\spacing##1{\def\defaultheight{\heightdepth{##1}}}
    \def\nextheight##1{\noalign{\gdef\tableheight{\heightdepth{##1}}}}
    \def\end{\cr\noalign{\gdef\tableheight{\defaultheight}}}
    \def\zerowidth##1{\omit\hidewidth ##1 \hidewidth}    
    \def\hline{\noalign{\hrule}}
    \def\skip##1{\noalign{\vskip##1}}
    \def\bskip##1{\noalign{\hbox to \twidth{\vrule height##1 depth 0pt \hfil
        \vrule height##1 depth 0pt}}}
    \def\header##1{\noalign{\hbox to \twidth{\hfil ##1 \unskip\hfil}}}
    \def\bheader##1{\noalign{\hbox to \twidth{\vrule\hfil ##1 
        \unskip\hfil\vrule}}}
    \def\spanloop{\span\omit \advance\mscount by -1}
    \def\extend##1##2{\omit
        \mscount=##1 \multiply\mscount by 2 \advance\mscount by -1
        \loop\ifnum\mscount>1 \spanloop\repeat \ \hfil ##2 \unskip\hfil}
    \def\vbar{&\vrule&}
    \def\nobar{&&}
    \def\hdash##1{ \noalign{ \relax \gdef\tableheight{\heightdepth{0pt}}
        \toks0={} \count0=1 \count1=0 \putout##1\end 
        \toks0=\expandafter{\the\toks0 &\end} \xdef\piggy{\the\toks0} }
        \piggy}
    \let\e=\expandafter
    \def\putspace{\ifnum\count0>1 \advance\count0 by -1
        \toks0=\e\e\e{\the\e\toks0\e&\e\multispan\e{\the\count0}\hfill} 
        \fi \count0=0 }
    \def\putrule{\ifnum\count1>0 \advance\count1 by 1
        \toks0=\e\e\e{\the\e\toks0\e&\e\multispan\e{\the\count1}\leaders\hrule\hfill}
        \fi \count1=0 }
    \def\putout##1{\ifx##1\end \putspace \putrule \let\next=\relax 
        \else \let\next=\putout
            \ifx##1- \advance\count1 by 2 \putspace
            \else    \advance\count0 by 2 \putrule \fi \fi \next}   }
\def\tablespec#1{
    \def\vdimens{\noexpand\tableheight}
    \def\tabby{\tabskip=0pt plus100pt minus100pt}
    \def\r{&################\tabby&\hfil################\unskip}
    \def\c{&################\tabby&\hfil################\unskip\hfil}
    \def\l{&################\tabby&################\unskip\hfil}
    \edef\templ{\noexpand\vdimens ########\unskip  #1 
         \unskip&########\tabskip=0pt&########\cr}
    \tabledelim
    \edef\body##1{ \vbox{
        \tabskip=0pt \offinterlineskip
        \halign to \twidth {\templ ##1}}} }

\def\t#1{#1} 
\def\t#1{\empty}
    
  \def\ba{\kern -1pt}
\parskip = 0pt % default is \parskip = 0pt+1pt
\def\ts{\thinspace} \def\cl{\centerline}
\def\ni{\noindent}

\def\nhi{\noindent  \hangindent=1.0truecm}

\def\nhi2{\noindent \hangindent=2.79cm}
  
\def\nhhj{\noindent \hangindent=9.33truecm}  
\def\nhi1{\indent \hangindent=0.8truecm}

\def\makeheadline{\vbox to 0pt{\vskip-30pt\line{\vbox to8.5pt{}\the
                               \headline}\vss}\nointerlineskip}
\def\toppageno{\headline={\hss\tenrm\folio\hss}}
\def\footnoterule{\kern-3pt \hrule width \hsize \kern 2.6pt \vskip 3pt}
\output={\plainoutput}    \pretolerance=10000   \tolerance=10000

\def\sup1{$^{\rm 1}$} \def\sup2{$^{\rm 2}$}
\def\r0{$\rho_0$}   
 \def\0{\phantom{0}} \def\bb{\kern -2pt}
\def\1{\phantom{1}}         
\def\etal{{et~al.\ }}
\def\gapprox{$_>\atop{^\sim}$} \def\lapprox{$_<\atop{^\sim}$}

% The following are from Scott:

\def\spose#1{\hbox to 0pt{#1\hss}}
\def\lesssim{\mathrel{\spose{\lower 3pt\hbox{$\mathchar"218$}}
     \raise 2.0pt\hbox{$\mathchar"13C$}}}
\def\gtrsim{\mathrel{\spose{\lower 3pt\hbox{$\mathchar"218$}}
     \raise 2.0pt\hbox{$\mathchar"13E$}}}

% End of Scott's definitions

\newdimen\sa  \def\sd{\sa=.1em \ifmmode $\rlap{.}$''$\kern -\sa$
                               \else     \rlap{.}$''$\kern -\sa\fi}
\newdimen\sb\def\degd{\sa=.15em\ifmmode $\rlap{.}$^\circ$\kern -\sa$
                               \else     \rlap{.}$^\circ$\kern -\sa\fi}
\def\ss{\ifmmode ^{\prime\prime}$\kern-\sa$ \else $^{\prime\prime}$\kern-\sa\fi}
\def\mm{\ifmmode ^{\prime}$\kern-\sa$ \else $^{\prime}$\kern-\sa \fi}
  
\def\mbh{$M_{\bullet}$~}  
\def\m31{M{\ts}31} \def\mm32{M{\ts}32} \def\mmm33{M{\ts}33} \def\M87{M{\ts}87}

%Luis's definitions
\def\lax{{$\mathrel{\hbox{\rlap{\hbox{\lower4pt\hbox{$\sim$}}}\hbox{$<$}}}$}}
\def\gax{{$\mathrel{\hbox{\rlap{\hbox{\lower4pt\hbox{$\sim$}}}\hbox{$>$}}}$}}
\input psfig.tex

\font\big=cmbx12 scaled 1100
\font\bigau=cmr12 scaled 1200
\font\bigbig=cmr12 scaled 2000

% Big style: (omitting redundant definitions above) (note: This works but is probably WRONG.  Expect trouble!)

\vfill\eject

\def\ARRed{\textColor{.23 1. 1. .17}}

\def\ARDiscard{\textColor{0.0 1.0 0.0 0.3}}

\cl{\null} \vskip -41pt  % (This is a cheat to get more lines)
 
\cl{\bigbig Coevolution (Or Not) of} \vsss

\cl{\bigbig Supermassive Black Holes and Host Galaxies}

\vsl
 
\cl{\bigau John Kormendy$^1$ and Luis C.~Ho$^{2,3}$}
\vsl
\cl{$^1$Department of Astronomy, University of Texas at Austin,}
\cl{2515 Speedway C1400, Austin, TX 78712-1205; email: kormendy@astro.as.utexas.edu}

\vs
\cl{$^2$The Observatories of the Carnegie Institution for Science,}
\cl{813 Santa Barbara Street, Pasadena, CA 91101; email: lho@obs.carnegiescience.edu}

\vs
\cl{$^3$Kavli Institute for Astronomy and Astrophysics,}
\cl{Peking University, Beijing 100871, China (Starting 2014 January)}
 
\vs
 
\cl{\null} \vskip 0pt  % (This is a cheat to get more lines)
\cl {\big\ARRed SUPPLEMENTAL MATERIAL}\textBlack
\vs

      Section S1 is a supplement to main paper Section 3 on machinery to measure BH masses 
$M_\bullet$.  It summarizes indirect methods that are used to estimate $M_\bullet$ in AGNs.
These are used sparingly in main paper
Section 6.9,
Sections 7.1{\ts}--{\ts}7.3, 
Sections 8.3, 8.5, and 8.6, and
Section 9.

      Section S2 lists the observational criteria that we use to classify classical and
pseudo bulges.  Section\ts4 in the main paper introduces the physical distinction 
between remnants of major mergers and pseudo (``fake'') bulges that were grown 
slowly out of disks, not made rapidly in violent events.  This distinction is
central to Sections 5\ts--\ts9.  However, the (pseudo)bulge classifications used in this
paper are not based on interpretation; they are based on the Section S2 observational criteria.

      Section S3 supplements the BH database in Section 5 of the main paper and Section S4 here.  
It discusses corrections to galaxy and BH parameters, most importantly to 2MASS $K$-band apparent 
magnitudes.~They are needed because 2MASS misses light at large radii when the images of galaxies 
subtend large angles on the sky or have shallow outer brightness gradients.  These corrections are
included in {\bf Tables 2} and {\bf 3}, which are numbered consistently here and in the main paper.

      Section S4 reproduces essentially verbatim the first part of Section 5 in the main paper, 
the BH database, including the list of BH and host-galaxy properties ({\bf Tables 2} and {\bf 3}).  
Its purpose~is~to provide notes on the properties given in the tables, especially 
all of the notes on individual objects.  To save space, only a few examples of these notes are 
included in the main paper.

\vs
\cl {\big\ARRed S1.~BH MASS MEASUREMENTS FOR ACTIVE GALAXIES}\textBlack
\vs

\hsize=15.truecm  \hoffset=0.0truecm  \vsize=21.5truecm  \voffset=1.2truecm

The methods reviewed in Sections 3.1{\ts}--{\ts}3.3 provide the most direct BH 
masses~({\bf Tables 2}~and~{\bf 3}) and lead to most of the demographic 
results discussed in this paper.  But these methods have two limitations.  It 
is difficult to study late-type galaxies, starbursting galaxies, and AGNs.  
This makes it harder to connect BH demographics with theories of galaxy 
formation.  And these methods require us to spatially resolve the sphere-of-influence 
radius $r_{\rm infl}$ of the BH.  Success is limited to nearby galaxies.  Most AGNs are 
so distant that we cannot resolve $r_{\rm infl}$.  In any case, the bright glare from 
the active nucleus often overwhelms the stellar features in host-galaxy spectra.
Also, the kinematics of circumnuclear ionized gas can easily be perturbed by 
non-gravitational forces such as radiation pressure or shocks.  The unfortunate 
consequence is that the objects whose current activity most directly tells us about 
BH-host coevolution are excluded from direct dynamical study.  

Much work has gone into using AGNs to solve these problems.  This section 
summarizes the machinery used to estimate BH masses in AGNs.  We cover only
the bare essentials and do not attempt a comprehensive review of this extensive subject.  
Recent reviews can be found in Peterson (2011) and Shen (2013).

\vs
\cl {\big\ARRed S1.1.~Reverberation Mapping}\textBlack
\vsss

      This technique is based on the assumption that the broad emission 
lines observed in the UV and optical spectra of quasars and Seyfert 1 galaxies
come from gravity-dominated regions close to the BH.  If the velocity widths 
of the emission lines trace the virial velocity $\Delta V$ of the broad-line 
region (BLR) at radius $r$, then
\vskip -16pt

$$
M_\bullet = {{f(\Delta V)^2{\ts}r}\over{G}}. \eqno{\rm(S1)}
$$
\vskip -4pt

\noindent Most of our astrophysical ignorance is lumped into the factor $f$, which 
depends on the geometry and detailed kinematics of the BLR.

      BLRs have radii of a few light days to a few light weeks, so we cannot resolve~them
spatially.  However, we can resolve them temporally using a technique known 
as reverberation mapping (Blandford \& McKee 1982).  Its origins trace back to 
Bahcall, Kozlovsky, \& Salpeter (1972), 
Lyutyi \& Cherepashchuk (1972), and 
Cherepashchuk \& Lyutyi (1973).  
The physical basis~is~simple. The broad emission lines arise from clouds that are
photoionized by the UV radiation from~the~hot, inner parts of the 
accretion disk.  When this ionizing continuum varies in brightness, the broad 
emission lines vary (``reverberate'') in response, but their brightness variations are
delayed by a time lag $\tau = r/c$ equal to the light travel time between the 
continuum source and the location of the line-emitting gas.  To date, reverberation
mapping experiments have yielded line-continuum lag measurements for approximately 50 
type 1 AGNs 
(Kaspi \etal 2000; 
Peterson \etal 2004; 
Bentz \etal 2008, 2009b; 
Denney \etal 2010; 
Barth \etal 2011; 
Grier \etal 2012).  
Most observations concentrate on H$\beta$ (and to 
a lesser extent H$\alpha$); much less data exist for Mg{\ts}{\sc II} or 
C{\ts}{\sc IV}.  The current sample is strongly biased toward relatively 
low-luminosity AGNs, mostly nearby Seyfert 1 galaxies.  A number of Palomar-Green 
(Schmidt \& Green 1983) quasars are included, but the highest-redshift source 
studied is only at $z = 0.29$ (PG 1700$+$518).  This reflects the 
historical roots of how samples were assembled and the practical challenges of 
spectroscopically monitoring higher-$z$ quasars 
(Kaspi \etal 2007; 
Trevese \etal 2007; 
Woo 2008). 

\pageno=2 \toppageno 

      The resulting BLR sizes can be combined with velocity width measures to calculate
$M_\bullet$.  However, implementing Equation{\ts}S1 involves a practical problem.~What 
part of the \hbox{emission-line} profile should we use? And how does it measure the virial 
velocity $\Delta V$? Popular choices are $\Delta V${\ts}={\ts}FWHM and 
$\Delta V = \sigma_{\rm line}$, i.{\ts}e., the second moment of the line profile.~Opinions 
vary as to which $\Delta V$ indicator is better (e.{\ts}g.,
Peterson{\ts}et{\ts}al.{\ts}2004; 
Collin{\ts}et{\ts}al.{\ts}2006; 
see Shen\ts2013~for~a~review).  
By far the biggest uncertainty is the virial coefficient~$f$.  It is unknown, and it
probably varies from source to source.  A spherical distribution of clouds on random, 
isotropic orbits has $f = 3/4$ for $\Delta V = {\rm FWHM}$ and $f = 3$ for 
$\Delta V = \sigma_{\rm line}$ (Netzer 1990). However, this idealization is almost 
certainly too simplistic.  In principle, $f$ can be constrained using sufficiently
accurate velocity-resolved delay maps (e.{\ts}g., 
Denney \etal 2009; 
Bentz \etal 2010; 
Grier \etal 2013), 
but such measurements are still rare and uncertain.~Direct dynamical modeling 
may eventually bypass the need to adopt an $f$ factor~(Brewer \etal 2011; 
Pancoast \etal 2011, 2012).  
For the time being, a practical way forward is to calibrate an average value of $f$ 
for the sample of reverberation-mapped objects by requiring that they follow the same 
$M_\bullet - \sigma$ relation as inactive galaxies.  This is a reasonable but 
unproven assumption.   It is motivated by the empirical fact that 
reverberation-mapped AGNs{\ts}--{\ts}at least those that have bulge stellar 
velocity dispersion measurements{\ts}--{\ts}{\it do}\ seem to follow an
$M_\bullet - \sigma$ relation 
(Gebhardt \etal 2000a; 
Ferrarese \etal 2001; 
Nelson \etal 2004; 
Greene \& Ho 2006).  
Following this approach and adopting 
the $M_\bullet - \sigma$ relation of Tremaine \etal (2002) as the fiducial 
reference, Onken \etal (2004) used 14 reverberation-mapped AGNs to obtain 
$\langle f \rangle = 5.5\pm1.8$ for $\Delta V = \sigma_{\rm line}$.  The 
scatter in $M_\bullet$ for AGNs is surprisingly small, less than a factor of 
3.  Woo \etal (2010) enlarged the calibration sample to 24 objects and found
$\langle f \rangle = 5.2\pm1.2$ with respect to the $M_\bullet - \sigma$ 
relation of G\"ultekin \etal (2009c) and an intrinsic scatter of 
$0.44\pm0.07$ dex.  Park \etal (2012) discuss uncertainties resulting from 
different fitting methods and the use of different reference samples of 
inactive galaxies.

\includegraphics{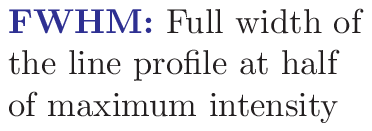}

%\hsize=15.5truecm  \hoffset=0.0truecm  \vsize=21.4truecm  \voffset=1.2truecm

      Equation{\ts}S1 depends on the assumption that BLR velocities are
controlled by BH gravity. (Other caveats are discussed in Krolik 2001 and in
Greene \& Ho 2009.) Outflows at velocities that are much larger than the 
escape velocity would cause us to overestimate $M_\bullet$.  Gas responds 
easily to non-gravitational forces.  An AGN is intrinsically a 
high-radiation-flux environment.  Winds and outflows are pervasive.  
Successful tests that BLR gas clouds move at velocities that are comparable to 
those in virial equilibrium are therefore central to our confidence in using 
this methodology.

      Velocity-resolved reverberation mapping provides this test.  The early study
by Gaskell (1988) established the principle.  If gas
flows outward, then the near side of the BLR moves toward us and the far side
moves away from us. Therefore, the blue wings of the emission lines should
vary first and the red wings should vary $\Delta \tau \simeq 2r \tan{i}/c$
later.  Infall results in the opposite behavior. From the emission lines
C{\ts}{\sc IV}\ts$\lambda$\ts1549 and Mg{\ts}{\sc II}\ts$\lambda$\ts2798,
Gaskell (1988) found that outflow is excluded in NGC 4151.  Shorter lags for
the red wings were most consistent with infall, although 
circular\ts$+${\ts}random motions were not strongly excluded.  The conclusion 
was that gravity controls the motions and that $M_\bullet \sim (5~{\rm to}~10) 
\times 10^7$\ts$M_\odot$.  Koratkar \& Gaskell (1989) got similar results for 
Fairall 9.  Gaskell (2010) reviews additional tests.  The cleanest infall 
signature is seen in Arp 151, observed as part of the Lick AGN Monitoring 
Project (LAMP: Bentz \etal 2008, 2009b, 2010).  Most reverberation 
observations are consistent with virialized motions or with some infall (Bentz 
\etal 2010).  A few exceptions (e.{\ts}g., Denney \etal 2009) emphasize the 
need for long or repeated observing campaigns.  Note that free fall from 
infinity in an $M_\bullet$-dominated potential results in a radial $V \propto
1/\sqrt{r}$ velocity field.  If AGNs have a distribution of small infall 
velocities in addition to mainly circular or random virial cloud motions, this 
infall will be ``calibrated out'' in the determination of the factor $f$.

      A critical consistency check is that different reverberation lags for different
emission lines in the same object should give the same $M_\bullet$.
BLR emission lines have different excitation potentials and so are 
emitted at different temperatures and radii.  If line widths measure Keplerian 
velocities, then different line widths at different radii should anticorrelate 
with lag $\tau$ as $\Delta V \propto 1/\sqrt{c\tau}$.  The best-observed 
galaxy, NGC 5548, shows this correlation convincingly, as do three other 
sources (Peterson 2011).  Recently, LAMP has substantially ``raised the bar''
on reverberation mapping and extended this test successfully to many more 
objects (Bentz \etal 2010).  

      On the other hand, a concern is raised by the observation that the BLR
structure changes with emission-line FWHM (Kollatschny \& Zetzl 2011).  They show 
that FWHM/$\sigma_{\rm line}$ is a strong, smooth function of FWHM, ranging from 
0.5 at FWHM $\simeq$ 1000 km s$^{-1}$ to 3 at FWHM $\simeq$ 12000 km s$^{-1}$.
These changes in line profile shapes suggest that the balance 
between rotation, random velocities, and in- or outflow changes with line 
width.  Hints of this were seen in earlier work 
(Collin \etal 2006; 
Marziani \& Sulentic 2012).  
Implications are worrisome:~$f$ must depend on FWHM and $\sigma_{\rm line}$.

      Another fundamental limitation is implied by the conclusion (Gaskell 2008, 2011) 
that AGN continua can flare in localized, off-center patches not far inside the
BLR radius.~This produces~noise.  Clearly $M_\bullet$ should be based 
on as many lines and as many observing campaigns as possible.

      {\it We emphasize an important shortcoming in all calibrations of $f$.  They
use $M_\bullet$\ts--\ts$\sigma$ 
relations that do not differentiate between classical and pseudo bulges.  We 
now know (Section\ts6.8\kern 0.2pt) that $M_\bullet$ does not correlate with pseudobulges.  
Their BH masses are similar to those of classical bulges but scatter 
to lower $M_\bullet$ values.  The reverberation-mapped AGNs that are used to 
calibrate $f$ contain a high proportion of pseudobulges.  A systematic error 
is introduced when $f$ is derived by comparing this sample with one that 
consists mostly of classical bulges.  The calibration needs to be improved
by classifying the (pseudo)bulges of AGN
galaxies.}\footnote{$^3$}{{\kern -4pt}Graham \etal (2011) check whether barred 
and unbarred galaxies have different normalizations~$f$.  This is not the same 
as differentiating classical and pseudo bulges: some unbarred galaxies contain 
pseudobulges and some barred galaxies contain classical bulges (Kormendy \& 
Kennicutt 2004).  But pseusobulges are more common in barred galaxies, so 
Graham's check does implicate pseudobulges.  Still, so much is different in 
their analysis that we cannot interpret the results.  For their combined 
(barred$+$unbarred) sample, they derive $f = 3.8^{+0.7}_{-0.6}$, roughly 
half of the canonical value.  For barred AGNs only, they get 
$f = 2.3^{+0.6}_{-0.5}$, and for unbarred AGNs, they get 
$f = 7.0^{+1.8}_{-1.4}$.  However, Xiao \etal (2011:~Fig.~8) find no 
significant differences in BH masses in barred and unbarred galaxies.}
%
%Absent this recalibration, it is premature to interpret differences in $f$ 
%values for different types of AGNs.
This work is in progress.

\hsize=15.0truecm  \hoffset=0.0truecm  \vsize=21.4truecm  \voffset=1.2truecm

\vs
\cl {\big\ARRed S1.2.~BH Masses from Single-Epoch Spectroscopy}\textBlack
\vsss\vskip -1pt

This method was pioneered by Dibai (1977, 1984) and is sometimes known as the 
Dibai~method (Bochkarev \& Gaskell 2009; Gaskell 2009).  Another name, 
photoionization modeling, has become a historical anachronism.  Early on, the 
radius~$r$ used in Equation S1 was calculated~by modeling the photoionization 
structure of the BLR (e.{\ts}g., Dibai 1977, 1984; Wandel \& Yahil 1985).  
Reverberation mapping showed that these model radii (Osterbrock \& Mathews 
1986) were too large by factors of $\sim$\ts10 (e.{\ts}g., Gaskell \& Sparke 
1986; Netzer \etal 1990).  Model radii have now been replaced with radii 
given by an observed correlation between $r$ (from reverberation 
mapping) and the continuum luminosity of the AGN (Kaspi \etal 2000, 2005; 
Bentz \etal 2006). Correcting for contributions to the luminosity from the 
underlying galaxy, $r \propto L^{0.519^{+0.063}_{-0.066}}$, where $L$ is the 
optical AGN luminosity conventionally measured at 5100 \AA\ (Bentz \etal 
2009a).  Luminosity and emission-line width together provide $M_\bullet$, 
after calibration to the reverberation-mapped AGNs.  An advantage 
of the technique is that it is inexpensive in telescope time.  A single 
spectrum yields a mass measurement.  We shall use the clumsy but accurate names 
``single-epoch spectroscopy mass''  or ``BH virial mass.''
\lineskip=-30pt \lineskiplimit=-30pt

\includegraphics{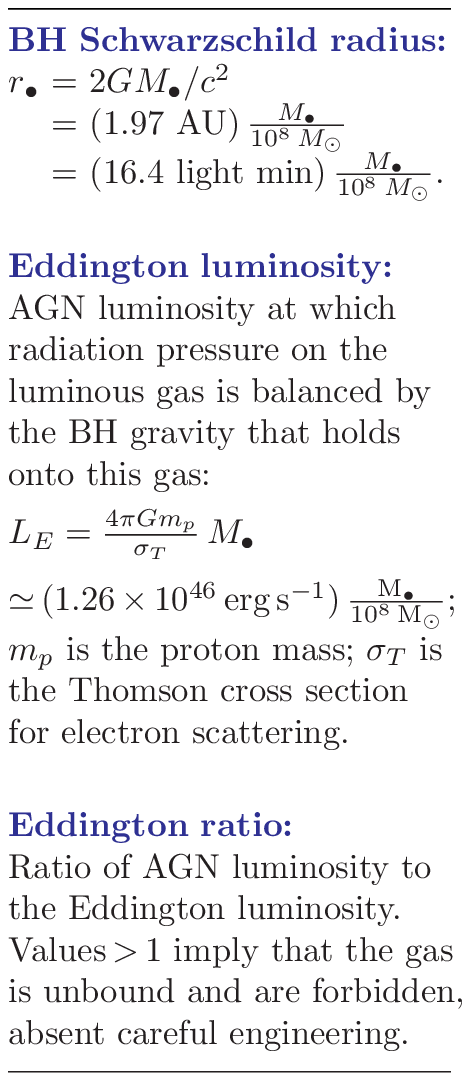}

%\special{psfile=schwarzschild.ps hoffset=373 voffset=-482 hscale=84 vscale=84}

In spite of its essentially empirical nature, single-epoch spectroscopy has 
become a popular tool to measure $M_\bullet$.  Whereas the most widely used 
mass estimator, especially at low redshifts $z \lesssim 0.75$, is based on 
the H$\beta$ line, a variety of alternatives using other lines have been 
developed.  Greene \& Ho (2005) advocate the intrinsically
stronger H$\alpha$ line as a better choice for lower-luminosity AGNs.  To help 
mitigate contamination from host-galaxy light, they further recommend using 
the H$\alpha$ line luminosity instead of the optical continuum luminosity 
to estimate the BLR size.  At intermediate redshifts, $0.75 \lesssim z 
\lesssim 2$, McLure \& Jarvis (2002) introduced a mass estimator based on  
Mg{\ts}{\sc II} $\lambda$2800, whereas at $z \gtrsim 2$, one often has to resort 
to C{\ts}{\sc IV} $\lambda$1549 (Vestergaard 2002).  These various mass 
estimators have been revised and updated by
Vestergaard \& Peterson (2006), 
Wang \etal (2009), 
Shen \etal (2011),
Xiao \etal (2011), 
and others.  Typical uncertainties are claimed to be $\sim 0.3-0.5$ dex.  The robustness 
of the rest-frame UV lines has been studied extensively (e.{\ts}g., 
McLure \& Dunlop 2004; 
Baskin \& Laor 2005; 
Vestergaard \& Peterson 2006; 
Netzer \etal 2007; 
McGill \etal 2008; 
Shen \etal 2008; 
Denney 2012; 
Ho \etal 2012; and
Shen \& Liu 2012).  
Mg{\ts}{\sc II} is widely viewed as an acceptable and unbiased 
surrogate for the Balmer lines, but many people regard C{\ts}{\sc IV} as potentially 
problematic.  Fortunately, the increasing availability of near-infrared 
spectroscopy obviates the need to rely on the UV lines.

The virial mass method is more indirect than reverberation mapping, and it 
contains fewer checks that we measure virialized motions.  We emphasize that 
application of single-epoch spectroscopy to quasars rests on the untested 
assumption that machinery which is calibrated for sub-Eddington BHs with 
$M_\bullet \sim 10^{7.5 \pm 1}$ $M_\odot$ still works for BHs with 
$M_\bullet \sim 10^{9 \pm 1}$ $M_\odot$ that radiate near the Eddington limit.  
In particular, work on quasars (e.{\ts}g., Section 8.6) requires us to 
extrapolate calibrations based on local AGNs of lower $L$, lower $M_\bullet$, 
and somewhat lower $L_{\rm bol}/L_{\rm Edd}$.  Croom (2011) argues that BLR 
line widths provide little additional information for quasars beyond what 
is given by the correlation of $M_\bullet$ with quasar luminosity.  More 
optimistically, Netzer \& Marziani (2010) provide a calibration of virial 
masses that takes radiation pressure into account.  Overall, the virial mass 
method is still controversial for quasars.  

At the other extreme, to study the demographics of low-mass BHs (Sections 7.1\ts--\ts7.2),
we need to extend the same mass-measurement formalism to $M_\bullet \approx 
10^5-10^6\,M_\odot$, $\sim 1-2$ orders of magnitude below the typical mass and 
luminosity of the reverberation-mapped AGNs.

\vs\vskip -1pt
\cl {\big\ARRed S1.3.~BH Masses from the X-Ray--Radio--M\lower.3ex\hbox{$\bullet$} ``Fundamental Plane''}\textBlack
\vsss\vskip -1pt

      Different physics controls accretion-disk radiation at different wavelengths. 
So it is better news than we might expect that the correlation between radio luminosity, 
X-ray luminosity, and $M_\bullet$ shows enough regularity so that it can be used -- albeit with large
uncertainties -- to estimate $M_\bullet$.  
The correlation between $L_R$, $L_X$, and $M_\bullet$ is called the ``fundamental plane of BH activity''
(e.{\ts}g., Merloni \etal 2003;
Falcke \etal 2004;
K\"ording 2008; 
Ho 2008;
Yuan \etal 2009; 
Hardcastle \etal 2009; 
G\"ultekin \etal 2009a). 
Here, we use this technique only in Section 7.3 on Henize 2-10.

\vfill\eject

\vs
\cl {\big\ARRed S2. CLASSIFICATION CRITERIA FOR PSEUDOBULGES}\textBlack
\vskip 3pt

      The observations that are used to distinguish between classical and pseudo bulges 
are listed below.~They are slightly refined from criteria listed in Kormendy \& Kennicutt 
(2004) based~on~new data (Kormendy \& Bender 2013).~Justifications are given in the
above reviews.~For all pseudobulges listed in {\bf Table 3}, 
at least two and as many as five classification criteria have been used.

      1 -- Pseudobulges often have disky morphology: their apparent flattening 
is similar to that of the outer disk, or they contain spiral structure all the way in to
the center.  Classical bulges are much rounder than their disks
unless the galaxy is almost face-on.  They cannot have spiral structure.

      2 -- In face-on galaxies, the presence of a nuclear bar shows that
a pseudobulge dominates the central light.  Bars are disk phenomena.  Triaxiality in
giant ellipticals involves different physics.  

      3 -- In edge-on galaxies, boxy bulges are edge-on bars; seeing one is sufficient 
grounds to identify a pseudobulge.  Boxy-nonrotating-core ellipticals (e.{\ts}g., 
Kormendy \etal 2009) cannot be confused with boxy, edge-on bars, because boxy bars rotate rapidly, 
whereas boxy ellipticals rotate slowly.

      4 -- Most pseudobulges have S\'ersic (1968) indices $n < 2$; almost all classical 
bulges have $n \geq 2$.  The processes that determine S\'ersic indices are not completely
understood, but the correlation of small $n$ with other pseudobulge indicators is so good 
that this has become a convenient classification criterion.  Note, however, that some 
pseudobulges do have S\'ersic indices as big as 4.

      5 -- Pseudobulges are slightly more rotation-dominated than classical bulges 
in the $V_{\rm max}/\sigma$ -- $\epsilon$ diagram; e.{\ts}g., $(V_{\rm max}/\sigma)^* > 1$. 
In two-dimensional velocity fields, pseudobulges generally appear as distinct, rapidly rotating,
and dynamically cold (see 6), central disk-like components.

      6 -- Many pseudobulges are low-$\sigma$ outliers in the Faber-Jackson 
(1976) correlation between (pseudo)bulge luminosity and velocity dispersion, or $\sigma$ 
decreases from the disk into the pseudobulge.

      7 -- If the center of the galaxy is dominated by Population I material (young stars, 
gas, and dust), but there is no sign of a merger in progress, then the bulge is at least
{\it mostly\/} pseudo.  

      8 -- Classical bulges fit the fundamental plane correlations for elliptical
galaxies.  Some pseudobulges do, too, and then these correlations are not helpful for
classification.  But extreme pseudobulges have larger effective radii $r_e$ and fainter 
effective surface brightnesses $\mu_e$ at $r_e$.  Also, Kormendy \& Bender (2012) found
some pseudobulges that are more compact than classical bulges of the same luminosity.  
These pseudobulges can be identified using fundamental plane correlations.

      9 -- Small bulge-to-total luminosity ratios do not guarantee that a bulge
is pseudo, but $B/T$\ts\gapprox\ts0.5 implies that the bulge is classical.

      We emphasize again that classifications are much more robust if they are based on many
criteria.

\vs 
\cl {\big\ARRed S3. CORRECTIONS TO PARAMETERS LISTED IN TABLES 2 AND 3}\textBlack
\vskip 3pt

      Papers that report BH detections necessarily put great effort into estimating uncertainties in BH 
masses $M_\bullet$ and related quantities such as bulge mass-to-light ratios.   
They~also~discuss~$\sigma$~because of the interest in the $M_\bullet$\ts--\ts$\sigma$ relation. 
They discuss galaxy absolute magnitudes in much~less~detail.  Often, little information is available 
on the provenance and reliability~of~photometry~and~bulge- disk decompositions.~However, results depend 
critically on the accuracy of bulge magnitudes. (Pseudo)bulge classification and bulge-disk decomposition 
are carried out in Kormendy \& Bender (2013).  Here, Section S3.1 focuses on corrections needed to get
accurate apparent magnitudes.  The sections that follow discuss corrections to other parameters.  
All these corrections are incorporated in {\bf Tables 2} and {\bf 3}, which are reproduced from 
the main paper in Section S4.

\vs 
\cl {\big\ARRed S3.1.~Corrections to 2MASS Apparent Galaxy Magnitudes}\textBlack
\vskip 3pt

      Almost all magnitudes used in this paper are $K_s$-band magnitudes (abbreviated as $K$) from 
the 2MASS sky survey (Skrutskie \etal 2006) and Large Galaxy Atlas (Jarrett \etal 2003). The reasons 
are well known: infrared magnitudes are less affected by dust absorption and young stars.  Both are 
special problems for pseudobulges; indeed, ongoing star formation is 
one of several pseudobulge classification criteria.  Luminosities are used as surrogates for 
stellar~masses; $K$-band mass-to-light ratios vary little and in a calibrated way with stellar 
population age and color (Section\ts6.6.1).  

      Integrated magnitudes and stellar masses for (pseudo)bulge components of disk galaxies require
detailed surface photometry and bulge-disk decompostion.  This work for all disk galaxies with dynamical BH
detections is reported in Kormendy \& Bender (2013).  Ellipticals are simpler: we use~total magnitudes from the
2MASS Large Galaxy Atlas and the online Extended Source Catalog.  These magnitudes are remarkably accurate
(Jarrett \etal 2003).  However, sanity checks reveal the need for small corrections in a few cases.  
These checks are based on $V$-band magnitudes, as follows:

\vs
\noindent{\bf \ARRed S3.1.1 Summary}\textBlack~A useful consistency check on $V$- and $K$-band 
apparent magnitudes is provided by the observation that galactic-absorption-corrected $(V - K)_0$ 
colors are exceedingly well behaved for almost all galaxies.  Exceptions are the most internally 
absorbed or starbursting galaxies, but these are not relevant here.~For other galaxies, we find a 
tight correlation between $(V - K)_0$ and $(B - V)_0$ for 0.6 \lapprox \ts$(B - V)_0$ \lapprox \ts1.05
({\bf Figure S1}).  Classical bulges and ellipticals have $(V - K)_0 \simeq 2.98$ with a total scatter 
(not a dispersion) of about $\pm 0.25$.  Both colors are tabulated in {\bf Tables\ts2}~and~{\bf 3}.  We use 
them to check the apparent magnitudes.  For a few, usually faint galaxies, discrepant colors suggest that 
the $V$ magnitudes in NED and Hyperleda are wrong,~usually by a few tenths of a magnitude.  For these, 
2MASS is more accurate, and we correct the $V$ magnitudes to make $(V - K)_0 = 2.98$.  More often -- 
usually for the largest galaxies on the sky -- the 2MASS magnitudes are the problem.  How we correct 
them is summarized here and discussed in detail in Section S3.1.2.

\vfill

\includegraphics{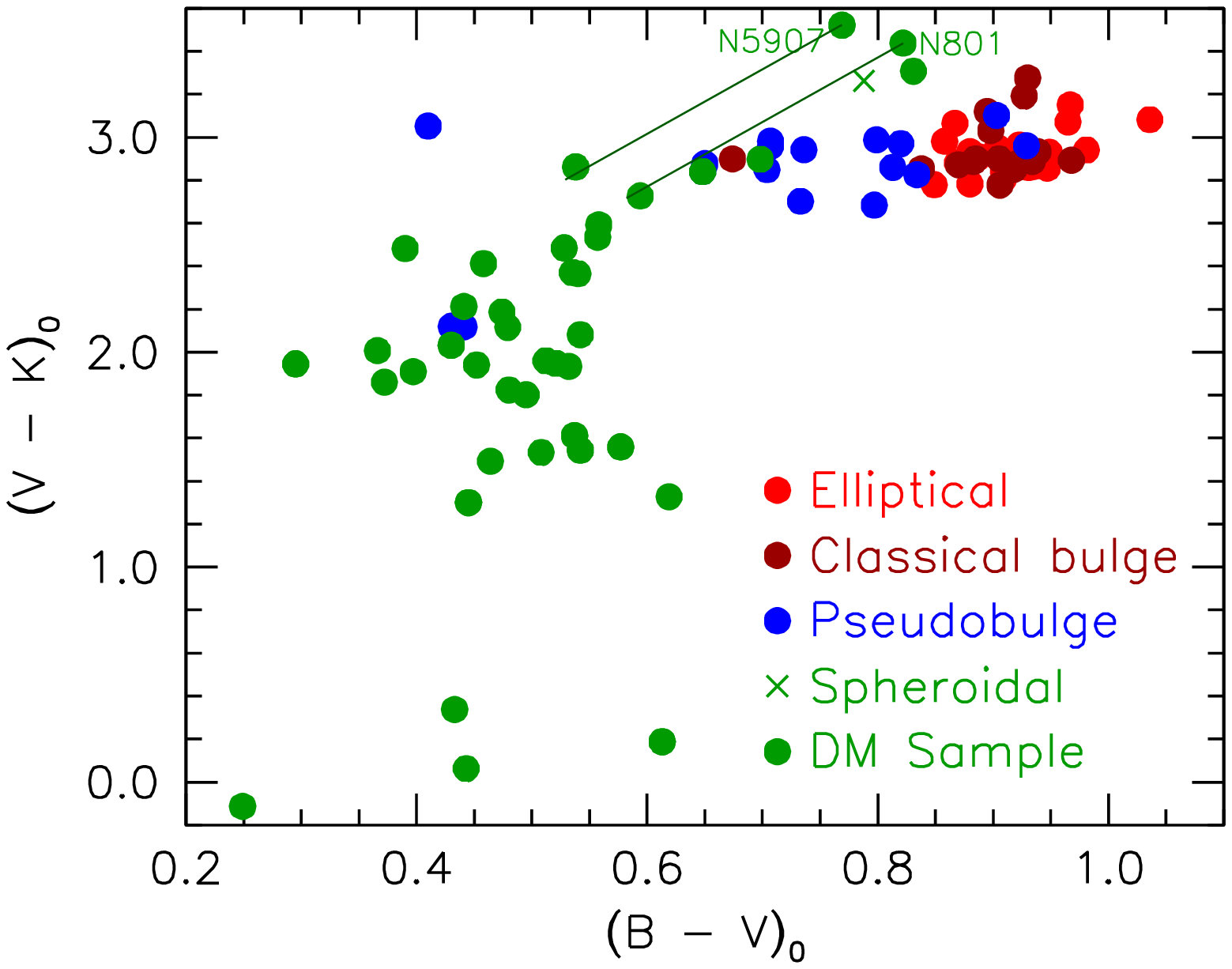}

\ni {\bf \textBlue Figure S1}\textBlack 

\vskip 1pt
\hrule width \hsize
\vskip 2pt

\ni Correlation between Galactic-reddening-corrected, total $(B - V)_0$ and $(V - K)_0$ colors as derived
by Kormendy \& Freeman (2013).  The ellipticals, classical bulges, and pseudobulges mostly are BH host
galaxies omitting the apparently largest galaxies (M{\ts}31 and NGC 4594) and Virgo ellipticals with
S\'ersic indices $n \gg 4$.  The filled green circles are the Sc{\ts}--{\ts}Im galaxies whose dark matter
halos are studied by Kormendy \& Freeman (2013); their large scatter at $(B - V)_0$ \lapprox \ts0.6 is
due to a combination -- details are not relevant here -- of magnitude errors, heterogeneous young stellar 
populations, and internal absorption.  The latter is illustrated for the edge-on galaxies
NGC 801 and NGC 5907 by straight lines that connect the observed color values to ones that are corrected
approximately for internal absorption.  The unlabeled green circle with $(V - K)_0 > 3$ is NGC 253; it
is very dusty and also needs correction for internal absorption.  The cross is for NGC 205; it probably
has an underestimated $V$-band luminosity.  The take-away message for this paper is that 
$(V - K)_0$ colors are well behaved for old stellar populations.  Therefore, when we observe a
color that is inconsistent with the illustrated correlation (usually by being too blue), and when we
trust the $V$-band magnitude, we correct the $K$-band magnitude to be consistent with the correlation.

\eject

      The 2MASS photometric system is very accurate; Jarrett \etal (2003) state 
that the photometric zero-point calibration is accurate to 2\ts\% -- 3\ts\% across the sky.  To the extent
that we have been able to check them (Kormendy \& Bender 2013), their integrated magnitudes are correspondingly
accurate at least within the radii out to which they have data.  The survey is somewhat shallow; the 1-sigma
sky noise is 20.0 mag arcsec$^{-2}$ in $K$, although profiles can be derived fainter than this by
averaging over many pixels.  When there is a problem, it is with the extrapolation to total magnitudes, called
$K_{\rm tot}$ in Jarrett \etal (2003) and {\tt k\_m\_ext} in the online catalog.  The extrapolation is made by
fitting a S\'ersic function to the parts of the profile that are relatively safe from noise and from the PSF
and then integrating the extrapolated function out to ``about four disk scale lengths.''  This procedure 
works best for disk galaxies, i.{\ts}e., ones that have nearly exponential outer profiles.  For ellipticals,
the radial range is too small to yield an accurate S\'ersic index (Section A2 in Kormendy \etal 2009), and 
their tabulated S\'ersic indices are too small.  The result is to underestimate the total brightnesses of 
ellipticals, particularly giant ellipticals that have $n \gg 4$.  The derivation of this conclusion and how 
we correct for it are the subjects of this section.  

      {\bf Figure S1} shows that, when our BH hosts are \underbar{not} very large on the sky and when
they \underbar{do not} have $n \gg 4$, then their $(V - K)_0$ colors are well enough behaved so that 
we can use them and the $V$ magnitudes to correct the $K$ magnitudes in problem cases.  We are 
conservative in making this correction -- we  make it only when we observe a color 
$(V - K)_0 < 2.75$ that is much too blue.  That is, if the correction is less than 0.230 mag, 
we do not use it.  Also, in some marginal cases, it is not clear whether the $V$ or the $K$ magnitude is 
the problem. In all these cases, we use the total magnitude from 2MASS, even for ellipticals.  In fact, 
so little is known about some of the more distant BH hosts that the 2MASS magnitudes are much more accurate 
than any available $V$-band measurements.  Finally, in a few cases, our own photometry yields a composite, 
$K$-band profile with 2MASS zeropoint whose integral is more accurate than the 2MASS result because our 
measurements reach out to much larger radii.  We adopt these total magnitudes also.

      The largest correction, $\Delta K = -0.411$, is derived for M{\ts}31.  Jarrett \etal (2003) warn us
that ``The Andromeda result should be viewed with caution, as we are not fully confident that the total 
flux of M{\ts}31 has been captured, because of the extreme angular extent of the galaxy and the associated 
difficulty with removing the infrared background.''  This problem happens in $V$ band, too.  We adopt 
$V_T = 3.47 \pm 0.03$ from Walterbos \& Kennicutt (1987; the error estimate is optimistic).  Then the 
2MASS magnitude $K = 0.984$ implies that $(V - K)_0 = 2.335$.  This is much too blue for an Sb galaxy 
with $(B - V)_0 = 0.87$ (see {\bf Figures S1} and {\bf S2}).  As in other such cases, we correct 
$K \rightarrow K - 0.411$ to make $(V - K)_0 = 2.98$ for the bulge only.  The $V$-band bulge magnitude 
is determined from the $V$-band $B/T = 0.25 \pm 0.01$.  The correction gives us a $K$-band magnitude for 
the bulge only.  The infrared $B/T = 0.31 \pm 0.01$ then gives us the total observed magnitude of the
galaxy, $K_T = 0.573$.  Including both the red bulge and the bluer disk, the total colors of the galaxy 
are $(B - V)_0 = 0.87$ and $(V - K)_0 = 2.75$ (cf.~{\bf Figure S1}).

      In summary, 2MASS quoted errors on $K$-band magnitudes, usually 0.02\ts--\ts0.04~mag, are most 
reliable for disk galaxies but can be too optimistic for giant Es.  We believe that our corrected 
magnitudes are generally accurate to $\sim 0.1$ mag, although the values for M{\ts}31 and for galaxies 
whose $(V - K)_0$ colors are very different from 2.98 are more uncertain.  We quote $K$ to higher precision 
because we do not wish to lose precision in arithmetic.  

\vs
\noindent{\bf \ARRed S3.1.2 Detailed Discussion: Dependence of $K$ Magnitude Corrections on
                            S\'ersic Index}\textBlack

      Two circumstances make it feasible to check $K$-band magnitudes using $V$-band magnitudes.  
The first is the tight correlation between $(B - V)_0$ and $(V - K)_0$ shown in {\bf Figure S1}.  
Second, we have $V$-band total magnitudes based on detailed surface photometry for all Virgo cluster 
ellipticals from Kormendy \etal (2009, hereafter KFCB).  These are more thoroughly checked and should 
be more accurate than other photometry in the literature.  KFCB conservatively estimate errors in 
total magnitudes as $\pm 0.07$ mag for ellipticals with S\'ersic (1968) index $n < 4$ and $\pm 0.10$ 
for those with $n > 4$.  The agreement with Hyperleda is actually better than this.  KFCB paid 
particular attention to the extrapolation of $V_T$ to infinite radius, so these data allow us
to look for faint outer light that may have been missed by 2MASS.  How we do this is illustrated 
in {\bf Figures S2} and {\bf S3}.

      {\bf Figure S2} correlates $(V - K)_0$ color with $(B - V)_0$ color and galaxy apparent magnitude~$V_T$. 
Here, $K_T$ is from 2MASS and $V_T$ is from KFCB for the green points and, in order of preference, from 
Kormendy \& Bender (2013), Hyperleda, or RC3 otherwise.  The right-hand panels are designed to 
compare $(V -K)_0$ to a raw observational parameter that should contain little physics but 
that can easily correlate with measurement errors. But first, we need to check whether $(V - K)_0$ 
correlates with stellar population age using the more sensitive indicator $(B - V)_0$ in the left panels.  
Ellipticals have only a small range in $(B - V)_0$ color (panel\ts{\it a}).  Bulges show a large range 
(panel\ts{\it c}), mostly in pseudobulges, and bluer pseudobulges live in galaxies that are bluer overall 
(Drory~\&~Fisher~2007).  Over this wide range in $(B - V)_0$, (pseudo)bulges show remarkably little 
variation in $(V - K)_0$, consistent with the shallow correlation shown at 
0.6 \lapprox \ts$(B - V)_0$ \lapprox \ts1.05 in {\bf Figure{\ts}S1}.  In contrast, ellipticals have 
a large range in $(V - K)_0$, and this results from the fact that $(V - K)_0$ correlates with $V_T$ 
(panel {\it b}). Note that, unlike {\bf Figure S1}, {\bf Figure S2} does not omit large-S\'ersic-index 
($n \gg 4$) galaxies.  This suggests that the large range in $(V - K)_0$ color for ellipticals is not 
a stellar population effect but rather is a result of a measurement problem.  

      KFCB total magnitudes $V_T$ were tested thoroughly and agree with Hyperleda to 0.06 mag with a slight
dependence on S\'ersic index $n$.  We therefore conclude that the correlation of the green points in 
{\bf Figure S2\it b} signals a problem with the $K$ magnitudes.  The black points for BH host galaxies share
this correlation with almost the same scatter:~the problem is not $V$ for them, either.  
As noted above, 2MASS could have trouble with sky subtraction for large galaxy images 
or trouble in adding up the faint outer light in very shallow brightness profiles.  Importantly, the 
problem is not shared by disk galaxies: panel {\it d\/} shows no significant correlation.  Because disks
have steep outer profiles ($n \simeq 1$) whereas bright ellipticals have shallow profiles
($n > 4$), the hint is that the problem is associated with profile shape.   This is confirmed in 
{\bf Figure S3}.  

      The exceptions among bulges are M{\ts}31 and the Sombrero Galaxy, NGC 4594.  In both galaxies,
$(B - V)_0$ is normal, but $(V - K)_0$ is very discrepant.  The problem with M{\ts}31 is its large
apparent size (see Section S3.1.1).  The problem with NGC 4594 is that it, too, has a large apparent
diameter, and moreover, it is bulge-dominated with large $n$ at large radii.  We therefore correct the
total $K$ magnitudes for these two galaxies.

      Returning to the ellipticals, we need to further investigate the correlation in 
{\bf Figure S1\it b} in order to decide how we should correct their $K_T$ magnitudes.  If the problem is 
mainly galaxy size as measured by $V_T$, then we should assume that all bright-$V_T$ galaxies need correction,
and we should carry out that correction using the fit shown by the green line in panel {\it b\/}.  On the 
other hand, if the problem is associated with S\'ersic index, then we should use either $n$ or the discrepant 
$(V - K)_0$ color but not $V_T$ to determine which galaxies to correct.

      If profile shape $n$ affects the measurement of $K_T$ and hence $(V - K)_0$, then we might expect 
that $n$ is a ``second parameter'' which controls the scatter about the correlation in {\bf Figure S2\it b\/}.
{\bf Figure~S3\it a\/} confirms that this is the case.  Largely independent of $V_T$ (which is encoded in 
symbol size), deviations $\Delta(V - K)_0$ from the correlation in {\bf Figure S2\it b} correlate with $n$.  
Galaxies with larger $n$ (that is, shallower outer profiles) have more negative deviations, i.{\ts}e., 
smaller $(V - K)_0$.  Therefore, if $V$ magnitudes are not a problem and if all bulges have approximately
the same color, then more negative $\Delta(V - K)_0$ implies that the $K_T$ magnitudes are too faint.
There are a few exceptions.  Two of these galaxies are faint and difficult for 2MASS.  In particular,
VCC 1627 is the faintest known elliptical in Virgo; it is less luminous than M{\ts}32.
The point for NGC 4636 is correct; we checked the 2MASS and KFCB magnitudes.  The KFCB and Hyperleda total 
magnitudes $V_T$ agree to 0.01 mag.  So a few exceptions are real.  But by and large, our suspicions are 
confirmed: the problem with $K_T$ magnitudes happens mostly for large-$n$ galaxies.

% 3 more lines on this page.

\vfill\eject

\cl{\null} \vfill

\vskip 11.2truecm

%\special{psfile=VK.ps hoffset=-96 voffset=-262 hscale=90   vscale=90}

\includegraphics{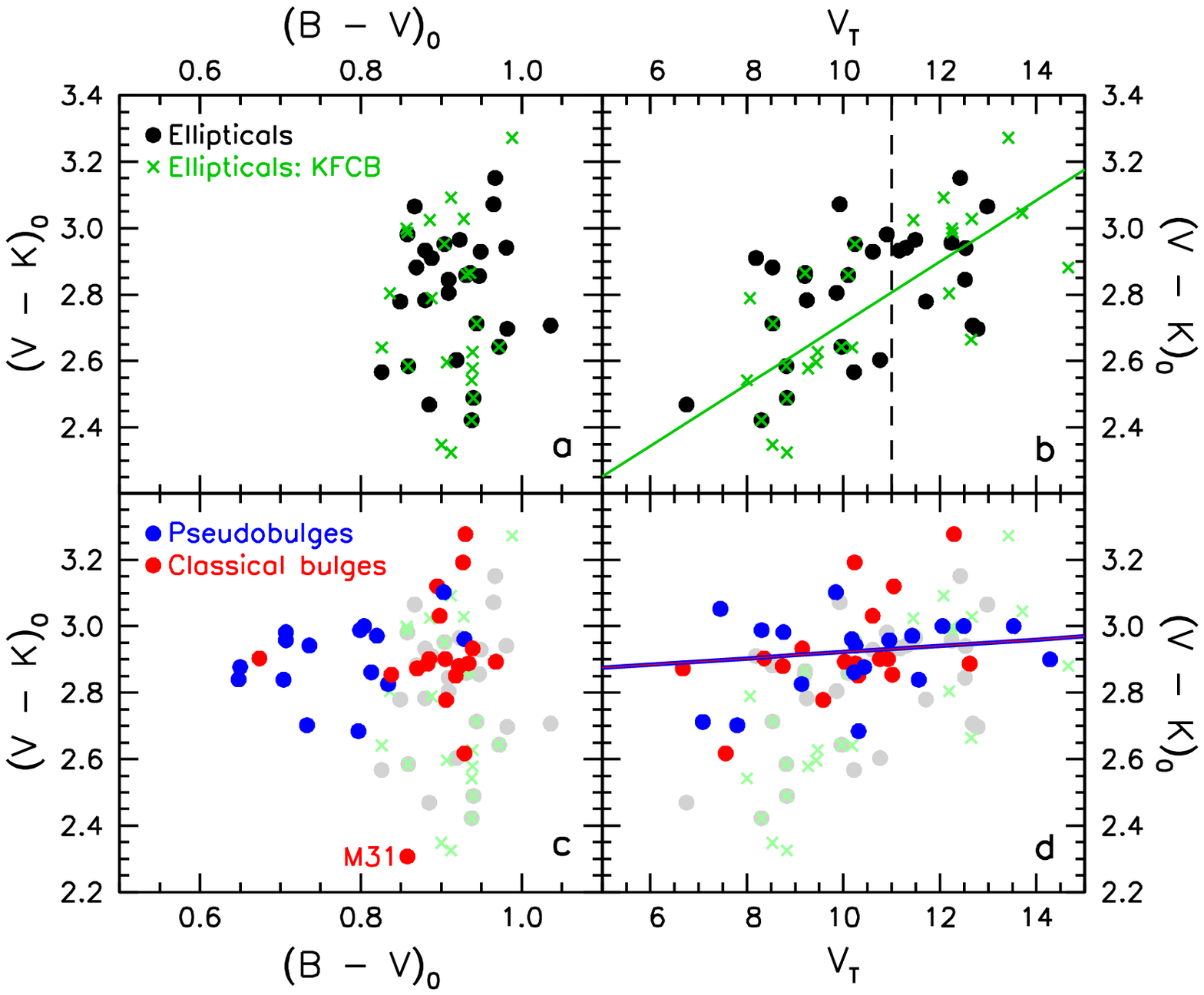}

\ni {\bf \textBlue Figure S2}\textBlack 

\vskip 1pt
\hrule width \hsize
\vskip 2pt
\ni Correlations of total galaxy $(V - K)_0$ color with ({\it left\/}) total $(B - V)_0$ color and 
({\it right\/}) total apparent magnitude $V_T$.  Subscript 0 means that colors are corrected for Galactic 
absorption (Schlegel \etal 1998).  The upper panels are for ellipticals; the lower panels are for classical 
and pseudo bulges.  The green points are for galaxies from Kormendy \etal (2009: KFCB).  Other points are 
for BH host galaxies, eight of which are also in KFCB.  The ellipticals are repeated in ghostly light colors 
in the lower panels to facilitate comparison.  In panel {\it b\/}, the green line is a least-squares fit
to the KFCB points assuming negligible errors in $V_T$.  The RMS deviations of the green points about this 
line and the RMS deviations of the black points about a similar fit ({\it not illustrated\/}) to the BH Es 
are both 0.16 mag.  These deviations correlate with galaxy S\'ersic index ({\bf Figure~S3\it a}).  For this 
reason, and because $(V - K)_0$ does not vary systematically over the small $(B - V)_0$ color range shown, 
we argue that both the panel {\it b\/} trend with $V_T$ and much of the scatter about this trend result 
either from sky subtraction errors in the apparently largest galaxies or from errors in adding up faint 
surface brightnesses over large areas in ellipticals with shallow profiles (large $n$).  {\bf Figure S3} 
shows that $n$ is the primary factor.  We therefore correct the 2MASS $K_T$ magnitudes for galaxies with 
$(V - K)_0 < 2.75$ to the mean color for KFCB galaxies with $V_T > 11$ ({\it vertical dashed line\/}), 
$<$\null$(V - K)_0$\null$>$ =  $2.98 \pm 0.05$.  In panel {\it d\/}, the red and blue line is a least-squares
fit to the classical and pseudo bulges; RMS = 0.13 mag.  They show no significant trend of $(V - K)_0$ 
with $(B - V)_0$ or $V_T$ and have the same mean $(V - K)_0$ color as the faint ellipticals.  In the 
lower panels, only the brightest galaxy M{\ts}31 ($V_T = 3.5$) and the bright, bulge-dominated galaxy
NGC 4594 have very anomalously blue colors.  We correct only their $K$-band magnitudes to make 
$(V - K)_0 = 2.98$.

\vskip 6pt

\vfill\eject

\cl{\null}

\vskip 6.8truecm

%\special{psfile=DVKn.ps  hoffset=0 voffset=-90 hscale=50 vscale=50}

\includegraphics{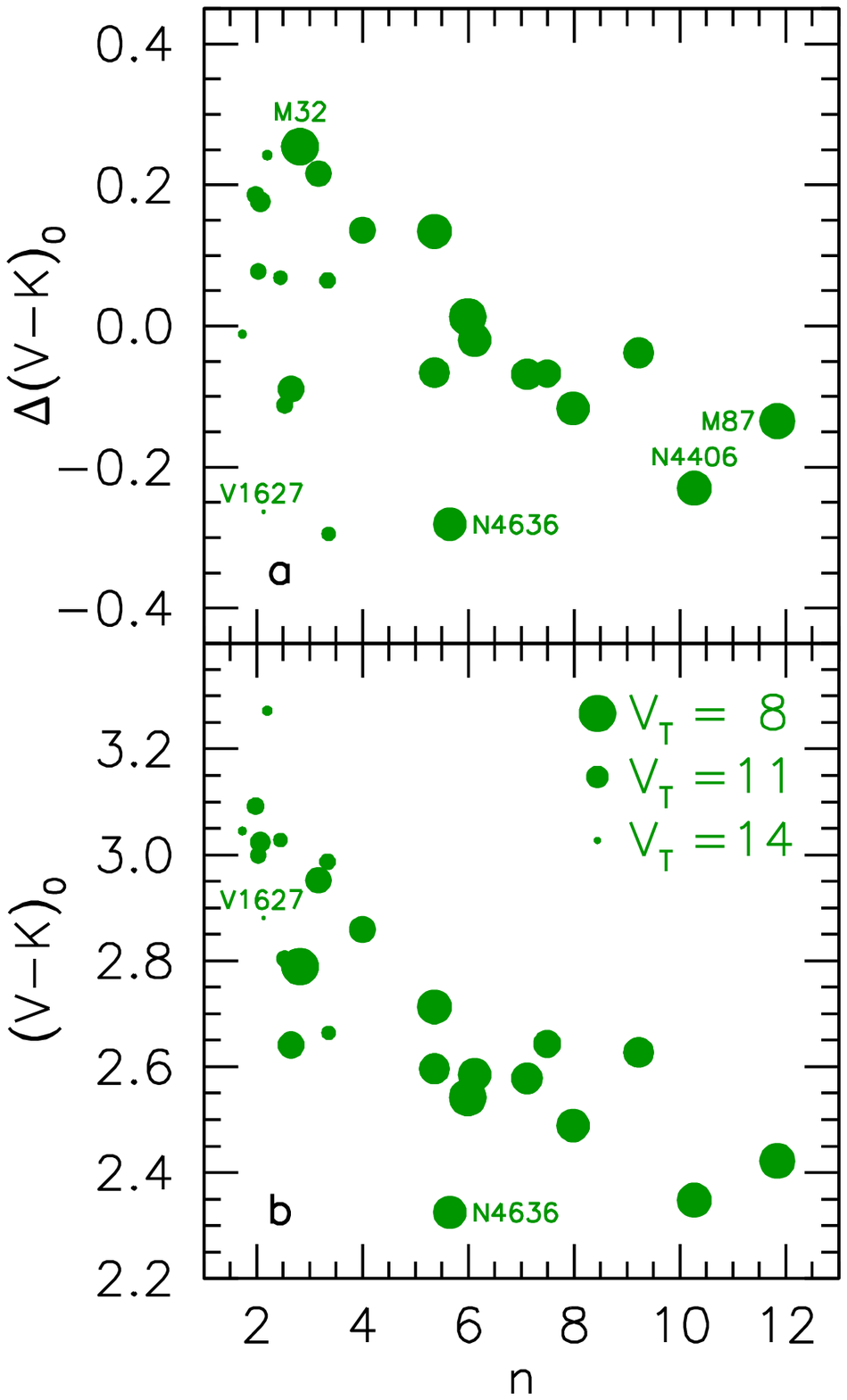}

\cl{\null}

\vskip -8.0truecm

\parindent=230pt
\def\hi{\indent\hangindent=230pt}
\hi {\bf \textBlue Figure S3}\textBlack 
                             \vskip -6pt

\hi{---------------------------------------------------------}

                             \vskip -5pt
\hi ({\it a\/}) Correlation of the $(V - K)_0$ deviations of the green points from the green~line~in
    {\bf Figure S2\it b\/} versus the S\'ersic index $n$ of the elliptical
    galaxy's brightness profile.  Point sizes are linearly proportional to apparent 
    magnitude $V_T$ (see~the~key).  Some points are labeled with galaxy~names.
    With few exceptions (mostly faint galaxies~but~also NGC\ts4636),
    $K$ brightnesses~are underestimated with respect to the correlation -- i.{\ts}e.,
    $\Delta(V - K)_0$ is negative -- by larger amounts for ellipticals that have
    shallower outer brightness profiles (larger $n$). This suggests that $n$ is the
    main factor that controls the \hbox{$(V - K)_0$\ts--\ts$V_T$} correlation in {\bf Figure{\ts}S2\it b}.~This 
    is confirmed here in panel~{\it b\/}.  Even~the~tiniest~Virgo~elliptical, VCC 1627,
    which is an exception in {\it a\/}, participates in the strong correlation~in {\it b}.
    Thus $n$ and not image area (represented by $V_T$) is the
    primary factor that controls the $K$-band magnitude errors.  We therefore correct
    $K_T$ and $M_K$ for BH ellipticals in {\bf Figure S2} that have $(V - K)_0 < 2.75$ until
    their colors agree with the mean color $<$\null$(V - K)_0$\null$>$ = $2.98 \pm 0.05$ for bright 
    ellipticals in {\bf Figure S2{\it b}} or for ellipticals with small $n$ in panel {\it b\/} here.

\parindent=10pt

\vskip 30pt

     Large-$n$ galaxies are also the most luminous galaxies (KFCB).  The range of distances represented in
{\bf Figure S2} is not very large; among green points, it is especially small.  Therefore $V_T$ is not too
different from $M_{VT}$.  This opens up the possibility that the whole correlation in {\bf Figure~S2\it b} and
not just the scatter is caused mainly by the 2MASS response to different $n$.  {\bf Figure S3\it b} confirms
that this is the case.  There is a very good, nonlinear correlation between $(V - K)_0$ and~$n$ such that
galaxies with larger $n$ have $K$-band luminosities that are more underestimated.  This correlation
has substantially smaller scatter than the one in {\bf Figure S2\it b}.  VCC 1627 is no longer discrepant.
We emphasize: 2MASS photometry of apparently fainter galaxies is usually excellent.  The errors uncovered 
in {\bf Figures~S2} and {\bf S3} are mostly $< 0.2$ to 0.5 mag, and they apply mainly to the biggest giant 
ellipticals and to M{\ts}31.  Painstaking, deep photometry was required in KFCB to measure the outer profiles 
of Virgo giant ellipticals.  It is not surprising that the somewhat shallow 2MASS survey misses some outer 
light in these large and shallow-surface-brightness-gradient galaxies.

      A related, more pessimistic discussion of 2MASS galaxy magnitude errors is in Schombert\ts(2011).

      The above results suggest a practical strategy for correcting the $K$-band magnitudes on which many
of our results depend.  We cannot make corrections based on S\'ersic indices, because these
have not reliably been measured for most BH ellipticals.  But the black and green points behave very
similarly in {\bf Figure S2}.  The faint, non-Virgo BH ellipticals that have anomalously blue colors also
are giant ellipticals.  We therefore correct $K_T$ for ellipticals that have $(V - K)_0 \leq 2.75$ to make
their resulting colors equal the mean color $<$\null$(V - K)_0$\null$>$ = $2.98 \pm 0.05$ for the faint KFCB 
ellipticals in {\bf Figure S2}.  That is, we correct ellipticals if their colors are ``too blue'' by 
$\sim$\ts1.5 sigma or more.  These corrections are included in {\bf Tables 2} and {\bf 3} in the main paper 
and below.

\vs
\cl {\big\ARRed S3.2.~Corrections to Stellar Dynamical BH Masses for Core Galaxies}\textBlack
\vs

Gebhardt \& Thomas (2009),
Shen \& Gebhardt (2010),
Schulze \& Gebhardt (2011), and
Rusli \etal (2013)
demonstrate convincingly that stellar dynamical mass models that do not include dark matter halos 
underestimate $M_\bullet$ by factors that can be as large as 2 or more when the BH sphere of influence is not
well resolved.  The effect is small for coreless galaxies; we neglect it when models that include 
dark matter are not available.  Fortunately, almost all stellar dynamical $M_\bullet$ estimates for
core galaxies are now based on models that include dark matter.  For one exception, NGC 5576, 
we correct $M_\bullet$ as discussed in the notes on individual objects.  For NGC 3607, the correction 
is too uncertain and we omit the object (table notes; orange point in {\bf Figure 12}).

\def\nhi1{\indent \hangindent=0.8truecm}

\vs\vskip 1pt
\cl {\big\ARRed S3.3. Corrections to Effective Velocity Dispersions}\textBlack
\vs

      This section enlarges on Section 5.1 in the main paper.  {\bf Figure 11} is reproduced from there.

       The velocity dispersion $\sigma_e$ that we correlate with $M_\bullet$ is more heterogeneously 
defined in different papers and less consistently measured within these definitions than we usually suppose. 
This is a stealth ``can of worms'' that could be a bigger problem than the more obvious uncertainties 
in BH mass measurements that preoccupy authors.  We check values when we can and fix a few problems.  
But the necessary data are not available for all objects.  Fortunately, we can show that this problem 
is not too severe.

      The worries are these: \vs\vskip -1pt

\nhi1 1.~{\it A priori}, we do not know how best to define $\sigma_e$ so that we learn important physics 
         from the $M_\bullet$\ts--\ts$\sigma_e$ correlation.  Clearly we should not include data at 
         $r$ \lapprox \ts$r_{\rm infl}$ in the average.  But inside what fraction of $r_e$ should we 
         average $\sigma(r)$?  We usually claim that we average inside $r_e$ and call the result $\sigma_e$.  
         However:  \vs

\nhi1 2.~Accurate values of $r_e$ are known for only a few galaxies.  KFCB demonstrate via high-dynamic-range 
         photometry that brightness profiles of giant ellipticals extend farther out than we have thought.  
         The $r_e$ values derived in KFCB are more accurate than previous results, and they are larger 
         than previous values for almost all giant ellipticals.  We use them here.  But we do not have 
         such data for most BH hosts.  It is safe to assume that the $r_e$ values in common use are too small.
         For bulges, the situation has been worse.  Accurate decompositions have been available for only a few 
         galaxies,  As part of the writing of this review, they are being derived for the remaining disk-galaxy
         BH hosts in Kormendy \& Bender (2013). \vs
 
\nhi1 3.~Different fractions of $r_e$ are used by different authors.  For example, Ferrarese \& Merritt (2000) 
         use $r_e/8$, whereas the Nuker team uses $r_e$. \vs
 
\nhi1 4.~Different authors perform the radial averaging differently.  We follow the Nuker team practice 
         (e.{\ts}g., Pinkney \etal 2003; G\"ultekin~et~al.~2009c) and use \hbox{$\sigma_e^2$ = average of 
         $V(r)^2 + \sigma(r)^2$}, weighting by $I(r) dr$.  Usually, we adopt $\sigma_e$ from the 
         $M_\bullet$ source paper or from G\"ultekin \etal (2009c).  When we calculate it, we perform 
         the average inside $r_e/2$, because our $r_e$ values tend to be larger than those in the
         literature, and we wish to be consistent with published $\sigma_e$ values.
         When we recalculate $\sigma_e$, we include a comment in the notes on individual objects.  \vs

\nhi1 $\phantom{5.}$~However (Section 5.1 of the main paper), our practice can be 
         contrasted with $\sigma_e$ values from the SAURON team: They add spectra that sample the 
         galaxy in two dimensions inside $r_e$ or inside the SAURON field, whichever is smaller 
        (Emsellem \etal 2007).  
         Because $\sigma$ and $V$ rather than $\sigma^2 + V^2$ are averaged and because the 
         weighting of different radii is essentially by $2 \pi r I(r) dr$ rather than by $I(r) dr$, 
         the resulting $\sigma_e$ values are smaller than the ones that we use.  {\it No study has 
         proved that the Nuker definition produces a tighter $M_\bullet$\ts--\ts$\sigma_e$ correlation 
         or that a tighter correlation is more meaningful physically.}  However, because it is most
         commonly used and therefore most widely available, we use the Nuker definition here.  \vs

      In view of these concerns, it is prudent to check how much our results might depend on the 
definition of $\sigma_e$ or on how well its measurement follows that definition.  {\bf Figure 11} 
(reproduced from the main paper for the convenience of readers) compares our $\sigma_e$ values with central 
velocity dispersions tabulated in HyperLeda and with $\sigma_e$ values calculated as described above 
by the SAURON/ATLAS3D team.  We conclude that adopting either alternative would have minimal effect on the 
scatter in the $M_\bullet$\ts--\ts$\sigma_e$ correlation and essentially no effect on qualitative conclusions.

      HyperLeda includes a few bad measurements.  These are particularly expected for pseudobulges, which 
can have small dispersions that are undersampled by low wavelength resolution.  Pseudobulge dispersion
measurements can also have serious problems with dust and star formation.  But it is relatively easy 
to find and discard these problems and get improved central $\sigma$ values that should agree with our 
$\sigma_e$ values as well as do the velocity dispersions of classical bulges and ellipticals.

      SAURON/ATLAS3D $\sigma_e$ values mostly agree very well with ours.  Some SAURON values are smaller, 
as expected.  But conclusions would not be changed if we had SAURON $\sigma_e$ values for all galaxies. 

      {\it Important:} The average shift 
$\Delta \log{\sigma_e} = \log ({\rm Nuker~\sigma_e}) - \log ({\rm SAURON~\sigma_e}) = 0.0299$
or a factor of 1.07 will be relevant in Section 8 when we compare the 
$z \simeq 0$ $M_\bullet$\ts--\ts$\sigma_e$ relation with ones derived for galaxies at large redshifts.  
All high-$z$ observations necessarily add spectra (not $V^2 + \sigma^2$ values) inside apertures that are large 
in kpc.  The SAURON $\sigma_e$ values are the closest match at $z \simeq 0$. {\it We~therefore use the above factor 
to correct our least-squares fit to the local $M_\bullet$\ts--\ts$\sigma_e$ correlation when we make comparisons
to high-$z$ objects.}

\cl{\null}

\vfill

% \special{psfile=Figure11.ps hoffset=-170  voffset=-490  hscale=115  vscale=115}

  \includegraphics{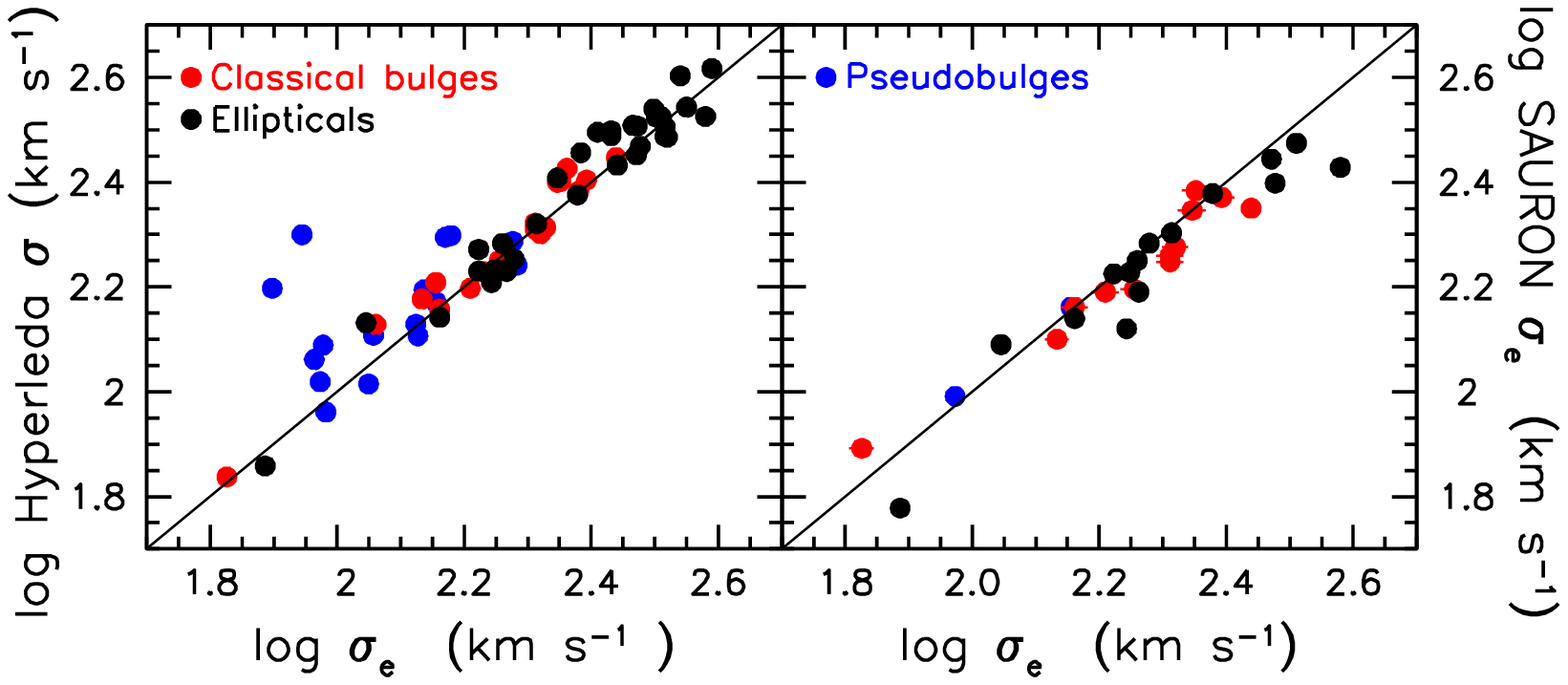}

\ni {\bf \textBlue Figure 11}\textBlack 

\vskip 1pt
\hrule width \hsize
\vskip 2pt

\ni Comparison of our adopted $\sigma_e$ ({\bf Tables 2} and {\bf 3}) to ({\it left}) the central 
    velocity dispersion as tabulated in HyperLeda and to ({\it right\/}) the $\sigma_e$ value tabulated 
    by the SAURON/ATLAS3D teams (Emsellem \etal 2007; Cappellari \etal 2013).  The straight
    lines are not fits; they indicate equality.  Given our definition, it is reassuring that our 
    $\sigma_e$ is approximately the geometric mean of the HyperLeda central and SAURON/ATLAS3D $\sigma_e$
    values.

\vs

\eject.

\hsize=16.99truecm  \hoffset=-0.5truecm  \vsize=25.5truecm  \voffset=-0.5truecm

\vs
\cl {\big\ARRed S4. BH DATABASE}\textBlack
\vs\vskip -2pt

      This section reproduces essentially verbatim the first part of Section 5 in the main paper.  Its 
purpose is to include the notes to {\bf Tables 2} and {\bf 3}.  For the convenience of readers, we do
not fragment the discussion by including the tables only in the main text and the notes (except for brief
examples) only here.  Rather, we include all of main paper Section 5 to the end of the table notes. 

      This section is an inventory of galaxies that have BH detections and $M_\bullet$ measurements based on 
spatially resolved stellar dynamics, ionized gas dynamics, CO molecular gas disk dynamics, or maser disk dynamics.
{\bf Table 2\/} lists ellipticals, including mergers-in-progress that have not yet relaxed into equilibrium. 
{\bf Table 3\/} lists disk galaxies with classical bulges (upper part of table) and pseudobulges (lower 
part of table).  The demographic results discussed in Sections 6\ts--\ts9 of the main paper are based on 
these tables.  Both tables are provided in machine-readable form in the electronic edition of this paper.

      We review the $M_\bullet$ measurements in Sections 2 and 3 of the main paper.~This Supplemental Material~provides 
more detail.  It includes all the notes on individual objects in Sections S4.1\ts--{\ts}S4.4.   
Derivations of host galaxy properties are relatively straightforward for ellipticals.  They are more complicated for 
disk-galaxy hosts, because (pseudo)bulge classification is crucial and because (pseudo)bulge{\ts}--{\ts}disk 
photometric decomposition is necessary.  This work is too long to fit here; it is published in a satellite
paper written in parallel with this review (Kormendy \& Bender 2013).  Some details are repeated here 
in the table notes for the convenience of readers.  

      Implicit in the tables are decisions about which published $M_\bullet$ measurements are reliable 
enough for inclusion. No clearcut, objective dividing line separates reliable and questionable measurements.  
Our decisions are personal judgments.  Our criteria are similar to those in G\"ultekin \etal (2009c); 
when we made a different decision, this is explained in the notes on individual objects.  We try to be 
conservative.~With a few exceptions that are not included in correlation fits, stellar dynamical masses 
are retained only if they are based on three-integral models.  Nevertheless (Section 3), it is likely that 
systematic errors -- e.{\ts}g., due to the neglect of triaxiality in giant ellipticals -- are still 
present in some data.  For this reason, we do not discuss correlation scatter in much detail.  We~do, 
however, include the important conclusion that the intrinsic scatter in the correlation of $M_\bullet$
with bulge stellar mass is as small as the intrinsic scatter in the correlation of $M_\bullet$ with $\sigma$.  
And we derive the most accurate correlations that we can with present data (Section 6.6).

      The sources of the adopted $M_\bullet$ measurements are given in the last column of each table, and 
earlier measurements are discussed in the notes on individual objects.  The $M_\bullet$ error bars 
present a problem, because different authors present error bars with different confidence intervals.  For
consistency, we use approximate one-sigma standard deviations, i.{\ts}e., 68\ts\% confidence intervals.
When authors quote two-sigma or three-sigma errors, we follow Krajnovi\'c \etal (2009) and estimate that
one-sigma errors are $N$ times smaller than $N$-sigma errors.  Flags in Column 12 of {\bf  Table\ts2} and 
Column 19 of {\bf Table 3} encode the method used to determine $M_\bullet$, whether the galaxy has a core 
(ellipticals only), and whether $M_\bullet$ was derived with models that include triaxiality or dark matter 
and large orbit libraries.  Only the models of M{\ts}32, NGC\ts1277, and NGC\ts3998 include all three.

      We use a distance scale (Column 3) based mainly on surface brightness fluctuation measurements at 
small distances and on the WMAP 5-year cosmology at  large distances 
($H_0$\ts=\ts$70.5${\ts}km{\ts}s$^{-1}${\ts}Mpc$^{-1}$; Komatsu \etal 2009).  Details are in the notes 
that follow {\bf Tables 2} and {\bf 3}.  Velocity dispersions~$\sigma_e$ (Colums 11 and 17 of
{\bf Tables 2} and {\bf 3}) are problematic; they are discussed in Section S3.3.

      We discuss luminosity correlations only in terms of $K_s$-band absolute magnitudes.  However,
we also provide $V$ magnitudes for the convenience of readers and because we use them to check~the~$K_s$
magnitudes.  Readers should view $(B - V)_0$ as a galaxy color that contains physical information but 
$(V$\null$-$\null$K_s)_0$ mainly as a sanity check of the independent $V$ and $K_s$ magnitude systems
(Section S3.1).  Our $K_s$ magnitudes are on the photometric system of the Two Micron All Sky Survey 
(2MASS: Skrutskie \etal 2006); the effective wavelength is $\sim 2.16$ $\mu$m.  To good approximation, 
$K_s = K - 0.044$ (Carpenter 2001; Bessell 2005), where $K$ is essentially Johnson's (1962) 2.2 $\mu$m 
bandpass.  Except in Section S3.1, in this paragraph, and in the tables, we abbreviate $K_s$ as $K$ for 
convenience.~{\bf Tables 2}~and~{\bf 3} list $K_s$ apparent magnitudes of the galaxies from the 2MASS 
Large Galaxy Atlas (Jarrett \etal 2003) or from the online Extended Source Catalog.  Corrections 
(usually a few tenths of a mag) have been made for some of the brightest or angularly largest galaxies 
as discussed in Section S3.1.

\vfill\eject

%  Maximal portrait table page size

%\hsize=19.truecm  \hoffset=-1.5truecm  \vsize=25.5truecm  \voffset=-0.5truecm

\magnification = \magstep0

%  Normal text page size

%\hsize=15.0truecm  \hoffset=0.0truecm  \vsize=20.1truecm  \voffset=1.5truecm

\font\sc=cmr8
\def\scbaselines{\baselineskip=8pt    \lineskip=0pt   \lineskiplimit=0pt}

\def\vsl{\vskip\baselineskip}   \def\vs{\vskip 6pt} \def\vsss{\vskip 3pt}
\parindent=10pt \nopagenumbers

\input colordvi

\def\t#1{#1} 
\def\t#1{\empty}
    
  \def\ba{\kern -1pt}
\parskip = 0pt % default is \parskip = 0pt+1pt
\def\ts{\thinspace} \def\cl{\centerline}
\def\ni{\noindent}  

\def\nhi{\noindent \hangindent=1.0truecm}
  
\def\ihi{\indent\hangindent=1.85truecm}
\def\ihij{\indent\null{\kern -4.7pt}\hangindent=1.85truecm}

\def\makeheadline{\vbox to 0pt{\vskip-30pt\line{\vbox to8.5pt{}\the
                               \headline}\vss}\nointerlineskip}
\def\toppageno{\headline={\hss\tenrm\folio\hss}}
\def\footnoterule{\kern-3pt \hrule width \hsize \kern 2.6pt \vskip 3pt}
\output={\plainoutput}    \pretolerance=10000   \tolerance=10000

\def\sup1{$^{\rm 1}$} \def\sup2{$^{\rm 2}$}
\def\r0{$\rho_0$}   
 \def\0{\phantom{0}} \def\bb{\kern -2pt}
\def\1{\phantom{1}}         
\def\etal{{et~al.\ }}
\def\gapprox{$_>\atop{^\sim}$} \def\lapprox{$_<\atop{^\sim}$}
          
\newdimen\sa  \def\sd{\sa=.1em \ifmmode $\rlap{.}$''$\kern -\sa$
                               \else \rlap{.}$''$\kern -\sa\fi}
\def\ss{\ifmmode ^{\prime\prime}$\kern-\sa$ \else $^{\prime\prime}$\kern-\sa\fi}
\def\mm{\ifmmode ^{\prime}$\kern-\sa$ \else $^{\prime}$\kern-\sa \fi}
  
\def\mbh{$M_{\bullet}$~}  
\def\m31{M{\ts}31} \def\mm32{M{\ts}32} \def\mmm33{M{\ts}33} \def\M87{M{\ts}87} 
\def\mbh{$M_\bullet$}

\def\nhi2{\noindent \hangindent=2.79cm}

\font\big=cmbx12 scaled 1100
\font\bigau=cmr12 scaled 1200
\font\bigbig=cmr12 scaled 2000

\font\small=cmr8

% Tenpoint style:

% Ninepoint style:

\def\ninepoint{
  \font\fiverm=cmr5
  \font\sevenrm=cmr7
  \font\ninerm=cmr9
  \font\fivei=cmmi5
  \font\seveni=cmmi7
  \font\ninei=cmmi9
  \font\fivesy=cmsy5
  \font\sevensy=cmsy7  
  \font\ninesy=cmsy9
  \font\it=cmti9
  \font\bf=cmbx9
  \font\sl=cmsl9
  \textfont0=\ninerm \scriptfont0=\sevenrm     
    \scriptscriptfont0=\fiverm                 %text fonts
  \def\rm{\fam0 \ninerm}   
  \textfont1=\ninei  \scriptfont1=\seveni  
    \scriptscriptfont1=\fivei                  %math italic fonts
  \def\mit{\fam1 } \def\oldstyle{\fam1 \ninei}
  \textfont2=\ninesy \scriptfont2=\sevensy 
    \scriptscriptfont2=\fivesy                 %math symbol fonts
}

% Eightpoint style:

\def\eightpoint{
  \font\fiverm=cmr5
  \font\sevenrm=cmr7
  \font\eightrm=cmr8
  \font\fivei=cmmi5
  \font\seveni=cmmi7
  \font\eighti=cmmi8
  \font\fivesy=cmsy5
  \font\sevensy=cmsy7  
  \font\eightsy=cmsy8
  \font\rm=cmr8
  \font\it=cmti8
  \font\bf=cmbx8
  \font\sl=cmsl8
  \textfont0=\eightrm \scriptfont0=\sevenrm     
    \scriptscriptfont0=\fiverm                 %text fonts
  \def\rm{\fam0 \eightrm}   
  \textfont1=\eighti  \scriptfont1=\seveni  
    \scriptscriptfont1=\fivei                  %math italic fonts
  \def\mit{\fam1 } \def\oldstyle{\fam1 \eighti}
  \textfont2=\eightsy \scriptfont2=\sevensy 
    \scriptscriptfont2=\fivesy                 %math symbol fonts
}

\ninepoint
\small

\def\tblbaselines{\baselineskip=10pt    \lineskip=0pt   \lineskiplimit=0pt}
\scbaselines

\hfuzz=100pt

%  Maximal portrait table page size

\hsize=19.8truecm  \hoffset=-1.8truecm  \vsize=26.15truecm  \voffset=-1.25truecm

\def\x{\textBlack}
\def\y{\textPurple}
\def\y{\ARDiscard}
\def\g{\textColor{1. .1 1. .1}}

\cl{\null}

\vskip -40pt

\cl{\null}

\table
\tblbaselines
\hskip -10pt
\tablewidth{20truecm} 
\tablespec{\l\l\c\c\c\c\c\c\c\c\c\c\l}
\body{
\header{\bf \Blue{Table 2 \quad  Supermassive black holes detected dynamically in 44 elliptical galaxies}\textBlack}
\skip{5pt}
\hline
%\skip{.2truecm}\hline
\skip{3pt}
&   Galaxy &\llap{T}ype& Distance &$  K_s   $&$\0M_{KsT}$ &$ M_{VT}$ & ($V$\null$-$\null$K_s$)\rlap{$_0$}~\ts &($B$\null$-$\null$V$)\rlap{$_0$}&\0$\log{M_{\rm bulge}}$&  \mbh (low \mbh\ -- high \mbh)     &  $\sigma_e$  &       Flags       &  Source                \end %
&          &        &     (Mpc)   &          &            &          &             &   & ($M_\odot$)&       ($M_\odot$)                & (km s$^{-1}$)&~M{\ts}C{\ts}$M_\bullet$ &                  \end %
& (1)      & (2)    &      (3)    &   (4)    &     (5)    &  (6)     &     (7)     &    (8)    &        (9)       &              (10)                    &     (11)     &       (12)        &  (13)                  \end %
\skip{3pt}	        	     	                	   				        
\hline		        	     	                	   				        
\skip{3pt}	        	     	                	   				        
&   M{\ts}32 &  E2  & \0\00.805 7 &$ 5.10   $&$  -19.45  $&$ -16.64 $&$  2.816    $&$ 0.895   $&$\09.05 \pm 0.10 $&$  2.45(   1.43-$~$3.46 ) \times 10^6 $&$\077\pm\03  $&   1     0     1   &  van{\ts}den{\ts}Bosch{\ts}+{\ts}2010  \end \skip{-2pt} %
&\g NGC 1316 &  E4  &  20.95 1    &$ 5.32   $&$  -26.29  $&$ -23.38 $&$  2.910    $&$ 0.871   $&$ 11.84 \pm 0.09 $&$  1.69(   1.39-$~$1.97 ) \times 10^8 $&$ 226\pm\09  $&   1     0     0   &  Nowak + 2008        \x \end \skip{-2pt} % -0.268
&   NGC 1332 &  E6  &  22.66 2    &$ 7.05   $&$  -24.73  $&$ -21.58 $&$  3.159    $&$ 0.931   $&$ 11.27 \pm 0.09 $&$  1.47(   1.27-$~$1.68 ) \times 10^9 $&$ 328\pm\09  $&   1     0     0   &  Rusli + 2011           \end \skip{-2pt} % 0
&   NGC 1374 &  E0  &  19.57 1    &$ 8.16   $&$  -23.30  $&$ -20.43 $&$  2.874    $&$ 0.908   $&$ 10.65 \pm 0.09 $&$  5.90(   5.39-$~$6.51 ) \times 10^8 $&$ 167\pm03   $&   1     0     1   &  Rusli\ts2012,Rusli+2013\end \skip{-2pt} % Rusli: 1-sig errors
&   NGC 1399 &  E1  &  20.85 1    &$ 6.31   $&$  -25.29  $&$ -22.43 $&$  2.863    $&$ 0.948   $&$ 11.50 \pm 0.09 $&$  8.81(   4.35-$\null$17.81  ) \times 10^8 $&$ 315\pm03   $&   1     1     0   &  see notes              \end \skip{-2pt} %
&   NGC 1407 &  E0  &  29.00 2    &$ 6.46   $&$  -25.87  $&$ -22.89 $&$  2.980    $&$ 0.969   $&$ 11.74 \pm 0.09 $&$  4.65(   4.24-$~$5.38 ) \times 10^9 $&$ 276\pm\02  $&   1     1     1   &  Rusli\ts2012,Rusli+2013\end \skip{-2pt} % deltaK=-.240
&   NGC 1550 &  E1  &  52.50 9    &$ 8.77   $&$  -24.87  $&$ -21.89 $&$  2.974    $&$ 0.963   $&$ 11.33 \pm 0.09 $&$  3.87(   3.16-$~$4.48 ) \times 10^9 $&$ 270\pm 10  $&   1     1     1   &  Rusli\ts2012,Rusli+2013\end \skip{-2pt} % deltaK=0
&\y NGC 2778 &  E2  &  23.44 2    &$ 9.51   $&$  -22.34  $&$ -19.39 $&$  2.955    $&$ 0.911   $&$ 10.26 \pm 0.09 $&$  1.45(   0.00-$~$2.91 ) \times 10^7 $&$ 175\pm\08  $&   1     0     1   &  Schulze + 2011      \x \end \skip{-2pt} %
&\g NGC 2960 &  E2  &  67.1\0 9   &$ 9.78   $&$  -24.36  $&$ -21.30 $&$  3.068    $&$ 0.880   $&$ 11.06 \pm 0.09 $&$  1.08(   1.03-$~$1.12 ) \times 10^7 $&$ 166\pm 16  $&   3     0     0   &  Kuo + 2011          \x \end \skip{-2pt} %
&   NGC 3091 &  E3  &  53.02 9    &$ 8.09   $&$  -25.54  $&$ -22.56 $&$  2.980    $&$ 0.962   $&$ 11.61 \pm 0.09 $&$  3.72(   3.21-$~$3.83 ) \times 10^9 $&$ 297\pm 12  $&   1     1     1   &  Rusli\ts2012,Rusli+2013\end \skip{-2pt} %  deltaK=0 !
&   NGC 3377 &  E5  &  10.99 2    &$ 7.16   $&$  -23.06  $&$ -20.08 $&$  2.980    $&$ 0.830   $&$ 10.50 \pm 0.09 $&$  1.78(   0.85-$~$2.72 ) \times 10^8 $&$ 145\pm\07  $&   1     0     1   &  Schulze + 2011         \end \skip{-2pt} % -0.286 SAURON sige
&   NGC 3379 &  E1  &  10.70 2    &$ 6.27   $&$  -23.88  $&$ -21.01 $&$  2.867    $&$ 0.939   $&$ 10.91 \pm 0.09 $&$  4.16(   3.12-$~$5.20 ) \times 10^8 $&$ 206\pm 10  $&   1     1     1   &  van{\ts}den{\ts}Bosch{\ts}+{\ts}2010  \end \skip{-2pt} % SAURON sige
&\y NGC 3607 &  E1  &  22.65 2    &$ 6.99   $&$  -24.79  $&$ -21.92 $&$  2.872    $&$ 0.911   $&$ 11.26 \pm 0.09 $&$  1.37(   0.90-$~$1.82 ) \times 10^8 $&$ 229\pm 11  $&   1     1     0   &  G\"ultekin + 2009b  \x \end \skip{-2pt} %
&   NGC 3608 &  E1  &  22.75 2    &$ 7.62   $&$  -24.17  $&$ -21.19 $&$  2.980    $&$ 0.921   $&$ 11.01 \pm 0.09 $&$  4.65(   3.66-$~$5.64 ) \times 10^8 $&$ 182\pm\09  $&   1     1     1   &  Schulze + 2011         \end \skip{-2pt} % -0.477
&   NGC 3842 &  E1  &  92.2\0 9   &$ 8.84   $&$  -25.99  $&$ -23.01 $&$  2.980    $&$ 0.941   $&$ 11.77 \pm 0.09 $&$\y9.09(   6.28-$\null$11.43  ) \times 10^9$\x&$270\pm 27  $&   1     1     1   &  McConnell+2011a,2012   \end \skip{-2pt} % -0.241
&\y NGC 4261 &  E2  &  32.36 2    &$ 6.94   $&$  -25.62  $&$ -22.64 $&$  2.980    $&$ 0.974   $&$ 11.65 \pm 0.09 $&$  5.29(   4.21-$~$6.36 ) \times 10^8 $&$ 315\pm 15  $&   2     1     0   &  Ferrarese + 1996    \x \end \skip{-2pt} % -0.301
&   NGC 4291 &  E2  &  26.58 2    &$ 8.42   $&$  -23.72  $&$ -20.76 $&$  2.954    $&$ 0.927   $&$ 10.85 \pm 0.09 $&$  9.78(   6.70-$\null$12.86  ) \times 10^8 $&$ 242\pm 12  $&   1     1     1   &  Schulze + 2011         \end \skip{-2pt} %
&   NGC 4374 &  E1  &  18.51 1    &$ 5.75   $&$  -25.60  $&$ -22.62 $&$  2.980    $&$ 0.945   $&$ 11.62 \pm 0.09 $&$  9.25(   8.38-$\null$10.23  ) \times 10^8 $&$ 296\pm 14  $&   2     1     0   &  Walsh + 2010           \end \skip{-2pt} % -0.471 1sig
&\g NGC 4382 &  E2  &  17.88 1    &$ 5.76   $&$  -25.51  $&$ -22.53 $&$  2.980    $&$ 0.863   $&$ 11.51 \pm 0.09 $&$  1.30(   0.00-$\null$22.4\0 ) \times 10^7 $&$ 182\pm\05  $&   1     1     0   &  G\"ultekin + 2011   \x \end \skip{-2pt} % -0.381
&\y NGC 4459 &  E2  &  16.01 1    &$ 7.15   $&$  -23.88  $&$ -20.91 $&$  2.975    $&$ 0.909   $&$ 10.88 \pm 0.09 $&$\y6.96(   5.62-$~$8.29 ) \times 10^7$\x&$167\pm\08  $&   2     0     0   &  Sarzi + 2001           \end \skip{-2pt} %  0
&   NGC 4472 &  E2  &  16.72 1    &$ 4.97   $&$  -26.16  $&$ -23.18 $&$  2.980    $&$ 0.940   $&$ 11.84 \pm 0.09 $&$  2.54(   2.44-$~$3.12 ) \times 10^9 $&$ 300\pm\07  $&   1     1     1   &  Rusli\ts2012,Rusli+2013\end \skip{-2pt} % deltaK = -.430
&   NGC 4473 &  E5  &  15.25 1    &$ 7.16   $&$  -23.77  $&$ -20.89 $&$  2.874    $&$ 0.935   $&$ 10.85 \pm 0.09 $&$  0.90(   0.45-$~$1.35 ) \times 10^8 $&$ 190\pm\09  $&   1     0     1   &  Schulze + 2011         \end \skip{-2pt} %
&   M{\ts}87 &  E1  &  16.68 1    &$ 5.27   $&$  -25.85  $&$ -22.87 $&$  2.980    $&$ 0.940   $&$ 11.72 \pm 0.09 $&$  6.15(   5.78-$~$6.53 ) \times 10^9 $&$ 324^{+28}_{-12}\0$& 1 1     1   &  Gebhardt + 2011        \end \skip{-2pt} % -0.546
&   NGC 4486A&  E2  &  18.36 1    &$ 9.49   $&$  -21.83  $&$ -18.85 $&$  2.980    $&$ \dots   $&$ 10.04 \pm 0.09 $&$  1.44(   0.92-$~$1.97 ) \times 10^7 $&$ 111\pm\05  $&   1     0     0   &  Nowak + 2007           \end \skip{-2pt} % +0.480
&   NGC 4486B&  E0  &  16.26 1    &$10.39\0 $&$  -20.67  $&$ -17.69 $&$  2.980    $&$ 0.991   $&$\09.64 \pm 0.10 $&$\y6.\0\0(4.\0\0-$~$9.\0\0)\times 10^8 $&\x$185\pm\09$&   1     0     0   &  Kormendy + 1997        \end \skip{-2pt} % +0.469 sig from Gult
&   NGC 4649 &  E2  &  16.46 1    &$ 5.49   $&$  -25.61  $&$ -22.63 $&$  2.980    $&$ 0.947   $&$ 11.64 \pm 0.09 $&$  4.72(   3.67-$~$5.76 ) \times 10^9 $&$ 380\pm 19  $&   1     1     1   &  Shen+Gebhardt\ts2010   \end \skip{-2pt} % -0.254
&   NGC 4697 &  E5  &  12.54 1    &$ 6.37   $&$  -24.13  $&$ -21.33 $&$  2.799    $&$ 0.883   $&$ 10.97 \pm 0.09 $&$  2.02(   1.52-$~$2.53 ) \times 10^8 $&$ 177\pm\08  $&   1     0     1   &  Schulze + 2011         \end \skip{-2pt} %
&   NGC 4751 &  E6  &  32.81 2    &$ 8.24   $&$  -24.38  $&$ -21.22 $&$  3.158    $&$ 0.983   $&$ 11.16 \pm 0.09 $&$  1.71(   1.52-$~$1.81 ) \times 10^9 $&$ 355\pm 14  $&   1     0     1   &  Rusli\ts2012,Rusli+2013\end \skip{-2pt} % 0.
&   NGC 4889 &  E4  & 102.0\0~9\1 &$ 8.41   $&$  -26.64  $&$ -23.63 $&$  3.007    $&$ 1.031   $&$ 12.09 \pm 0.09$&\y$\02.08(  0.49-$~$3.66  ) \times 10^{10}$\x&$347\pm\05$&   1     1     1   &  McConnell+2011a,2012   \end \skip{-2pt} %  0
&   NGC 5077 &  E3  &  38.7\0 9   &$ 8.22   $&$  -24.74  $&$ -21.66 $&$  2.949    $&$ 0.987   $&$ 11.28 \pm 0.09 $&$  8.55(   4.07-$\null$12.93  ) \times 10^8 $&$ 222\pm 11  $&   2     1     0   &  De{\ts}Francesco{\ts}+{\ts}2008 \end \skip{-2pt} %
&\g NGC 5128 &  E   & \03.62 6    &$ 3.49   $&$  -24.34  $&$ -21.36 $&$  2.980    $&$ 0.898   $&$ 11.05 \pm 0.09 $&$  5.69(   4.65-$~$6.73 ) \times 10^7 $&$ 150\pm\07  $&   1     1     0   &  Cappellari + 2009   \x \end \skip{-2pt} % -0.452 3sig -> 1sig
&   NGC 5328 &  E2  &  64.4\0 9   &$ 8.49   $&$  -25.58  $&$ -22.61 $&$  2.966    $&$ 1.004   $&$ 11.65 \pm 0.09 $&$  4.75(   2.81-$~$5.63 ) \times 19^9 $&$ 333\pm\02  $&   1     1     1   &  Rusli + 2013           \end \skip{-2pt} %
&   NGC 5516 &  E3  &  55.3~\ts~9 &$ 8.31   $&$  -25.47  $&$ -22.50 $&$  2.970    $&$ 0.993   $&$ 11.60 \pm 0.09 $&$  3.69(   2.65-$~$3.79 ) \times 10^9 $&$ 328\pm 11  $&   1     1     1   &  Rusli\ts2012,Rusli+2013\end \skip{-2pt} %  0
&   NGC 5576 &  E3  &  25.68 2    &$ 7.83   $&$  -24.23  $&$ -21.29 $&$  2.939    $&$ 0.862   $&$ 11.00 \pm 0.09 $&$  2.73(   1.94-$~$3.41 ) \times 10^8 $&$ 183\pm\09  $&   1     1     0   &  G\"ultekin + 2009b     \end \skip{-2pt} %
&   NGC 5845 &  E3  &  25.87 2    &$ 9.11   $&$  -22.97  $&$ -19.73 $&$  3.238    $&$ 0.973   $&$ 10.57 \pm 0.09 $&$  4.87(   3.34-$~$6.40 ) \times 10^8 $&$ 239\pm 11  $&   1     0     1   &  Schulze + 2011         \end \skip{-2pt} %
&   NGC 6086 &  E   & 138.0\0~9\1 &$ 9.97   $&$  -25.74  $&$ -22.84 $&$  2.884    $&$ 0.965   $&$ 11.69 \pm 0.09 $&$  3.74(   2.59-$~$5.50 ) \times 10^9 $&$ 318\pm\02  $&   1     1     1   &  McConnell + 2011b      \end \skip{-2pt} % was -0.283 now 
&\y NGC 6251 &  E1  & 108.4\0~9\1 &$ 9.03   $&$  -26.18  $&$ -23.18 $&$  2.998    $&$ \dots   $&$ 11.88 \pm 0.09 $&$  6.14(   4.09-$~$8.18 ) \times 10^8 $&$ 290\pm 14  $&   2     1     0   &  Ferrarese + 1999    \x \end \skip{-2pt} %  0
&   NGC 6861 &  E4  &  28.71 2    &$ 7.71   $&$  -24.60  $&$ -21.42 $&$  3.179    $&$ 0.962   $&$ 11.25 \pm 0.09 $&$  2.10(   2.00-$~$2.73 ) \times 10^9 $&$ 389\pm\03  $&   1     0     1   &  Rusli\ts2012,Rusli+2013\end \skip{-2pt}% 0.
&\y NGC 7052 &  E3  &  70.4~\ts~9 &$ 8.57   $&$  -25.70  $&$ -22.86 $&$  2.841    $&$ 0.86\0  $&$ 11.61 \pm 0.10 $&$  3.96(   2.40-$~$6.72 ) \times 10^8 $&$ 266\pm 13  $&   2     1     0   &  van{\ts}der{\ts}Marel{\ts}+{\ts}1998    \x\end \skip{-2pt} %
&   NGC 7619 &  E3  &  53.85 2    &$ 8.03   $&$  -25.65  $&$ -22.83 $&$  2.821    $&$ 0.969   $&$ 11.65 \pm 0.09 $&$  2.30(   2.19-$~$3.45 ) \times 10^9 $&$ 292\pm\05  $&   1     1     1   &  Rusli\ts2012,Rusli+2013\end \skip{-2pt} %  0
&   NGC 7768 &  E4  & 116.0\0~9\1 &$ 9.34   $&$  -26.00  $&$ -23.19 $&$  2.811    $&$ 0.906   $&$ 11.75 \pm 0.09 $&$  1.34(   0.93-$~$1.85 ) \times 10^9 $&$ 257\pm 26  $&   1     1     1   &  McConnell + 2012       \end \skip{-2pt} %  0
&   IC 1459  &  E4  &  28.92 2    &$ 6.81   $&$  -25.51  $&$ -22.42 $&$  3.081    $&$ 0.966   $&$ 11.60 \pm 0.09 $&$  2.48(   2.29-$~$2.96 ) \times 10^9 $&$ 331\pm\05  $&   1     0     0   &  Cappellari + 2002      \end \skip{-2pt} %
&\g IC 1481  &  E1.5&  89.9\0 9   &$10.62\0 $&$  -24.17  $&$  \dots $&$  \dots    $&$ \dots   $&$       \dots    $&$  1.49(   1.04-$~$1.93 ) \times 10^7 $&$ \dots      $&   3     0     0   &  Hur\'e + 2011       \x \end \skip{-2pt} % 
&\y A1836 BCG&  E   & 152.4\0~9\1 &$ 9.99   $&$  -25.95  $&$ -22.64 $&$  3.310    $&$ 1.043   $&$ 11.81 \pm 0.10 $&$  3.74(   3.22-$~$4.16 ) \times 10^9 $&$ 288\pm 14  $&   2     1     0   &  Dalla{\ts}Bont\'a{\ts}+\ts2009\x\end \skip{-2pt} % -0. omit (gassig)
&\y A3565 BCG&  E   &  49.2~\ts~9 &$ 7.50   $&$  -25.98  $&$ -23.03 $&$  2.948    $&$ 0.956   $&$ 11.78 \pm 0.09 $&$  1.30(   1.11-$~$1.50 ) \times 10^9 $&$ 322\pm 16  $&   2     1     0   &  Dalla{\ts}Bont\'a{\ts}+\ts2009\x\end \skip{-2pt} %     omit (gassig)
&\y Cygnus A &  E   & 242.7\0~9\0 &$10.28\0 $&$  -26.77  $&$ -23.23 $&$  3.54\0   $&$ \dots   $&$       \dots    $&$  2.66(   1.91-$~$3.40 ) \times 10^9 $&$ 270\pm 90  $&   2     1     0   &  Tadhunter + 2003    \x \end \skip{-2pt} % Omit(V-K)     omit(gassig)
\skip{3pt}											    
\hline												    
}												    
\endtable											    
												    
%&NGC 3706  &  E4  &   38.1\0 9   &$ 7.902   $&$  -25.031  $&$ -22.15 $&$  2.883    $&$ 0.959   $&$  $&$  4.97(   4.22-\05.63 ) \times 10^8 $&$ 325\pm\05  $&   1     1     1   &  G\"ultekin + 2012     \end \skip{-2pt} %  0.

\scbaselines

\null
\vskip -15pt
\null

{\sc\parindent=39pt\eightpoint
\def\nhi{\noindent \hangindent=1.4truecm}
\lineskip=-20pt \lineskiplimit=-20pt
\noindent Column 1 is the galaxy name; BCGs are brightest cluster galaxies in the Abell clusters named. \y Purple listings are not included in fits (Section 6.3).\par\x
\nhi Column 2 is Hubble type (mostly RC3). \g Green lines are for mergers in progress (Section 6.4). \hfill \y If only $M_\bullet$ is purple, we accept it but do not include it\x\par
\nhi Column 3 is the assumed distance from the following sources, starting with the highest-priority sources: \hfill \y in the correlation fits (see Section 6.6).\x        \par
\ihi            (1) Blakeslee \etal (2009) surface-brightness fluctuation (SBF) distances for individual galaxies in the Virgo and Fornax clusters; \par
\ihi            (2) Tonry \etal (2001) SBF corrected via Equation A1 in Blakeslee \etal (2010);                                                     \par
\ihi            (3) Mei \etal (2007) SBF mean distance to the Virgo W$^\prime$ cloud (de Vaucouleurs 1961);                                         \par
\ihi            (4) Mei \etal (2007) SBF mean distance to the Virgo cluster (no W$^\prime$ cloud);                                                  \par
\ihi            (5) Thomsen \etal (1997) SBF distance to NGC 4881 in the Coma cluster;                                                              \par
\ihi            (6) mean of distance determinations adopted in Kormendy \etal (2010); sources are given there;                                      \par
\ihi            (7) Monachesi \etal (2011); agrees with (8);                                                                                        \par
\ihi            (8) mean of many determinations listed on NED, using mainly SBF distances and those based on Cepheid, TRGB, and RR Lyrae stars.     \par
\ihi            (9) As a last resort, we adopt D (Local Group) given by NED for the recession velocity of the galaxy 
                    (if isolated) or its group (if in a group or cluster) and for the WMAP 5-year cosmology parameters (Komatsu \etal 2009).        \par
\ihij          (10) van den Bosch \etal (2012).                                                                                                     \par
\nhi Column \04 is the 2MASS $K_s$ total magnitude.  When $(V - K_s)_0 = 2.980$ in Column 7, $K_s$ has been corrected as discussed in Section S3.1. \par
\nhi Columns 5 and 6 are the $K_s$- and $V$-band absolute magnitudes based on the adopted distances and Galactic absorption corrections from Schlegel
                \etal (1998) as recalibrated by Schlafly \& Finkbeiner (2011).  The $V$-band magnitudes are taken, in order of preference, from KFCB, from RC3, or from Hyperleda 
                (usually ``integrated photometry'' but sometimes the main table if it implies a more realistic $(V - K_s)_0$ color).        \par
\nhi Columns \ts7 and 8 are the $V - K_s$ and $B - V$ colors of the galaxy corrected for Galactic reddening.                                             \par
\nhi Column \09 is the base-10 logarithm of the bulge mass (Section 6.6.1).\par
\def\nhi{\noindent \hangindent=1.55truecm}
\nhi Column 10 is the measured BH mass with 1-sigma range in parentheses from sources in Column 13.\par
\nhi Column 11 is the stellar velocity dispersion $\sigma_e$.  We adopt the usual convention that $\sigma_e^2$ is the intensity-weighted mean of $V^2 + \sigma^2$ 
               out to a fixed fraction of the effective radius $r_e$ that contains half of the light of the galaxy.  As discussed briefly in Section 5.1 of the main paper
               and in more detail in Section S3.3 here,
               we adopt $r_e/2$ when we calculate $\sigma_e$ from photometry and published kinematics (see notes on individual objects).  When no note is given,
               $\sigma_e$ is from the $M_\bullet$ source paper or from G\"ultekin \etal (2009c).  \par
\nhi Column 12 lists three flags: ``M'' encodes the method used to measure $M_\bullet$, using 1 for stellar dynamics, 2 for ionized gas dynamics, and 3 for maser dynamics.
                 ``C'' = 1 implies that the galaxy has a core (e.{\ts}g., Lauer \etal 1995).
                 ``$M_\bullet$'' = 1 implies that the BH mass has been ``corrected'' by making dynamical models that include large orbit libraries and triaxiality (M{\ts}32 and NGC 3379) 
                 or dark matter halos.\par
}

\parindent=10pt

\vfill\eject

\vfill\eject

\hsize=19.6truecm  \hoffset=-1.5truecm  \vsize=24.7truecm  \voffset=-0.5truecm

\cl{\null}

\vskip 1truecm
\includegraphics{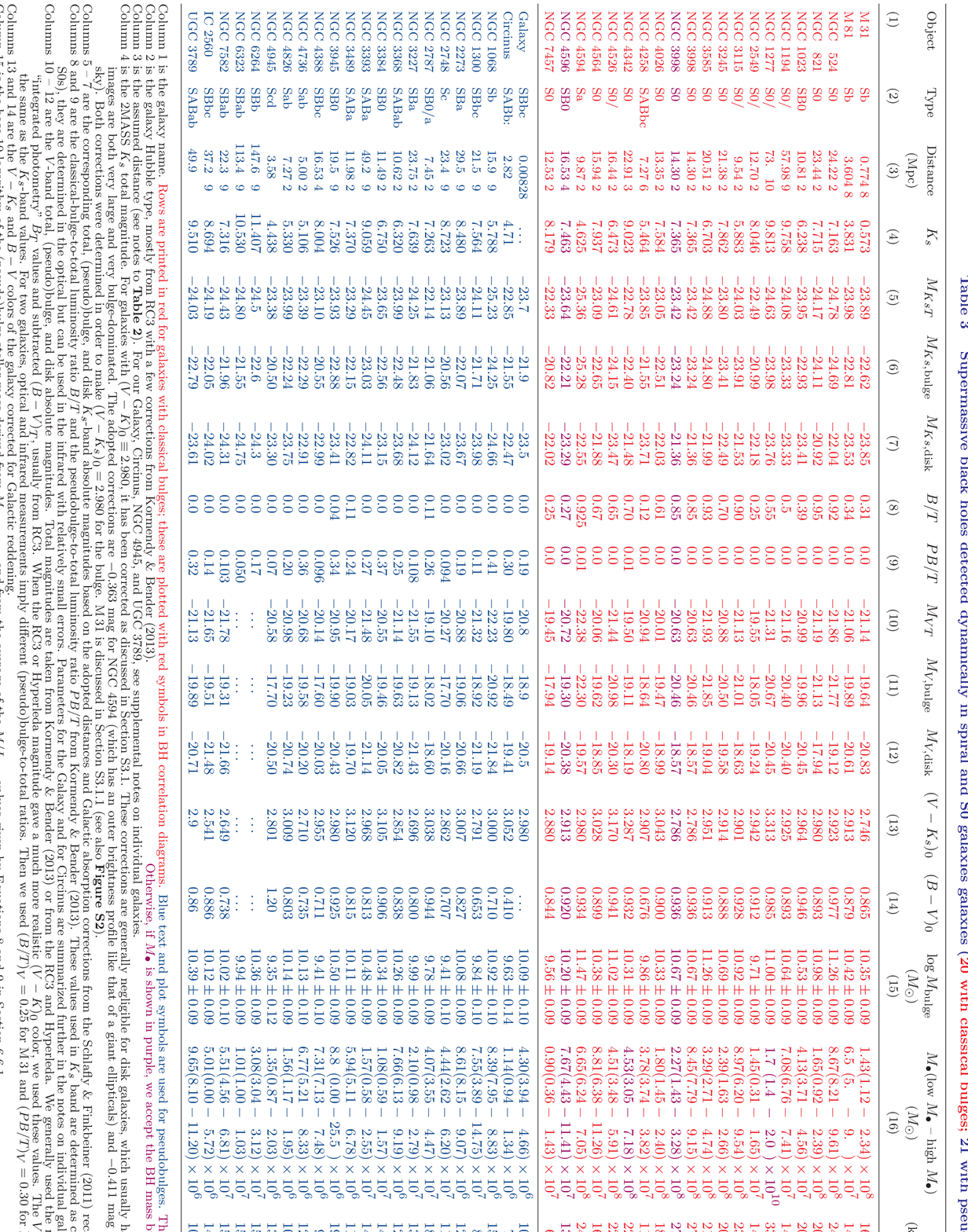}
%\special{psfile=annrev4K-S.ps angle=-90 hoffset=570 voffset=-660 hscale=71 vscale=71}

\cl{\null}
\vfill\eject

{\eightpoint\small\smedbaselines

\def\vsn{\vskip 4pt}

\lineskip=-30pt \lineskiplimit=-30pt

\hsize=19.truecm  \hoffset=-1.5truecm  \vsize=24.7truecm  \voffset=-0.5truecm

\vs
\cl {\big\ARRed S4.1.~Notes on Individual Elliptical Galaxies}\textBlack
\vs

M{\ts}32: The BH discovery papers and history of $M_\bullet$ measurements are discussed in \S\ts2.2.1. We adopt $M_\bullet$
          from the triaxial models of van den Bosch \& de Zeeuw (2010).   We calculate $\sigma_e = 77 \pm 3$ km s$^{-1}$ from 
          our photometry ($r_e/2 = 38^{\prime\prime}/2$) and kinematics in Simien \& Prugniel (2002).  The value averaged inside 
          $r_e$ is 1 km s$^{-1}$ smaller.  This is in good agreement with $\sigma_e = 75 \pm 3$ km s$^{-1}$ in 
          Tremaine \etal (2002),
          G\"ultekin \etal (2009c), and
          McConnell \& Ma (2013).
          In contrast, Graham \& Scott (2013) adopt $\sigma_e = 55$ km s$^{-1}$, in part because they do not include $V(r)$.

\vsn

NGC 2778, NGC 3608, NGC 4291, NGC 4473, NGC 4649, NGC 4697, and NGC 5845:~The BH discovery is by Gebhardt \etal (2003). \vsn

NGC 1316: This galaxy is bluer than the typical giant E, with $(B - V)_0 = 0.87$ (NED).  Based on our experience with
          published total magnitudes of bright galaxies (e.{\ts}g., KFCB), we adopt $B_T = 9.17$ from the Hyperleda 
          integrated photometry table.  This yields $M_{VT} = -23.38$.  The 2MASS $K = 5.587$ magnitude then implies
          that $(V - K)_0 = 2.64$.  This is implausible for $(B - V)_0$ = 0.871, indicating that the $K$ luminosity
          is underestimated.  We use the $(V - K)_0$ versus $(B - V)_0$ correlation to derive $(V - K)_0 = 2.91$ and 
          hence to correct the $K$ magnitude by $-0.268$ magnitudes to $K = 5.319$.  
          This is a typical correction for a nearby giant elliptical. \vsn

NGC 1332 and NGC 4751: These galaxies initially presented us with an interpretation dilemma.  We believe that it is solved and
          that both galaxies are best interpreted as (rather extreme) ellipticals.  However, realizing this required us to learn
          something new about elliptical galaxies.  This note explains our conclusions and summarizes the consequeunces
          of the more canonical alternative that these are S0 galaxies. 

          Both galaxies are highly flattened (E6) and contain prominent, almost-edge-on nuclear dust disks.  It is remarkable how
          many BH host ellipticals contain nuclear dust disks: NGC 1332, NGC 3379 (faintly), NGC 3607, NGC 4261,
          NGC 4374, NGC 4459, NGC 4486A and NGC 5845 (in which much of the nuclear gas disk has formed stars), NGC 4697, NGC 4751, 
          NGC 6251, NGC 6861, NGC 7052, NGC 7768, A1836 BCG, A3565 BCG, and probably IC 1459.  Several more contain nuclear disks
          of stars that plausibly formed out of gas-and-dust disks in the manner illustrated by NGC 4486A (Kormendy \etal 2005) and
          NGC 5845 (Lauer \etal 1995).  Of course, this does not prove that all these objects -- especially NGC 1332 and NGC 4751 --
          are ellipticals; bona fide S0$_3$ galaxies contain nuclear dust disks, too (e.{\ts}g., NGC 5866 in the {\sit Hubble Atlas},
          Sandage 1961).  One reason why dust-lane E and S0 galaxies are preferentially found among BH hosts is that 
          seeing a dust lane motivates authors to measure an emission-line rotation curve.  The prevalence of central gas
          disks in BH hosts is interesting from a BH feeding point of view, but their importance in this note is that they
          tell us that NGC 1332 and NGC 4751 are almost edge-on.
 
          Images of NGC 1332 and NGC 4751 suggest that both galaxies contain two components, a central one that is relatively round
          and that has a steep brightness gradient and an outer one that looks flatter and that has a shallower brightness gradient.
          These are defining features of S0 galaxies.  If the central component is interpreted as a bulge and the outer one as a disk,
          then plausible decompositions are possible and give $B/T = 0.43$ for NGC 1332 (Rusli \etal 2011) and $B/T = 0.55 \pm 0.05$
          for NGC 4751 (Kormendy \& Bender 2013).  We emphasize:  {\sit If these interpretations are correct, then NGC 1332 closely
          resembles the S0 galaxies NGC 1277 and MGC 4342 in having a bulge that contains an abnormally high-mass~BH (see {\bf Figure 15}
          for illustration).  NGC 4751 is similar but less extreme.  Like NGC 1277 and NGC 4486B, both galaxies then also have high 
          velocity dispersions that are well outside the scatter in the Faber-Jackson (1976) correlation between E or bulge luminosity 
          and velocity dispersion.  That is, if NGC 1332 and NGC 4751 are S0s, then they are further examples of the high-$M_\bullet$
          BH monsters discussed in Section 6.5.}

          However, a compelling argument suggests that these galaxies are extremely flattened extra-light ellipticals: 

          KFCB present
          and review evidence that Virgo cluster ellipticals are naturally divided into two kinds, $M_{VT} < -21.6$ galaxies that 
          have cores and $M_{VT} \geq -21.5$ galaxies that have central extra light above the inward extrapolation of the outer 
          $n \simeq 3 \pm 1$
          S\'ersic-function main body (see Sections 6.7 and 6.13 in the main paper).  They suggest that and (e.{\ts}g.)~Hopkins
          \etal (2009) model how extra-light ellipticals form in wet mergers, such that the main body of the galaxy is the scrambled-up
          remnant of the pre-merger disk and bulge stars and the extra-light component is manufactured by a starburst during the merger.
          In Virgo ellipticals, the fraction of the stellar mass that is in the extra-light component is $\sim$\ts5\ts\% (if just the 
          extra light is counted: KFCB) or a few 10s of percents (if a standard S\'ersic-S\'ersic decomposition is applied:
          Hopkins \etal 2009).  Both kinds of ellipticals are represented among our BH hosts; NGC 3377 and NGC 4459 are typical 
          extra-light ellipticals.

          Huang \etal (2013) make a similar study of ellipticals in field environments.
          They show that field ellipticals are different from cluster ellipticals in three ways that are relevant here: (1) extra light 
          (their ``two inner components'') makes up a larger fraction $\sim$\ts20\ts\% to 40\ts\% of the galaxies, (2) extra-light 
          ellipticals extend to higher luminosities in the field than in the Virgo cluster, and (3) field ellipticals
          can be as flat as E6 (their Figure\ts1).  In the context of these results, NGC 1332 and NGC 4751 are more plausible interpreted
          as extra-light ellipticals, not S0s.  In fact, even though NGC 1332 is exceedingly close to edge-on, its isophotes are more
          rectangular than those of an edge-on thin disk; this motivated Sandage \& Bedke (1994) to emphasize (their italics) that the
          galaxy contains a ``{\sit thick disk\/}''.  

          We now believe that there may be almost a continuum in the properties (although not a seamless overlap in numbers) of outer bodies
          of these galaxies from E5 ellipticals with a modest amount of extra light (NGC 3377) to E6 ellipticals that are roughly half extra
          light (NGC 4751 and NGC 6861) to E6 galaxies whose outer parts resemble thickened disks (NGC 1332) to true S0s with thin disks
          (NGC 5866).  Objects like NGC 4751 and NGC 1332 may be rare, and it appears that they are confined to the field, perhaps because 
          this environment favors formation by a small number of gentle mergers that involve progenitors with large gas fractions.  This is 
          the interpretation of NGC 1332 and NGC 4751 that we adopt when we construct BH correlation diagrams and least-squares fits.
          {\sbf Figure 15} illustrates both interpretations for NGC 1332. \vsn

NGC 1399: As discussed in Section 3.1, we adopt the mean $M_\bullet$ measured by Houghton \etal (2006) and by Gebhardt \etal (2007).
          Conservatively, we adopt 1-$\sigma$ errors that span the complete range obtained in both measurements.  Also,
          $\sigma_e = 315$ km s$^{-1}$ is calculated using kinematic data from Graham \etal (1998), intensity-weighting $V^2 + \sigma^2$
          out to $0.5 r_e = 56^{\prime\prime}$ using our photometry.

\vsn

NGC 2778 had a BH detection in Gebhardt \etal (2003) but only a $M_\bullet$ upper limit in Schulze \& Gebhardt (2011). However,  
         $M/L_K = 3.3$ is too big for an old stellar population minus dark matter, implying that the $M_\bullet$
          limit is too small.  We illustrate it in {\bf Figure 12} and then omit it. \vsn

NGC 2960:  The BH discovery paper is Henkel \etal (2002); a reliable BH mass was determined by Kuo \etal (2011).
           NGC 2960 has frequently been classified as Sa?~(RC3, UGC, NED), but Kormendy \& Bender (2013) and {\bf Figure 13} in the    
           main paper show that it is a merger in
           progress.  The galaxy is therefore listed here, with the ellipticals. The dispersion is from Greene \etal (2010). \vsn

NGC 3377: The BH was found by Kormendy \etal (1998), whose measurements $M_\bullet = (2.1 \pm 0.9) \times 10^8$ $M_\odot$ and 
          $M/L_V = 2.0 \pm 0.2$ agree well with $M_\bullet = (1.9 \pm 1.0) \times 10^8$ $M_\odot$ and $M/L_V = 2.3 \pm 0.4$ in 
          Schultze \& Gebhardt (2011).  The reasons are (1) that the resolution of the CFHT spectroscopy was very good,
          (2) that the assumption by Kormendy of an isotropic velocity distribution in this low-luminosity, 
          extra-light (Kormendy 1999), and rapidly rotating (Emsellem \etal 2004) elliptical was close enough to the truth, and (3) that 
          this E5 galaxy is essentially guaranteed to be edge-on.  The BH mass in NGC 3377 was also measured in Gebhardt \etal (2003).  
          The 2MASS $K$-band magnitude gives an implausible color of $(V - K)_0 = 2.69$.  We have corrected $K_s$ in {\bf Table 2} 
          to give the mean color for giant ellipticals, $(V - K)_0 = 2.980$.  \vsn

NGC 3379 is a core elliptical with dynamical models that include triaxiality (the only core galaxy that has such models)
         but not dark matter (van den Bosch \& de Zeeuw 2010).  However, the HST FOS spectra (Gebhardt \etal 2000b) 
         resolve the BH sphere of influence with $r_{\rm infl}/\sigma_* \simeq 7.1$.  Schulze \& Gebhardt (2011)
         and Rusli \etal (2013) show that $M_\bullet$ does not require a significant correction for dark matter 
         under these circumstances.   \vsn

\omit{NGC 3706 has a high-surface-brightness, counterrotating nuclear disk of stars (Lauer \etal 2002; G\"ultekin \etal 2013) that
         complicates both the central profile classification and the BH mass measurement.  However, it is clear that the inward 
         extrapolation of the profile at $r > 1^{\prime\prime}$ reaches a much higher central surface brightness than we observe.  
         Therefore this must have been a core elliptical before the nuclear disk was added, presumably (since it counterrotates) by 
         accretion. \vsn}

\omit{NGC 4291: Graham & Scott use the central sigma = 285 km/s even though sigma_e = 242 +- 12 km/s was published in
      Pinkney et nuk. 2003 and is used both by G\"ultekin and by McConnell.  Not careful.}

NGC 4374 (M{\ts}84) is the fifth-brightest elliptical in the Virgo cluster, but it is only 0.61 mag fainter than NGC\ts4472~(KFCB).  
         Like many radio galaxies, it (3C 272.1) has a nuclear gas and dust disk (Bower \etal 1997) which makes it feasible 
         to search for a BH relatively independently of the stellar mass distribution (Walsh, Barth \& Sarzi 2010) by measuring 
         the emission-line gas rotation curve.  NGC 4374 is the first galaxy in which a BH discovery was made using HST STIS 
         (Bower \etal 1998).  STIS's long-slit capability made it possible to see the prominent zig-zag in the emission lines 
         that is the signature of the BH.  

         However, as discussed in Section 3.2, the line profile is complicated by a two-component structure, and this has led to 
         some~uncertainty~in~$M_\bullet$.  Bower{\ts}et{\ts}al.{\ts}(1998) decompose the line profiles into slowly- and rapidly-rotating 
         components and use the latter to get $M_\bullet$\ts=\ts$1.63(0.99$\null$-$\null2.83)$\times$\kern -0.4pt$10^9$\ts$M_\odot$. 
         Maciejewski \& Binney (2001) suggest that the complicated line profile is caused by integrating the light of
         the nuclear disk inside a spectrograph slit that is broader (0\sd2 wide) than the telescope PSF.  They estimate
         $M_\bullet = 4.4 \times 10^8$ $M_\odot$.  The large difference between these two determinations has been a cause of
         concern, not only for NGC 4374 but also for other $M_\bullet$ determinations based on emission lines.  Recently, 
         Walsh, Barth \& Sarzi (2010) model the observations in much greater detail and -- importantly -- include the effects of the 
         velocity dispersion in the gas.  After application of this ``asymmetric drift'' correction, they get 
         $M_\bullet = 9.25(8.38~-~10.23) \times 10^8$ $M_\odot$.  We adopt their value but note (\S\ts3.2) that there are
         still significant uncertainties in emission-line $M_\bullet$ measurements.  {\it The folklore that gas rotation curves 
         give $M_\bullet$ easily and without the complications and uncertainties inherent in stellar dynamical modeling is 
         much too optimistic.} \vsn

NGC 4382:~The G\"ultekin{\ts}et{\ts}al.{\ts}(2011) $M_\bullet$ limit is based on stellar dynamical models that do not include dark matter.~For 
          $M_\bullet$\ts$\leq$\ts$1.3^{+5.2}_{-1.2}\ts\times\null10^7$\ts$M_\odot$, $r_{\rm infl}/\sigma_* \simeq 0.3$ is not well resolved.  
          Since this is a core galaxy, we need to ask whether an upward correction to the $M_\bullet$ limit is required.  
          Examination of G\"ultekin's analysis suggests that no correction is needed: (1) The core is unusually small
          (break radius $r_b = 81${\ts}pc, Lauer \etal 2005) for the high luminosity of the galaxy.  Our only BH host E that has a smaller 
          $r_b = 54$ pc in Lauer \etal (2005) is NGC 3608, for which Schulze \& Gebhardt (2011) find the same $M_\bullet$ with and without 
          dark matter. 
          (2) The kinematic measurements analyzed by G\"ultekin only reach 0.28\ts$r_e$, i.{\ts}e., the inner part of the galaxy that is most
          dominated by visible matter.  Large $M_\bullet$ corrections in Schulze \etal (2011) happen when the ground-based observations reach
          larger fractions of $r_e$.  In confirmation, (3) G\"ultekin \etal (2011) remark: ``Within $\sim$\ts2 kpc, about the outer 
          extent of our data, [Nagino \& Matsushita 2009, who] study the gravitational potential as revealed by X-ray emission from the
          interstellar medium, \dots~find a constant $B$-band mass-to-light ratio consistent with a potential dominated by stellar mass.''
          We therefore use G\"ultekin's BH mass limit in {\bf Figure 14}.  However, the Table 2 listing includes in the upper error bar
          a (probably too conservative) correction for the inclusion of halo dark matter.  \vsn

NGC 4459 is the second-brightest extra-light elliptical in the Virgo cluster and the brightest one that has S\'ersic $n < 4$
          ($n = 3.2 \pm 0.3$: KFCB).  Therefore -- as indicated by the normal color, $(V - K)_0 = 2.975$ -- the 2MASS $K_T$ 
          magnitude is accurate.   The galaxy has a prominent nuclear dust disk; in this sense, it closely resembles NGC 1332,
          NGC 4751, and many other BH host ellipticals discussed in the notes to those objects. \vsn

M{\ts}87: The early history of BH searches is reviewed in KR95.  We consider the BH discovery paper to be the HST gas-dynamical 
          study by Harms \etal (1994).~They derived $M_\bullet = (2.7 \pm 0.8) \times 10^9$ $M_\odot$ (all masses are corrected 
          to the SBF distance of 16.68 Mpc:~Blakeslee \etal 2009).   For many years, the definitive mass measurement -- also based 
          on HST gas kinematics -- was $M_\bullet = (3.6 \pm 1.0) \times 10^9$ $M_\odot$ (Macchetto \etal (1997).  Stellar dynamical 
          measurement of $M_\bullet$ is difficult, because the central brightness profile is shallow inside the break radius $r_b = 5\sd66$ 
          that defines the ``core'' (Lauer \etal 1992, 2007).  The result (Kormendy 1992a, b) is unfavorably small luminosity weighting 
          inside the BH sphere of influence, $r_{\rm infl} \simeq 3\sd1$.  Even a steep central increase in $\sigma(r)$ is strongly 
          diluted by projection.  Measuring LOSVDs helps if one can detect the resulting high-velocity wings (van der Marel 1994b).  
          M{\ts}87 remains too expensive for HST absorption-line spectroscopy, but high-$S/N$ ground-based spectroscopy is successful.  
          Gebhardt \& Thomas (2009) fit a variety of kinematic measurements, including two-dimensional spectroscopy~from~SAURON 
          (Emsellem \etal 2004) and higher-resolution, long-slit spectroscopy from van der Marel \etal (1994a:~seeing FWHM = 0\sd6; 
          slit width = 1$^{\prime\prime}$).  They for the first time include dark matter in the dynamical models; this is important because 
          the tradeoff in mass between dark and visible matter inevitably decreases the measured stellar mass-to-light ratio at large radii.  
          Analysis machinery is still based on the assumption that $M/L$ is independent of radius, so the consequence is to reduce $M/L$ near
          the center, too.  To maintain a good fit to the kinematics, $M_\bullet$ must be increased.  
          Gebhardt \& Thomas (2009) derive $M/L_V = 10.9 \pm 0.4$ and $M_\bullet = (2.1 \pm 0.6) \times 10^9$ $M_\odot$ without 
          including dark matter and        $M/L_V =  6.8 \pm 0.9$ and $M_\bullet = (6.0 \pm 0.5) \times 10^9$ $M_\odot$ 
          including dark matter.  The change is in the expected sense. 
          Recently, Gebhardt \etal (2011) add Gemini telescope integral-field spectroscopy aided by laser-guided~AO; the resulting PSF 
          has a narrow core (FWHM = 0\sd06) that contributes 14\ts--\ts45\ts\% of the PSF light.
          Such good resolution allows a substantial improvement in the reliability of the BH mass measurement.
          They get and we adopt $M_\bullet = 6.15(5.78~-~6.53) \times 10^9$ $M_\odot$. 

          As in many other galaxies in which stellar- and gas-dynamical $M_\bullet$ measurements can be compared, the stellar dynamical mass
          is substantially larger.  In M{\ts}87, it is a factor of $1.74 \pm 0.50$ larger than the Macchetto \etal (1997) value.  Such 
          comparisons are discussed further in Section 3.2. \vsn

NGC 4486A: This galaxy has a bright star 2\sd5 from its center that affects most published magnitude measurements.  KFCB measure the total
           $V_T = 12.53$ magnitude without this star, and this provides $M_{VT} = -18.85$.  The 2MASS magnitude appears to include the star, 
           so we do not~use~it.  Instead, we adopt the well determined $(V - K)_0 = 2.980$ color for old elliptical galaxies and derive 
           $M_{KT} = -21.83$ from $M_{VT}$.  For $M_\bullet$, we read 1-$\sigma$ error bars from the $\chi^2$ contour diagram in Figure 6 
           of Nowak \etal (2007). \vsn

NGC 4486B is one of the lowest-luminosity, normal ellipticals known.  The most accurate photometry (KFCB) gives $V_T = 13.42$ and, with the 
          present distance, $M_{VT} = -17.69$.  M{\ts}32 is about 1 mag fainter.  Again, we adopt $(V - K)_0 = 2.980$ in preference to the 
          2MASS $K$ magnitude to get $M_{KT} = -20.67$.  The BH mass $M_\bullet \simeq 6.1^{+3.0}_{-2.0} \times 10^8$ $M_\odot$ is the only 
          one based on spherical, isotropic stellar dynamical models that we use in this paper.  We use it (1) because the central velocity 
          dispersion $\sigma = 291 \pm 25$ km s$^{-1}$ is much larger than the upper envelope $\sigma \sim 160$ km s$^{-1}$ (for the galaxy's 
          luminosity) of the scatter in the Faber-Jackson (1976) correlation.  As in the case of NGC 1277 (q.{\ts}v.),~this is a strong 
          indicator of unusually high masses, even though anisotropic 
          models can formally fit the data without a BH.  However, we have already noted that low-luminosity, coreless ellipticals (e.{\ts}g.,
          M{\ts}32) are not generally very anisotropic.  NGC 4486B is a rapid rotator.~(2) We do not use $M_\bullet$ in
          any correlation fits.  Instead, we include NGC 4486B as the earliest discovery of a compact, early-type galaxy which deviates from
          the $M_\bullet$--host-galaxy correlations in the direction of abnormally high $M_\bullet$.  The most extreme such galaxy is
          NGC 1277 ({\bf Table 3} and {\bf Figure 15}). \vsn

NGC 5077: We use the BH mass that was calculated including emission-line widths in the analysis (De Francesco \etal 2008, see p.~361) and
          adopt the corresponding high-$M_\bullet$ error bar.  The low-$M_\bullet$ error bar is from the analysis that does not include line widths. 
          \vsn

NGC 5128: At a distance of 3.62 Mpc, NGC 5128 = Centaurus A is the second-nearest giant elliptical (after Maffei 1 at 2.85 Mpc) and the 
          nearest radio galaxy.  It is therefore very important for the BH search.  It is also a merger-in-progress, although that merger 
          may be relatively minor. It will prove to be important to our conclusion in Section 6.4 that mergers-in-progress often have BH masses
          that are small compared to expectations from the $M_\bullet$--host-galaxy correlations.

          The problem is that the many published $M_\bullet$ measurements show big disagreements (Section 3.2).
          Five measurements based on the rotation of a nuclear gas disk are available.  In recent years, as measurements and modeling
          have improved, the mass measurements have converged reasonably well:  We have \par\noindent
               $M_\bullet = 2.07(0.62~-~5.18) \times 10^8$ $M_\odot$ (Marconi \etal 2001)  based on ESO VLT observations without AO; \par\noindent
\nhhj          $M_\bullet = 1.14(0.60~-~1.24) \times 10^8$ $M_\odot$ (Marconi \etal 2006)  based on HST STIS spectroscopy and including
                                                                                           uncertainties from the poorly constrained
                                                                                           inclination of the gas disk in the error bars; \par\noindent
               $M_\bullet = 0.63(0.55~-~0.69) \times 10^8$ $M_\odot$ (H\"aring-Neumayer \etal 2006) based on VLT with AO and including the
                                                                                           gas velocity dispersion in the estimate; \par\noindent
\nhhj          $M_\bullet = 0.85(0.71~-~0.93) \times 10^8$ $M_\odot$ (Krajnovi\'c, Sharp, \& Thatte 2007) based on Gemini telescope spectroscopy without AO;
                                                                                           the Pa$\beta$ rotation curve is consistent
                                                                                           with zero velocity dispersion; and \par\noindent
\nhhj          $M_\bullet = 0.47(0.43~-~0.52) \times 10^8$ $M_\odot$ (Neumayer \etal 2007) based on VLT SINFONI spectroscopy with AO; the gas
                                                                                           dispersion is taken~into~account. 
          The convergence of $M_\bullet$ measurements based on gas dynamics is reassuring.  However, two stellar dynamical measurements
          agree poorly: \par
\nhhj          $M_\bullet = 2.49(2.28- 2.80) \times 10^8$ $M_\odot$ (Silge \etal 2005) based on Gemini observations without AO, and \par\noindent
\nhhj          $M_\bullet = 0.57(0.47- 0.67) \times 10^8$ $M_\odot$ (Cappellari \etal 2009) based on VLT SINFONI spectroscopy with AO. \par\noindent
          The last set of measurements has the highest resolution and $S/N$, and it agrees with the latest gas-dynamical results.  We adopt 
          this $M_\bullet$. \vsn

\vfill\eject

NGC 5576: This is a core elliptical (Lauer \etal 2007) with a BH detection and $M_\bullet$ measurement in G\"ultekin \etal (2009b).
          Halo dark matter was not included in the dynamical models, but the resolution $r_{\rm infl}/\sigma_* = 3.3$ is good
          enough so that we can apply a correction calibrated by Schulze \& Gebhardt (2011) and by Rusli \etal (2013).  The
          mean of the two corrections is a factor of $1.6 \pm 0.3$.  We applied this correction and added the uncertainty 
          in the correction to the uncertainty in G\"ultekin's $M_\bullet$ measurement in quadrature to give the estimated 1-$\sigma$
          error quoted in the table. \vsn

NGC 6861 is classified as an S0$_3$ galaxy in Sandage \& Tammann (1981) because of its prominent nuclear dust disk.
          M\'endez-Abreu \etal (2008) estimate that $B/T \simeq 0.64$.  However, our photometry shows little or no significant 
          departure from an $n \simeq 2$ S\'ersic-function main body with central extra light.  That is,
          the galaxy is similar to NGC 4459, another extra-light elliptical (KFCB) with a central dust disk that motivated an S0$_3$ 
          classification in Sandage \& Tammann (1981).  There could be a faint disk component in NGC 6861 such as the one in NGC 3115; if so,
          it makes no difference to any conclusions in this paper.  We therefore classify the galaxy as an extra-light elliptical.\vsn

IC 1459: The BH discovery by Verdoes Kleijn \etal (2000) is based on HST WFPC2 photometry, HST FOS spectroscopy through
         six apertures to measure the emission-line rotation curve, and ground-based, CTIO 4 m telescope spectroscopy with FWHM
         $\sim 1\sd9$ seeing to measure the stellar kinematics.  They get $M_\bullet \sim (5~\rm to~7) \times 10^9$ $M_\odot$ from
         stellar dynamical modeling but $M_\bullet \sim (0.2~\rm to~0.7) \times 10^9$ $M_\odot$ from the HST gas dynamics.
         In contrast, Cappellari \etal (2002) combine HST STIS spectroscopy with high-$S/N$ ground-based spectroscopy (CTIO 4 m 
         telescope; seeing FWHM $\approx 1\sd5$; slit width = 1\sd5; CCD scale = 0\sd73 pixel$^{-1}$) that provide both emission-line
         and absorption-line kinematics.  Again, the agreement between gas and star measurements is not superb: stellar dynamics
         give $M_\bullet = 2.48(2.29~-~2.96) \times 10^9$ $M_\odot$ (which we adopt), whereas gas dynamics give
         $M_\bullet \approx 3.34 \times 10^9$ $M_\odot$.  As noted in \S\ts3.2, the comparison between gas and stellar dynamics
         is not reassuring. 

         At $r \leq 43^{\prime\prime}/2 = r_e/2$, we measure $\sigma_e = 329$ km s$^{-1}$ from the kinematic data in Cappellari \etal (2002)
         and $\sigma_e = 332$ km s$^{-1}$ from the kinematic data in Samurovi\'c \& Danziger (2005).  We adopt $\sigma_e = 331 \pm 5$ km s$^{-1}$.
         The data reach far enough out to measure a value inside $r_e$; this would be only 2.6 km s$^{-1}$ smaller than the value that we adopt.
         Our photometry shows that this is an extra-light elliptical with $n \simeq 3.1^{+0.4}_{-0.3}$.  \vsn

IC 1481: The BH discovery paper is Mamyoda \etal (2009). They detect maser sources distributed along a line indicative of an edge-on 
         molecular disk, and they see a symmetrical rotation curve.  But $V(r) \propto r^{-0.19 \pm 0.04}$ is substantially sub-Keplerian.  
         They conclude that the maser disk is more massive than the BH.  Hur\'e \etal (2011) present an analysis method that is suitable 
         for a wide range of disk-to-BH mass ratios.  They measure $M_\bullet = (1.49 \pm 0.45) \times 10^7$ $M_\odot$ and a maser disk mass
         of about $4.1 \times 10^7$ $M_\odot$ (see \S\ts3.3.3, where these values are quoted for the distance $D = 79$ Mpc adopted by Hur\'e).
         We adopt these values.

         The host galaxy is illustrated in {\bf Figure 13}.   SDSS images show loops, shells, and dust lanes characteristic
         of a major merger in progress.  The overall light distribution is that of a normal extra-light elliptical with S\'ersic
         $n = 2.5^{+0.25}_{-0.2}$.  Consistent with this, the central $2^{\prime\prime} \times 2^{\prime\prime}$
         of the galaxy has an A{\ts}--{\ts}F, post-starburst spectrum (Bennert, Schulz \& Henkel 2004).  The ellipticity profile shows 
         that this object is turning into an E1.5 elliptical, so we list it in {\bf Table 2}.  In Section 6.4, we include 
         it among mergers in progress. 

\vs\vskip 1pt
\cl {\big\ARRed S4.2.~Notes on Individual Disk Galaxies with Classical Bulges}\textBlack
\vs

M{\ts}31: The BH mass measurements are discussed in \S\ts2.  We adopt $M_\bullet$ determined
          from P3, the blue cluster part of the triple nucleus (Bender \etal 2005).  We recomputed $\sigma_e = 169 \pm 8$ km s$^{-1}$
          from our photometry and kinematic data in Saglia \etal (2010).  Integrating to $r_e$ or $r_e/2$ gives
          the same result.  The agreement with $\sigma_e = 160 \pm 8$ km s$^{-1}$ in G\"ultekin \etal (2009c) is good. 
          Chemin, Carignan, \& Foster (2009) provide $V_{\rm circ}$.

          Photometry is difficult, because the galaxy is large.  The $V$- and $K$-band magnitudes are discussed in Section S3.1.1
          (see also {\bf Figure S2}) as an example of the correction of 2MASS magnitudes.  Since our BH correlations are derived in $K$ band, 
          we use an infrared measurement of the bulge-to-disk ratio.  Kormendy \& Bender (2013) measure $B/T = 0.31 \pm 0.03$ 
          in $L$ band ({\bf Table 3} here).  This agrees with the mean of four, $L$-band measurenets by 
          Seigar, Barth \& Bullock (2008),
          Tempel, Tamm \& Tenjes (2010),
          Kormendy \etal (2010), and
          Courteau \etal (2011). 
          However, $B/T = 0.25 \pm 0.02$ is smaller in $V$ band, and
          we use this smaller value in deriving the $V$-band bulge and disk magnitudes. \vsn

M{\ts}81: The BH discovery and $M_\bullet$\ts=\ts$6\ts(\pm\ts20\ts\%)$\null$\times$\kern -0.6pt$10^7$\ts$M_\odot$ measurement are 
          reported in Bower \etal (2000).  There are two concerns: this result is based on axisymmetric, two-integral stellar dynamical
          models, and it has never been published in a refereed journal.  Also, Devereux \etal (2003) measure $M_\bullet$
          using HST STIS spectroscopy to get the ionized gas rotation curve; the 
          problems here are that the [N\ts\sc II] emission lines are blended with and had 
          to be extracted from broad H$\alpha$ emission and that the width of the [N\ts\sc II] emission~lines~is~not~discussed.  
          But, whereas the danger is that the emission-line rotation curve will lead us to 
          underestimate $M_\bullet$, Devereux \etal (2003) get $M_\bullet = 7(6-9) \times 10^7$\ts$M_\odot$, 
          larger than Bower's value.  Both measurements are problematic, but they agree.
          Also, Bower's measurement of $M_\bullet$ in NGC 3998 in the same abstract agrees
          with a reliable stellar dynamical measurement in Walsh \etal (2012).  So we
          adopt the average of the Bower and Devereux $M_\bullet$ measurements.

          Available $B/T$ measurements in the visible and infrared agree within errors.~We
          adopt the mean $B/T = 0.34 \pm 0.02$ (Kormendy \& Bender~2013). \vsn

NGC 524: Krajnovi\'c \etal (2009) use $\sigma_e = 235$ km s$^{-1}$ from Emsellem \etal (2007), but this is from a 
          luminosity-weighted sum of spectra inside $r_e$.  It is therefore not consistent with the G\"ultekin \etal
          (2009c) definition.  For consistency, we computed $\sigma_e = 247$ km s$^{-1}$ from kinematic data in
          Simien \& Prugniel (2000) and our photometry.  \vsn

NGC 821, NGC 3384, NGC 4564, and NGC 7457: The BH discovery is by Gebhardt \etal (2003). \vsn

NGC 821 is usually considered to be an elliptical galaxy, but the shapes of the isophotes in 
        the image in the {\it Carnegie Atlas of Galaxies\/} (Sandage \& Bedke 1994) suggests that it is an almost-edge-on
        S0.  In fact, Scorza \& Bender (1995) did a bulge-disk decomposition and got $B/T = 0.943$.
        Kormendy \& Bender (2013) collect $V$-band photometry; they get $V_T = 10.96$ and $B/T = 0.969$.  We adopt
        $B/T = 0.95$ here.  We emphasize that no conclusions depend on $B/T$ or on our reclassification of the galaxy
        as an S0.  The bulge S\'ersic index is $\sim$ 4.9; under these circumstances, it is commonly necessary 
        to correct the 2MASS $K_T$ magnitude slightly.  We determine a correction of $\Delta K_T = -0.185$ and apply 
        it to derive the photometric parameters listed in {\bf Table 3}. \vsn

NGC 1023: The asymptotic outer rotation velocity $V_{\rm circ} = 251 \pm 15$ km s$^{-1}$ is from Column (12) of
          Table 1 in Dressler \& Sandage (1983). \vsn

NGC 1277: We adopt the Perseus cluster distance and NGC 1277 BH mass from van den Bosch \etal (2012).
          However, our analysis of the host galaxy (Kormendy \& Bender 2013) is different from that of van den Bosch,
          who decomposes the light distribution into four radially overlapping components.  This is operationally 
          analogous to a multi-Gaussian expansion in the sense that it forces the S\'ersic indices of all components
          to be small.  Partly for this reason, they concluded that the bulge is not classical.  We find that the
          ellipticity at large radii is similar to the ellipticity near the center; this is a sign also seen in many
          edge-on S0s in the Virgo cluster (Kormendy \& Bender 2012) and indicates that the bulge dominates 
          at both small and large radii.  We decomposed the galaxy into two components such that the bulge
          dominates at both small and large radii.  The decomposition is robust, the bulge has a S\'ersic 
          index of $3.5 \pm 0.7$, and $B/T = 0.55 \pm 0.07$.  Both results imply that the bulge is classical. \vsn

NGC 2549: Krajnovi\'c \etal (2009) find that $M_\bullet = (1.4^{+0.2}_{-1.3}) \times 10^7$ $M_\odot$ for $D = 12.3$ Mpc,
          quoting 3-$\sigma$ errors.  In this case, dividing the 3-$\sigma$ error bars by 3 would obscure the 
          fact that this is an unusually weak BH detection.  We therefore read the 1-$\sigma$ errors directly from the
          $\chi^2$ contours shown in their paper.  The result is approximate, $M_\bullet = 1.45(0.31 - 1.65) \times 10^7$ $M_\odot$
          for our adopted $D = 12.70$ Mpc, but more realistic. \vsn

NGC 3115: Kormendy \& Richstone (1992) discovered the BH and got $M_\bullet$\ts=\ts1.0(0.3$-$3.3)$\times$\kern -0.5pt$10^9$\ts$M_\odot$ 
          from isotropic models and a smallest possible $M_\bullet$ = 1$\times$\null$10^8$\ts$M_\odot$ from the most extreme anisotropic 
          model that fit their CFHT kinematic data.  These are consistent with $M_\bullet$=0.90(0.62$-$0.95)$\times$\kern -0.5pt$10^9$\ts$M_\odot$
          adopted from Emsellem, Dejonghe \& Bacon (1999).  Additional measurements have ranged from 5$\times$\kern -0.5pt$10^8$\ts$M_\odot$ to 
          2$\times$\kern -0.5pt$10^9$\ts$M_\odot$ (Kormendy \etal 1996b; Magorrian \etal 1998). \vsn

NGC 3585: The outer rotation velocity $V_{\rm circ}$ for the embedded disk is from Scorza \& Bender (1995). \vsn

NGC 3998 is listed twice in {\bf Table 3}, once with the BH mass that we adopt from stellar dynamical models (Walsh
          \etal 2012) and once with the smaller BH mass based on the emission-line rotation curve (De Francesco \etal 2006).
          We illustrate this in {\bf Figure 12} as an example of why we do not use $M_\bullet$ values that are determined from
          ionized gas rotation curves when line widths are not taken into account.

          We use $B/T = 0.85 \pm 0.02$ from bulge-disk decompositions in Kormendy \& Bender (2013) and in
          S\'anchez-Portal \etal (2004).  With this $B/T$, NGC 3998 is the most significant bulge outlier to the 
          $M_\bullet$\ts--\ts$M_{K,\rm bulge}$ correlation, as Walsh \etal (2012) concluded.  There is a possibility that 
          a three-component, bulge-lens-disk decomposition is justified; if so, $B/T$ would be smaller, $\sim 0.66$.  
          Then NGC 3998 would be a more significant outlier, in the manner of NGC 4342 and the galaxies discussed in Section 6.5.
          For $\sigma_e$, we adopt the mean of $\sigma_e = 270$ km s$^{-1}$ found by Walsh for $r_e \simeq 18^{\prime\prime}$ 
          from our photometry and $\sigma_e = 280$ km s$^{-1}$ which we find using G\"ultekin's definition, our photometry,
          and kinematic data from Fisher (1997). \vsn

NGC 4258: The discovery paper for the spectacular H$_2$O maser disk and consequent accurate BH mass measurement is Miyoshi \etal (1995).
          Herrnstein \etal (1999) uses the masers to measure a direct geometric distance $D = 7.2 \pm 0.3$ Mpc to NGC 4258.
          Herrnstein \etal (1999) interpret small departures from precise Keplerian rotation in terms of a warped gas disk
          and derive an improved BH mass.  Our adopted mass is based in large part on this result.  Section 3.3 provides the
          details.  Sources for $V_{\rm circ}$ are listed in Kormendy \etal (2010).

          Given a ``bomb-proof'' accurate BH mass in a conveniently inclined galaxy, NGC 4258 has been used to test both
          stellar dynamical and ionized-gas-dynamical $M_\bullet$ measurement machinery (Sections 3.1 and 3.2, respectively). \vsn

NGC 4526: This is the brightest S0 galaxy in the Virgo cluster and the first galaxy to have a BH discovered using
          the central CO rotation curve (Davis \etal 2013).  We use $B/T = 0.65 \pm 0.05$ from Kormendy \& Bender (2013) 
          and $\sigma_e = 222 \pm 11$ km s$^{-1}$ from Davis \etal (2013), but we checked that $\sigma_e$ is consistent 
          with our definition of how to average $V^2(r) + \sigma^2(r)$.
          The asymptotic circular velocity is from Pellegrini, Held, \& Ciotti (1997), but it is uncertain whether
          the measured rotation curve reaches far enough out in this and almost any bulge-dominated S0. \vsn

NGC 4594 = M{\ts}104 = the Sombrero Galaxy: The BH discovery paper was Kormendy (1988), who obtained
           $M_\bullet = 5.5(1.7~-~17) \times 10^8$ $M_\odot$.  The quoted error bar was conservative, 
           but the best-fitting mass was within 17\ts\% of the present adopted value,
           $M_\bullet = 6.65(6.24~-~7.05) \times 10^8$\ts$M_\odot$ (Jardel \etal 2011).   The BH
           detection was confirmed at HST resolution by Kormendy \etal (1996a), but the mass was
           estimated only by reobserving at HST resolution a set of models that were designed for ground-based data.
           As a result, $M_\bullet \sim 1.1 \times 10^9$ $M_\odot$ was not very accurate.  
           Emsellem \etal (1994) measured $M_\bullet \sim 5.3 \times 10^8$ $M_\odot$ and
           Magorrian \etal (1998) got $M_\bullet = 6.9(6.7~-~7.0) \times 10^8$ $M_\odot$ based
           on two-integral models.   The presently adopted BH mass is based on three-integral models.

           We adopt the total magnitude measurement $B_T = 8.71$ in Burkhead (1986) and correct the
           2MASS $K$ magnitude to give $(V - K)_0 = 2.980$.  Also, $V_{\rm circ}$ is from Faber \etal (1977) 
           and Bajaja \etal (1984). \vsn

NGC 4596: We adopt the $M_{\rm BH,fix}$ mass in Table 2 of Sarzi \etal (2001).  Also, $B/T = 0.27 \pm 0.04$
          comes from comparing Benedict's (1976) decomposition at surface brightnesses \lapprox \ts23.5 B mag
          arcsec$^{-2}$ with the adopted total magnitude $B_T = 11.37$, i.{\ts}e., the mean of values in RC3,
          the Hyperleda main table, and the Hyperleda integrated photometry table.  The rotation velocity
          corrected for asymmetric drift is from Kent (1990). \vsn

NGC 7457: We confirm G\"ultekin's value of $\sigma_e = 67 \pm 3$ km s$^{-1}$ with our photometry and kinematic measurements. 
          The outer disk circular velocity is from our kinematic data and those of 
          Cherepashchuk \etal 2010, corrected for asymmetric drift by them but for our assumed
          inclination of the galaxy, $i = 59^\circ \pm 2^\circ$.

\vs
\cl {\big\ARRed S4.3.~Notes on Individual Disk Galaxies with Pseudobulges}\textBlack
\vs

Our Galaxy: Photometric parameters are discussed in Kormendy \& Bender (2013).  Our Galaxy requires special procedures because 
we live inside it.  For the convenience of readers, we summarize the provenance of the photometric parameters here.  The Galaxy 
has a boxy bulge (e.{\ts}g., 
Weiland \etal 1994;
Dwek \etal 1995;
Wegg \& Gerhard 2013)
that is understood as an almost-end-on bar
(Combes \& Sanders 1981;
Blitz \& Spergel 1991).
It is therefore a pseudobulge -- a component built out of the disk.  There is no photometric or kinematic sign of a classical bulge (see 
Freeman 2008,
Howard \etal 2009, 
Shen \etal 2010, and
Kormendy \etal 2010 
for reviews and for some of the evidence).  We average pseudobulge-to-total luminosity ratios from
Kent, Dame \& Fazio (1991) and
Dwek \etal (1995)
to get $PB/T = 0.19 \pm 0.02$.  To get $M_{KsT} = -23.7$, we adopt the total $K$-band luminosity
$L_K = 6.7 \times 10^{10}$ $L_{K\odot}$ from Kent, Dame \& Fazio (1991) and convert it from their assumed
distance of 8 kpc to our assumed distance of 8.28 kpc from Genzel \etal (2010).
The disk and pseudobulge absolute magnitudes follow from $PB/T$.
Finally, $V$-band magnitudes are derived from $K$-band magnitudes by assuming that $(V - K)_0 = 2.980$.
The bulge absolute magnitude is reasonably accurate, because the bulge is old; the main effect of 
estimating $M_{V,\rm bulge}$ from $M_{K,\rm bulge}$ is to implicitly correct for internal extinction.
The disk magnitude is much more uncertain, because the assumed color does not take young stars into account.
However, this has only minimal effects on our conclusions.

      The adopted BH mass is now securely derived from the orbits of individual stars.
The history of the remarkable improvement in $M_\bullet$ measurements is reviewed in Genzel \etal (2010);
early stages were covered in KR95.  The velocity dispersion $\sigma_e$ is from G\"ultekin \etal (2009c).

\vsn

Circinus is like M{\ts}31 in structure and inclination, but it is a smaller galaxy with a gas-rich
pseudobulge, and it has a smaller BH than M{\ts}31.  It is a difficult case, because it is close to 
the Galactic plane.  The Galactic absorption is large, and our estimates of it are uncertain.  Kormendy \& Bender (2013)
measure the galaxy's photometric parameters; the total apparent magnitude is $K_T = 4.71$.  The pseudobulge classification and 
$PB/T = 0.30 \pm 0.03$ are from the same paper and from Fisher \& Drory (2010).  We adopt $V_T = 10.60 \pm 0.04$ 
as the average of values tabulated in the RC3 (de~Vaucouleurs \etal 1991) and Hyperleda (Paturel \etal 2003).  
Comparing $K_T$ and $V_T$ in the context of various published estimates of the Galactic absorption,  we adopt $A_V = 3.15$ 
from Karachentsev \etal (2004), because it gives the most reasonable total color for the galaxy, $(V - K)_0 = 3.05$.  This 
then determines the other photometric parameters, including the distance $D = 2.82$ Mpc (Karachentsev \etal 2004).  
 
      Greenhill \etal (2003) measure masers both in outflowing gas and in a
well-defined, essentially edge-on accretion disk.  The latter masers show a well-defined Keplerian rotation
curve which implies that $M_\bullet = (1.14 \pm 0.20) \times 10^6$ $M_\odot$. We adopt this value, although
Hur\'e \etal (2011) find hints that $M_\bullet$ may be smaller.  We are uncomfortable about the conflicting published
velocity dispersion measurements: Oliva \etal (1995) measure 168 km s$^{-1}$ consistently (RMS = 10 km s$^{-1}$)
from four infrared CO bands; their instrumental resolutions ($\sigma_{\rm instr} \simeq 80$ and 51 km s$^{-1}$
should be sufficient.  But Maiolino \etal (1998) measure $\sigma \simeq 79 \pm 3$ km s$^{-1}$ at
$\sigma_{\rm instr} \simeq 64$ km s$^{-1}$ in the 2.3\ts--\ts2.4 $\mu$m CO bands using an integral-field 
spectrograph and AO; there is little gradient in the central 1\sd2 except that the nucleus has a
bulge-subtracted velocity dispersion of $\sigma = 55 \pm 15$ km s$^{-1}$.  More recently, Mueller S\'anchez 
\etal (2006) use SINFONI AO integral-field spectroscopy on the VLT to measure $\sigma \simeq 80$
km s$^{-1}$ in the central 0\sd4 $\times$ 0\sd4.  We adopt $\sigma_e = 79 \pm 3$ km s$^{-1}$.  \vsn

NGC 1068 is a prototypical oval galaxy (Kormendy \& Kennicutt 2004) with an unusually massive pseudobulge that is more than
         a magnitude more luminous and a factor of $\sim$\ts4 more massive than the classical bulge of M{\ts}31.
         Kormendy \& Bender (2013) find that the pseudobulge-to-total luminosity ratio is quite different in 
         the optical and infrared; $PB/T \simeq 0.41$ at $H$ but $\simeq 0.3$ at $r$ and $i$.  We use these values at $K$ and
         $V$, respectively.  We adopt $\sigma_e = 151 \pm 7$ km s$^{-1}$ from G\"ultekin \etal (2009c) and $V_{\rm circ} = 283 \pm 9$
         km s$^{-1}$ from Hyperleda but note that the latter value is uncertain.  We know of no two-dimensional analysis of the outer
         velocity field that takes the two differently oriented nested ovals into account; for a galaxy that is close to face-on,
         this is very important.
 
         The BH discovery papers are Gallimore \etal (1996) and Greenhill \etal (1996) who found and
         measured positionally resolved H$_2$O maser emission with the VLA and with VLBA, respectively.  The
         case is not as clean as that in NGC 4258, because the rotation velocity in the non-systemic-velocity
         sources decreases with increasing radius more slowly than a Keplerian, $V(r) \propto r^{-0.31 \pm 0.02}$
         (Greenhill \etal 1996).  The simplest and most plausible explanation is that the mass of the masing
         disk is not negligible with respect to the BH.  Ignoring this, the above papers derive a first approximation
         to $M_\bullet$ of $1 \times 10^7$ $M_\odot$.  Greenhill \& Gwinn (1997a) report additional VLBI observations
         and refine the total mass to $1.54 \times 10^7$ $M_\odot$.  Lodato \& Bertin (2003) confirm this: they get $M_\bullet =
         (1.66 \pm 0.02) \times 10^7$ $M_\odot$ using the approximation of a Keplerian rotation curve. 
         However, both Lodato \& Bertin (2003) and Hur\'e (2002; see also Hur\'e \etal 2011) derive models that account for the disk mass,
         and we adopt the average of their results, $M_\bullet = (8.39 \pm 0.44) \times 10^6$ $M_\odot$ (Section 3.3.3).

         Note again the extreme misalignment of the maser disk, which is essentially edge-on, and the rest of the
         galaxy, which is $\sim$\ts21$^\circ$ from face-on. \vsn

NGC 1300: We adopt $D$ (Local Group) = 21.5 Mpc, consistent with the distances to neighbors NGC 1297 and NGC 1232, all members
          of grouping 51 $-7$ $+4$ (Tully 1988).~However, Tonry \etal (2001) find $D$\ts=\ts28.5{\ts}Mpc for NGC\ts1297.
          We cannot tell whether there is a problem~with~one of the distances or whether NGC 1297 is fortuitously
          close to NGC\ts1300 in the sky but behind it by half of the distance from us to the Virgo cluster.  This is one example of
          a general problem: Distances remain uncertain, and we do not fold these uncertainties into our error estimates.

          The effective radius of the pseudobulge is $r_e \simeq 4\sd5 \pm 0\sd1$ (Fisher \& Drory 2008; Weinzirl \etal 2009).
          For $\sigma_e$, we use the mean dispersion $88 \pm 3$ km s$^{-1}$ interior to 3\sd5 as shown in Figure 6 of
          Batcheldor \etal (2005). 
          \vsn

NGC 2273: We calculate $\sigma_e = 125 \pm 9$ km s$^{-1}$ from our photometry and from kinematic data in Barbosa \etal (2006).
          % Also, $V_{\rm max} = 196 \pm 5$ km s$^{-1}$ is from H{\ts}I data in Noordermeer \etal (2007). Inconsistent with our Table 3
          \vsn

NGC 2787 is an example of a phenomenon that must be moderately common -- a galaxy that contains both a classical
         and a pseudo bulge.~Erwin \etal (2003) make a decomposition with $B/T = 0.11$ and $PB/T = 0.26$.  We adopt this
         decomposition to make the~above~point.  However, at our present level of understanding, trying to separate bulges from 
         pseudobulges is risky.  In other galaxies, we identify the dominant component and assign all of the (pseudo)bulge
         light to it.  Here, too, Columns 6 and 11 list the magnitudes of the bulge and pseudobulge together, and we treat
         this as a pseudobulge galaxy.  The outer rotation velocity $V_{\rm circ}$ is from Shostak (1987) and from van Driel
         \& van Woerden (1991). \vsn

NGC 3227: We use the corrected SBF $D = 23.75$ Mpc for companion galaxy NGC\ts3226 (Tonry \etal 2001). Mundell \etal (1995) provide $V_{\rm circ}$.\vsn

NGC 3368 is a pseudobulge-dominated S(oval)ab spiral galaxy with a central decrease in $\sigma$ at $r < 1^{\prime\prime}$ (Nowak \etal 2010).
          As emphasized by these authors, different definitions give different values of $\sigma_e$ and this affects whether or not the BH
          falls within the scatter of the $M_\bullet$\ts--\ts$\sigma_e$ relation.  Luminosity-weighted within the VLT SINFONI field of view
          of $3^{\prime\prime} \times 3^{\prime\prime}$, $\sigma = 98.5$ km s$^{-1}$ and the BH is consistent with $M_\bullet$\ts--\ts$\sigma_e$.
          However, we use the definition that $\sigma_e^2$ is the luminosity-weighted mean of $V^2 + \sigma^2$ within approximately $r_e$ (the
          exact radius makes little difference).  Because rotation contributes and because $r_e \simeq 11\sd2 \pm 2\sd7$ for the pseudobulge,
          we get a substantially larger value of $\sigma_e \simeq 125 \pm 6$ km s$^{-1}$.  This is based on our photometry and on kinematic
          data in H\'eraudeau \etal (1999) corrected inside $r_e/2$ to agree with Nowak \etal (2010).  Sarzi \etal (2002) derived
          $\sigma_e = 114 \pm 8$ km s$^{-1}$ using dispersions only; this shows approximately how much difference rotation makes to the
          definition.  Many pseudobulge galaxies are similar in that central velocity dispersions are much smaller than the $\sigma_e$
          that is obtained from the $V^2 + \sigma^2$ definition. 

          Nowak \etal (2010) suggest that NGC 3368 contains a small classical bulge in addition to the dominant pseudobulge.  We add them together.\vsn

NGC 3384: G\"ultekin \etal (2009c) used $\sigma_e = 143 \pm 7$ km s$^{-1}$.  We essentially confirm this: With our photometry and kinematic
          measurements, we get $\sigma_e = 150 \pm 8$ km s$^{-1}$.  We adopt the mean. \vsn

NGC 3393 is another prototypical oval galaxy with a large pseudobulge.  {\bf Figure 10} is included to emphasize its similarity to NGC 1068: 
         It is only $\sim$\ts$13^\circ$ from face-on (Cooke \etal 2000), but it contains an edge-on, masing accretion disk.
         
         All measurements of this galaxy are somewhat uncertain.  Kormendy \& Bender (2013) find a preliminary $PB/T = 0.27 \pm 0.06$.
         The BH mass measurement is based on the rotation curve of a masing molecular disk (Kondratko, Greenhill \& Moran 2008).  
         The maser sources are well distributed along a line indicative of an edge-on disk, but they cover only a small radius range, 
         so they do not securely measure the rotation curve shape.  They are consistent with Keplerian; this gives an enclosed~mass of 
         $M_\bullet = (3.4 \pm 0.2) \times 10^7$ $M_\odot$ at $r \leq 0.40 \pm 0.02$ pc for our $D$\ts=\ts49.2{\ts}Mpc.  But there are signs that 
         the rotation curve is slightly flatter than Keplerian.  For their best-fitting sub-Keplerian rotation curve, Kondratko \etal (2008) 
         get $M_\bullet \simeq 3 \times 10^7$ $M_\odot$.  In contrast, Hur\'e \etal (2011) find a good solution with a maser disk that 
         is 6 times as massive as the BH.  Then $M_\bullet \simeq 0.6 \times 10^7$ $M_\odot$.  This is the only galaxy in our sample 
         in which two such analyses give substantially different results.  We adopt the mean of the two masses and half of the difference
         as our error estimate.

         The velocity dispersion $\sigma_e$ is securely measured by Greene \etal (2010), but $V_{\rm circ}$ is too
         uncertain, because the galaxy is too close to face-on. \vsn

NGC 3489 is a weakly barred S0 with a dominant (pseudo)bulge that contributes $\sim$\ts35\ts\% of the light of the galaxy (Nowak \etal 2010).
         These authors argue plausibly that about one-third of this component is a classical bulge.  We conservatively add them together.  
         Also, we derive $\sigma_e = 113 \pm 4$ km s$^{-1}$ from our photometry and kinematic data in McDermid \etal (2006). \vsn

NGC 4388: We have only central velocity dispersion data for this galaxy.  Greene \etal (2010) measure $\sigma = 107 \pm 7$ km s$^{-1}$;
          Ho \etal (2009) get $91.7 \pm 9.5$ km s$^{-1}$, and we adopt the average, $\sigma_e = 99 \pm 10$ km s$^{-1}$.  This is likely
          to be an underestimate of $\sigma_e$ as we define it, because it neglects rotation inside the half-light radius $r_e \simeq 3\sd0$
          of the pseudobulge. \vsn

NGC 4736 and NGC 4826: We are most grateful to Karl Gebhardt for making $M_\bullet$ available before publication (Gebhardt \etal 2013).
         For NGC 4736, we calculated $\sigma_e$ from our photometry (Kormendy \& Bender 2013) and kinematic data in M\"ollenhoff \etal (1995).
         For NGC 4826, $\sigma_e$ is from our photometry and kinematic data in Rix \etal (1995).  For both galaxies, the result 
         is not significantly different if we integrate inside the pseudobulge $r_e \simeq 9\sd7$ and $16\sd7$, respectively (Fisher \& Drory 2008)
         or inside $r_e/2$.  Sources for $V_{\rm max}$ are given in Kormendy \etal (2010). \vsn

NGC 4945 is an edge-on, dusty Scd with a small pseudobulge ($PB/T = 0.07$) that is heavily absorbed at optical wavelengths.  
         We use $D = 3.58$ Mpc, i.{\ts}e., the mean of two TRGB distances determined from the magnitude of the tip of the red giant
         branch in the stellar color-magnitude diagram (3.36 Mpc: Mouhcine \etal 2005 and 3.80 Mpc: Mould \& Sakai 2008).
         We adopt $K_T = 4.438$, i.{\ts}e., the integral of the surface brightness and ellipticity profiles measured in Kormendy \& Bender (2013). 
         For comparison, 2MASS lists $K_s = 4.483$.  Our total magnitude implies a slightly more plausible color $(V - K)_0 = 2.80$.  

         The BH detection in Greenhill, Moran \& Herrnstein (1997b) was rejected by G\"ultekin \etal (2009c) because the maser 
         rotation curve is asymmetric and because the maser disk inclination is only approximately constrained to be edge-on by its 
         linear distribution at PA $\approx 45^\circ$.  Still, because the disk is masing and masing requires a long line-of-sight
         path length, the assumption that the disk is edge-on is at least as secure as
         many other assumptions that we routinely make. And the rotation curve decreases cleanly with radius on one side of the center.
         Thus $M_\bullet$ is not more uncertain than the most problematic cases based on stellar and ionized gas dynamics. 

         The maser disk and the galaxy disk have similar PA, are similarly edge-on, and rotate in the same direction (Greenhill \etal 1997b). \vsn

NGC 6264 and NGC 6323 are the most distant disk galaxies with maser BH detections.  Kormendy \& Bender (2013)
      measure $r$- and $K$-band brightness profiles, respectively.  No HST imaging is available for either galaxy, 
      although CFHT images with PSF dispersion radii of $\sigma_* = 0\sd23$ are available for NGC 6323.  Both galaxies
      have small pseudobulges; NGC 6264 has $PB/T \simeq 0.17 \pm 0.03$ and NGC 6323 has $PB/T \simeq 0.05 \pm 0.01$.    
      For both galaxies, we adopt $\sigma_e$ equal to the central velocity dispersion measured by Greene \etal (2010). \vsn

IC 2560: Evidence for a BH based on the dynamics of a H$_2$O maser disk was reported in Ishihara \etal (2001) and
         refined with further observations in Yamauchi \etal (2012).  We adopt $M_\bullet$ and its upper error bar
         from the latter paper.  However, only one point in the rotation curve is observed from high-$|V|$ masers,
         so we cannot tell whether the rotation curve is Keplerian.  Centripetal acceleration of the systemic masers
         is accurately measured, but their velocity gradient with position along the major axis of the disk
         is not accurately enough known to give a second meaningful $V(r)$ point as discussed in Section  3.3.2.
         Since some maser disks have gas masses that are significant fractions of the BH masses, we must regard the
         BH mass determination for IC 2560 as an upper limit.  Nevertheless, $M_\bullet$ is clearly small enough   
         to support our conclusion that BHs do not correlate with pseudobulges in the same way as they do with
         classical bulges. 

         We have only a central velocity dispersion measurement (Greene \etal 2010).  The outer rotation velocity
         $V_{\rm circ}$ is from Hyperleda. \vsn

UGC 3789: We adopt the geometric distance $D = 49.9 \pm 7.0$ Mpc from Braatz \etal (2010).  It provides a
          check of our D (Local Group) distances ({\bf Table 3}, Column 3, source 9); this distance would have been 47.8 Mpc. 
          The $K$ magnitude is from 2MASS, but the $V$ magnitude is estimated from $K$ using a photoelectric measurement of
          $B - V = 0.92$ in a 50$^{\prime\prime}$ diameter aperture by Arkhipova \& Saveleva (1984).  The correlation
          of $(V - K)_0$ with $(B - V)_0$ then gives $(V - K)_0 \simeq 2.90$.

          UGC 3789 is another example of a common phenomenon for which we know no explanation: It is a 
          prototpical, almost-face-on oval galaxy with an edge-on molecular disk surrounding the BH ({\bf Figure 10}).
          The velocity dispersion is from Greene \etal (2010) and $V_{\rm circ}$ is from Hyperleda. \vsn

\vs
\cl {\big\ARRed S4.4.~Notes on Discarded Galaxies {\ARDiscard (Purple Lines in Tables 2 and 3})\ARRed:}\textBlack
\vs

NGC 2778 is discarded (1) because it provides only an upper limit on $M_\bullet$ (Schulze \& Gebhardt 2011)
         and (2) because the implied mass-to-light ratio $M/L_K = 3.3$ is too large for an old stellar population
         from which the dark matter has been subtracted and modeled separately (Section 6.6).  We conclude that
         dark matter is still included and therefore that the $M_\bullet$ upper limit is substantially too small.
         This galaxy is a good illustration of the importance of adding $M/L$ constraints to our mass measurements,
         something that has not heretofore been done \vsn

      NGC 3607 is a core elliptical with a BH detection and $M_\bullet$ measurement in G\"ultekin \etal (2009b).
The modeling did not include dark matter.  Schulze \& Gebhardt (2011) and Rusli \etal (2013) show
that this is not a large problem if $r_{\rm infl}$ is very well resolved, but they show that $M_\bullet$ is
systematically underestimated if $r_{\rm infl}/\sigma_*$ \lapprox \ts5.  Both papers provide calibrations of
how the $M_\bullet$ correction factor depends on  $r_{\rm infl}/\sigma_*$, but the two calibrations disagree
even though they both use variants of the Nuker code.  The disagreement may result from technical details such 
as the number of orbits used in the modeling.  But the problem is severe for NGC 3607, for which the apparent
value of $r_{\rm infl}/\sigma_* \simeq 1.5$.  Based on the Schulze calibration, we should multiply the
G\"ultekin $M_\bullet = 1.4 \times 10^8$ $M_\odot$ ({\bf Figure 12}) by a factor of $\sim$2.~The Rusli calibration
gives a factor~of~$\sim$4.7.  We~conclude that the BH detection is reliable but that we do not
know the BH mass well enough to retain NGC 3607 in our sample.  \vsn

NGC 4261, NGC 6251, NGC 7052, A1836 BCG, and A3565 BCG: All of these galaxies have valid BH detections based on
          optical emission-line kinematic observations in the papers listed in Column 13.  However, the widths
          of the emission lines are comparable to the rotation velocities near the center, and these widths were
          not taken into account in estimating $M_\bullet$.  As discussed in the BH discovery papers and in Section 3.2
          here, it is not guaranteed that line widths imply a ``contribution'' to $M_\bullet$ as they would for
          absorption-line velocity dispersions.  But it is likely that they are not ignorable.  In Section 6.2 ({\bf Figure 12}), 
          we compare these BH masses to the BH--host-galaxy correlations that we derive for the most reliable BH masses.
          We find that the above galaxies do indeed have anomalously small BH masses.  The conservative conclusion 
          therefore is that neglecting emission-line widths can result in underestimated BH masses.  We therefore 
          omit all such masses, even those for lower-luminosity bulges in which the measured $M_\bullet$ does not
          obviously deviate from the correlations (e.{\ts}g., NGC 4459). \vsn

Cygnus A: The BH discovery (Tadhunter \etal 2003) is based on the optical emission-line rotation curve measured with HST STIS.
          The authors note that the emission lines are very broad, but they do not include line widths in their $M_\bullet$
          determination.  Therefore the BH mass is probably underestimated.
          The velocity dispersion is from Thornton \etal (1999) and is very uncertain; it is based on the observed width of the
          Ca infrared triplet absorption lines but not on any of the standard methods of comparing the spectrum to a standard star.
          Finally, although we used the brightest $V$-band magnitude in the literature (from Paturel \etal 2000 as listed by Hyperleda),
          the $(V - K)_0 = 3.54$ color is implausibly large.  It may be affected by large internal absorption.  If so, the $K$-band 
          magnitude may be usable.  Nevertheless, for all of the above reasons, this galaxy is plotted in Section 6.2 ({\bf Figure 12})
          and thereafter is omitted from all correlations and fits.
          
}

\vs\vs
\cl {\big\ARRed ACKNOWLEDGMENTS}\textBlack
\vs\vskip -1pt

%\vfill\eject

%  Reset Normal text page size, font size, and line spacing.

%\hsize=15.0truecm  \hoffset=0.0truecm  \vsize=20.1truecm  \voffset=1.5truecm

\rm\refbaselines

      We warmly thank Mary Kormendy for help in proofreading this paper and in checking its references.

      We made extensive use of NASA's Astrophysics Data System bibliographic services and the 
NASA/IPAC Extragalactic Database (NED).
NED is operated by the Jet Propulsion Laboratory and the California Institute of Technology 
under contract with NASA.  We also used the HyperLeda electronic database (Paturel \etal 2003)
at {\bf http://leda.univ-lyon1.fr}.
And we made extensive use of the Two Micron All Sky Survey (2MASS: Skrutskie \etal 2006).  
2MASS is a joint project of the University of Massachusetts and the Infrared Processing and 
Analysis Center/California Institute of Technology, funded by NASA and the NSF.  

      JK's research was supported in part by NSF grant AST-0607490.  Also, this multi-year 
research project would not have been possible without the long-term support provided to JK 
by the Curtis T.~Vaughan, Jr.~Centennial Chair in Astronomy.  We are most sincerely grateful to
Mr.~and Mrs.~Curtis T.~Vaughan, Jr.~for their continuing support of Texas astronomy.  JK also warmly 
thanks Ralf Bender and the staffs of the University Observatory,
Ludwig-Maximilians-University, Munich, Germany and the Max-Planck-Institute for Extraterrestrial
Physics, Garching-by-Munich, Germany for their support and hospitality during several visits during 
which much of this review was written.  

     LCH's work is supported by the Carnegie Institution for Science
and by NASA grants from the Space Telescope Science Institute (operated by AURA, 
Inc., under NASA contract NAS5-26555).   LCH thanks the Chinese Academy of
Sciences and the hospitality of the National Astronomical Observatories, where
part of this review was written.

\vs\vs
\cl {\big\ARRed SUPPLEMENTARY LITERATURE CITED}\textBlack
\vs\vskip -1pt
\def\nhi{\noindent  \hangindent=1.0truecm}

\rm\refbaselines

\frenchspacing

\nhi Arkhipova, V. P., \& Saveleva, M. V. 1984, Trudy Gosudarstvennogo Astronomicheskogo Instituta P. K. Sternberga, 54, 33

\nhi Atkinson, J. W., Collett, J. L., Marconi, A., \etal 2005, MNRAS, 359, 504

\nhi Bahcall, J. N., Kozlovsky, B.-Z., \& Salpeter, E. E. 1972, ApJ, 171, 467

\nhi Bajaja, E., van der Burg, G., Faber, S. M., \etal 1984, A\&A, 141, 309

%\nhi Baldwin, J., Ferland, G., Korista, K., \& Verner, D. 1995, ApJ, 455, L119

\nhi Barbosa, F.~K.~B., Storchi-Bergmann, T., Cid Fernandes, R., Winge, C., \& Schmitt, H.~2006, MNRAS, 371, 170

\nhi Barth, A.~J., Pancoast, A., Thorman, S. J., et al. 2011, ApJ, 743, L4

\nhi Barth, A. J., Sarzi, M., Rix, H.-W., \etal 2001, ApJ, 555, 685  %  NGC 3245

\nhi Baskin, A., \& Laor, A. 2005, MNRAS, 356, 1029 

\nhi Batcheldor, D., Axon, D., Merritt, D., \etal 2005, ApJS, 160, 76

%\nhi Bell, E. F., \& de Jong, R. S. 2001, ApJ, 550, 212

\nhi Bender, R., Kormendy, J., Bower, G., \etal 2005, ApJ, 631, 280

\nhi Benedict, G. F. 1976, AJ, 81, 799

\nhi Bennert, N., Schulz, H., \& Henkel, C. 2004, A\&A, 419, 127

%\nhi Bentz, M.{\ts}C., Denney, K.{\ts}D., Cackett, E.{\ts}M., \etal 2007, ApJ, 662, 205 %  NGC 5548 study in our Figure 0

\nhi Bentz, M. C., Peterson, B. M., Netzer, H., Pogge, R. W., \& Vestergaard, M.   2009a, ApJ, 697, 160     % lag - L_5100 7 b/T for RM galx

\nhi Bentz, M.{\ts}C., Peterson, B.{\ts}M., Pogge, R.{\ts}W., Vestergaard, M., \& Onken, C.{\ts}A. 2006, ApJ, 644, 133      %  r-L correlation for AGNs

\nhi Bentz, M. C., Walsh, J. L., Barth, A. J., \etal  2008, ApJ, 689, L21      % Arp 151 cleanest infall signature from V-resolved RM

\nhi Bentz, M. C., Walsh, J. L., Barth, A. J., \etal  2009b, ApJ, 705, 199     %  Lick AGN project = reverb on 12 AGNs

\nhi Bentz, M. C., Walsh, J. L., Barth, A. J., \etal  2010, ApJ, 716, 993      %  LAMP data and V-resolved RM

\nhi Bessell, M. S. 2005, ARA\&A, 43, 293

\nhi Blakeslee, J. P., Cantiello, M., Mei, S., \etal 2010, ApJ, 724, 657

\nhi Blakeslee, J. P., Jord\'an, A., Mei, S., \etal 2009, ApJ, 694, 556

\nhi Blandford, R. D., \& McKee, C. F. 1982, ApJ, 255, 419    %     First suggestion of and name of reverberation mapping

\nhi Blitz, L., \& Spergel, D. N. 1991, ApJ, 379, 631

\nhi Bochkarev, N. G., \& Gaskell, C. M. 2009, ALett, 35, 287

\nhi Bower, G. A., Green, R. F., Bender, R., \etal 2001, ApJ, 550, 75    %  NGC 1023

\nhi Bower, G. A., Green, R. F., Danks, A., \etal  1998, ApJ, 492, L111  %  NGC 4374 BH

\nhi Bower, G. A., Heckman, T. M., Wilson, A. S., \& Richstone, D. O. 1997, ApJ, 483, L33   %  NGC 4374 emission-line disk

\nhi Bower, G. A., Wilson, A. S., Heckman, T. M., \etal 2000, BAAS, 32, 1566  %  MBH for M81 and NGC 3998

\nhi Braatz, J. A., Reid, M. J., Humphreys, E. M. L., \etal 2010, ApJ, 718, 657    % U3789

\nhi Brewer, B. J., Treu, T., Pancoast, A., et al. 2011, ApJ, 733, L33

\nhi Burkhead, M. S. 1986, AJ, 91, 777

\hsize=15.0truecm  \hoffset=0.0truecm  \vsize=21.4truecm  \voffset=1.5truecm

\nhi Cappellari, M., Neumayer, N., Reunanen, J., \etal 2009, MNRAS, 394, 660                     %  Cen A

\nhi Cappellari, M., Scott, N., Alatalo, K., \etal 2013, MNRAS, 432, 1709 

\nhi Cappellari, M., Verolme, E. K., van der Marel, R. P., \etal 2002, ApJ, 578, 787             %  IC1459

\nhi Carpenter, J. M. 2001, AJ, 121, 2851

\nhi Chemin, L., Carignan, C., \& Foster, T. 2009, ApJ, 705, 1395

\nhi Cherepashchuk, A. M., Afanas'ev, V. L., Zasov, A. V., \& Katkov, I. Yu. 2010, Astr. Reports, 54, 578

\nhi Cherepashchuk, A. M., \& Lyutyi, V. M. 1973, ApLett, 13, 165

\nhi Collin, S., Kawaguchi, T., Peterson, B. M., \& Vestergaard, M. 2006, A\&A, 456, 75

\nhi Combes, F., \& Sanders, R. H. 1981, A\&A, 96, 164

\nhi Cooke, A. J., Baldwin, J. A., Ferland, G. J., Netzer, H., \& Wilson, A. S. 2000, ApJS, 129, 517  %  i in NGC 3393

\nhi Courteau, S., Widrow, L. M., McDonald, M., \etal 2011, ApJ, 739, 20

\nhi Cretton, N., \& van den Bosch, F. C. 1999, ApJ, 514, 704   %   NGC 4342

\nhi Croom, S. M. 2011, ApJ, 736, 161  %  Pblms with virial masses for quasars

\nhi Dalla Bont\`a, E., Ferrarese, L., Corsini, E. M., \etal 2009, ApJ, 690, 537

\nhi Davies, R. I., Thomas, J., Genzel, R., \etal 2006, ApJ, 646, 754

\nhi Davis, T. A., Bureau, M., Cappellari, M., Sarzi, M., \& Blitz, L. 2013, Nature, 494, 328

\nhi De Francesco, G., Capetti, A., \& Marconi A. 2006, A\&A, 460, 439  %  NGC 3998 BH

\nhi De Francesco, G., Capetti, A., \& Marconi A. 2008, A\&A, 479, 355  %  NGC 5077 BH

\nhi Denney, K. D. 2012, ApJ, 759, 44

\nhi Denney, K. D., Peterson, B. M., Pogge, R. W., \etal 2009, ApJ, 704, L80  %  Outflows exist in reverb BLRs

\nhi Denney, K. D., Peterson, B. M., Pogge, R. W., \etal 2010, ApJ, 721, 715

\nhi de Vaucouleurs, G. 1961, ApJS, 6, 213

\nhi de Vaucouleurs, G., de Vaucouleurs, A., Corwin, H. G., \etal 1991, Third Reference Catalogue of Bright Galaxies 
     (Berlin: Springer) (RC3)

\nhi Devereux, N., Ford, H., Tsvetanov, Z., \& Jacoby, G. 2003, AJ, 125, 1226  %  M81 MBH via emission lines 

\nhi Dibai, \'E. A. 1977, Sov. Astron. Lett., 3, 1

\nhi Dibai, \'E. A. 1984, Sov. Astron., 28, 245

\nhi Dressler, A., \& Sandage, A. 1983, ApJ, 265, 664

\nhi Drory, N., \& Fisher, D. B. 2007, ApJ, 664, 640

\nhi Dwek, E., Arendt, R. G., Hauser, M. G., \etal 1995, ApJ, 445, 716

\nhi Emsellem, E., Cappellari, M., Krajnovi\'c, D., \etal 2007, MNRAS, 379, 401  %  Kinematic classification

\nhi Emsellem, E., Cappellari, M., Peletier, R. F., \etal 2004, MNRAS, 352, 721

\nhi Emsellem, E., Dejonghe, H., \& Bacon, R. 1999, MNRAS, 303, 495   %  NGC 3115

\nhi Emsellem, E., Monnet, G., Bacon, R., \& Nieto, J.-L. 1994, A\&A, 285, 739  %  NGC 4594

\nhi Erwin, P., Vega Beltr\'an, J. C., Graham, A. W., \& Beckman, J. E.~2003, ApJ, 597, 929  %  NGC 2787 

\nhi Faber, S. M., Balick, B., Gallagher, J. S., \& Knapp, G. R. 1977, ApJ, 214, 383

\nhi Faber, S. M., \& Jackson, R. E. 1976, ApJ, 204, 668

\nhi Falcke, H., K\"ording, E., \& Markoff, S. 2004, A\&A, 414, 895

\nhi Ferrarese, L., \& Ford, H. C. 1999, ApJ, 515, 583            %  NGC 6251

\nhi Ferrarese, L., Ford, H. C., \& Jaffe, W. 1996, ApJ, 470, 444   %  NGC 4261

\nhi Ferrarese, L., \& Merritt, D. 2000, ApJ, 539, L9

\nhi Ferrarese, L., Pogge, R.~W., Peterson, B.~M., et al. 2001, ApJ, 555, L79

\vfill\eject

\nhi Fisher, D. 1997, AJ, 113, 950

\nhi Fisher, D. B., \& Drory, N. 2008, AJ, 136, 773

\nhi Fisher, D. B., \& Drory, N. 2010, ApJ, 716, 942

\nhi Freeman, K. C. 2008, in Formation and Evolution of Galaxy Disks, ed. J. G. Funes \& E. M. Corsini 
     (San Francisco, CA: ASP), 3

\nhi Gallimore, J. F., Baum, S. A., O'Dea, C. P., Brinks, E., \& Pedlar, A. 1996, ApJ, 462, 740

\nhi Gaskell, C. M. 1988, ApJ, 325, 114  %  First reverberation mapping in NGC 4151

\nhi Gaskell, C. M. 2008, Rev. Mex. A\&A, 32, 1  %  Nice table of radii, temp, etc, for NGC 5548

\nhi Gaskell, C. M. 2009, NewAR, 53, 140  %  What broad emission lines tell us about how AGNs work

\nhi Gaskell, C. M. 2010, in ASP Conference Series, Vol. 427, Accretion and Ejection in AGNs: A Global View, 
     ed. L. Maraschi, G. Ghisellini, R. Della Ceca, \& F. Tavecchio (San Francisco: ASP), 68

%\nhi Gaskell, C. M. 2010b, ApJ, submitted (arXiv:1008.1057) 
%
%XX I would advise against citing papers that have clearly been rejected; 
%   the 2008 and 2011 references suffice

\nhi Gaskell, C. M. 2011, Baltic A, 20, 392

\nhi Gaskell, C. M, \& Sparke L. S. 1986, ApJ, 305, 175   %    Launched reverb mapping; we ref it for downward-revised r_BLR

\nhi Gebhardt, K., Adams, J., Richstone, D., \etal 2011, ApJ, 729, 119          % BH in M87 -- Gemini AO

\nhi Gebhardt, K., Kormendy, J., Ho, L. C., \etal 2000a, ApJ, 543, L5            %  MBH <-- reverberation (checked)

\nhi Gebhardt, K., Lauer, T. R., Pinkney, J., \etal 2007,  ApJ, 671, 1321       %  BH in NGC 1399 (checked)

\nhi Gebhardt, K., Richstone, D., Kormendy, J., \etal 2000b, AJ, 119, 1157      %  MBH in NGC 3379 (checked)

\nhi Gebhardt, K., Richstone, D., Tremaine, S., \etal 2003,  ApJ, 583, 92       %  models --> MBH (checked)

\nhi Gebhardt, K., \& Thomas, J. 2009, ApJ, 700, 1690 

\nhi Gebhardt, K., \etal 2013, in preparation                                   %  BHs in NGC 4736 and NGC 4826

\nhi Genzel, R., Eisenhauer, F., \& Gillessen, S. 2010, Rev. Mod. Phys., 82, 3121 

\nhi Graham, A. W., Colless, M. M., Busarello, G., Zaggia, S., \& Longo, G. 1998, A\&AS, 133, 325

\nhi Graham, A. W., Onken, C. A., Athanassoula, E., \& Combes, F. 2011, MNRAS, 412, 2211

\nhi Graham, A. W., \& Scott, N. 2013, ApJ, 764, 151

\nhi Greene, J. E., \& Ho, L. C. 2005, ApJ, 630, 122

\nhi Greene, J. E., \& Ho, L. C. 2006, ApJ, 641, L21

\nhi Greene, J. E., \& Ho, L. C. 2009, PASP, 121, 1167

\nhi Greene, J. E., Peng, C. Y., Kim, M., \etal 2010, ApJ, 721, 26  %  maser BHs, 10 authors

\nhi Greenhill, L. J., Booth, R. S., Ellingsen, S. P., \etal 2003, ApJ, 590, 162  %  Circinus MBH

\nhi Greenhill, L. J., \& Gwinn, C. R. 1997a, ApSS, 248, 261           %  N1068

\nhi Greenhill, L. J., Gwinn, C. R, Antonucci, R., \& Barvainis, R. 1996, ApJ, 472, L21 % N1068 Discover paper 2

\nhi Greenhill, L. J., Moran, J. M., \& Herrnstein, J. R. 1997b, ApJ, 481, L23           %  N4945 BH mass

\nhi Grier, C. J., Peterson, B. M., Horne, K., et al. 2013, ApJ, 764, 47

\nhi Grier, C. J., Peterson, B. M., Pogge, R. W., et al. 2012, ApJ, 755, 60

\nhi G\"ultekin, K., Cackett, E. M., Miller, J. M., \etal 2009a, ApJ, 706, 404                   %  BH FP

\nhi G\"ultekin,{\ts}K., Richstone,{\ts}D.{\ts}O., Gebhardt,{\ts}K., \etal 2009b, ApJ, 695, 1577 %  5 Detections

\nhi G\"ultekin,{\ts}K., Richstone,{\ts}D.{\ts}O., Gebhardt,{\ts}K., \etal 2009c, ApJ, 698, 198  %  Scatter paper

\nhi G\"ultekin, K., Richstone, D. O., Gebhardt, K., \etal 2011,  ApJ, 741, 38                   %  N4382

\nhi Hardcastle, M. J., Evans, D. A., \& Croston, J. H. 2009, MNRAS, 396, 1929                   %  BH FP

\nhi H\"aring-Neumayer, N., Cappellari, M., Rix, H.-W., \etal 2006, ApJ, 643, 226                %  Cen A

\nhi Harms, R. J., Ford, H. C., Tsvetanov, Z. I., \etal 1994, ApJ, 435, L35

\nhi Henkel, C., Braatz, J. A., Greenhill, L. J., \& Wilson, A. S. 2002, A\&A, 394, L23

\nhi H\'eraudeau, Ph., Simien, F., Maubon, G., \& Prugniel, Ph. 1999, A\&AS, 136, 509

\nhi Herrnstein, J. R., Moran, J. M., Greenhill, L. J., \etal 1999, Nature, 400, 539             %  Geometric D to NGC 4258

\nhi Ho, L. C. 2008, ARA\&A, 46, 475

\nhi Ho, L.~C., Goldoni, P., Dong, X.-B., Greene, J. E., \& Ponti, G. 2012, ApJ, 754, 11

\nhi Ho, L. C., Greene, J. E., Filippenko, A. V., \& Sargent, W. L. W. 2009, ApJS, 183, 1        %  sigma catalog

\nhi Hopkins, P. F., Cox, T. J., Dutta, S. N., \etal 2009, ApJS, 181, 135 %  extra light Es

\nhi Houghton, R. C. W., Magorrian, J., Sarzi, M., \etal 2006, MNRAS, 367, 2  %  NGC 1399

\nhi Howard, C. D., Rich, R. M., Clarkson, W., \etal 2009, ApJ, 702, L153

\nhi Huang, S., Ho, L. C., Peng, C. Y., Li, Z.-Y., \& Barth, A. J. 2013, ApJ, 766, 47

\nhi Hur\'e, J.-M. 2002, A\&A, 395, L21

\nhi Hur\'e, J.-M., Hersant, F., Surville, C., Nakai, N., \& Jacq, T. 2011, A\&A, 530, 145

\nhi Ishihara, Y., Nakai, N., Iyomoto, N., \etal 2001, PASJ, 53, 215

\nhi Jardel J. R., Gebhardt, K., Shen, J., \etal 2011, ApJ, 739, 21

\nhi Jarrett, T. H., Chester, T., Cutri, R., Schneider, S. E., \& Huchra, J. P. 2003, AJ, 125, 525

\nhi Johnson, H. L. 1962, ApJ, 135, 69

\nhi Karachentsev,{\ts}I.{\ts}D., Karachentseva,{\ts}V.{\ts}E., Huchtmeier,{\ts}W.{\ts}K., \& Makarov,{\ts}D.{\ts}I.~2004, AJ, 127, 2031 

\nhi Kaspi, S., Brandt, W. N., Maoz, D., \etal 2007, ApJ, 659, 997

\nhi Kaspi, S., Maoz, D., Netzer, H., et al. 2005, ApJ, 629, 61

\nhi Kaspi, S., Smith, P. S., Netzer, H., et al. 2000, ApJ, 533, 631

\nhi Kent, S. M. 1990, AJ, 100, 377

\nhi Kent, S. M, Dame, T. M., \& Fazio, G. 1991, ApJ, 378, 131

%\nhi Kollatschny, W. 2003, A\&A, 407, 461             % Keplerian lag-V relation in reverb. mapping

\nhi Kollatschny, W., \& Zetzl, M. 2011, Nature, 470, 366                % 

\nhi Komatsu, E., Dunkley, J., Nolta, M. R., \etal 2009, ApJS, 180, 330  % WMAP 5-yr cosmology

\nhi Kondratko, P. T., Greenhill, L. J., \& Moran, J. M. 2008, ApJ, 678, 87   %  NGC 3393 BH

\nhi Koratkar, A. P., \& Gaskell, C. M. 1989, ApJ, 345, 637

\nhi K\"ording, E. 2008, in Proceedings of the VII Microquasar Workshop: Microquasars and Beyond,
     Foca, Izmir, Turkey; published online at {\bf http://pos.sissa.it/cgi-bin/reader/conf.cgi?confid=62}, 23

\nhi Kormendy, J. 1988, ApJ, 335, 40

\nhi Kormendy, J. 1992a, in Testing the AGN Paradigm, ed. S. S. Holt, S. G. Neff, \& C. M. Urry (New York: AIP), 23

\nhi Kormendy, J. 1992b, in High Energy Neutrino Astrophysics, ed. V. J. Stenger, J. G. Learned, S. Pakvasa, 
     \& X. Tata (Singapore: World Scientific), 196

\nhi Kormendy, J. 1999, in ASP Conference Series, Vol. 182, Galaxy Dynamics: A Rutgers Symposium, 
     ed. D. Merritt, J. A. Sellwood, \& M. Valluri (San Francisco: ASP), 124

\nhi Kormendy, J., \& Bender, R. 2012, ApJS, 198, 2

\nhi Kormendy, J., \& Bender, R. 2013, Disk Galaxy Hosts of Dynamically Detected Supermassive 
     Black Holes, in preparation

\nhi Kormendy, J., Bender, R., Ajhar, E. A., \etal 1996a, ApJ, 473, L91                                %   BH in NGC 4594

\nhi Kormendy, J., Bender, R., Evans, A. S., \& Richstone, D. 1998, AJ, 115, 1823                      %  BH in NGC 3377 

\nhi Kormendy, J., Bender, R., Magorrian, J., \etal 1997, ApJ, 482, L139                               %   BH in NGC 4486B

\nhi Kormendy, J., Bender, R., Richstone, D., \etal 1996b, ApJ, 459, L57                               %   BH in NGC 3115

\nhi Kormendy, J., Drory, N., Bender, R., \& Cornell, M. E. 2010, ApJ, 723, 54 

\nhi Kormendy, J., Fisher, D. B., Cornell, M. E., \& Bender, R. 2009, ApJS, 182, 216 (KFCB)

\nhi Kormendy, J., \& Freeman, K. C., 2013, Scaling Laws for Dark Matter Halos in Late-Type and
     Dwarf Spheroidal Galaxies, in preparation

\nhi Kormendy, J., Gebhardt, K., Fisher, D. B., \etal 2005, AJ, 129, 2636                               %  NGC 4486A

\nhi Kormendy, J., \& Kennicutt, R. C. 2004, ARA\&A, 42, 603

\nhi Kormendy, J., \& Richstone, D. 1992, ApJ, 393, 559 %(KR92)

\nhi Krajnovi\'c, D., McDermid, R. M., Cappellari, M., \& Davies, R. L. 2009, MNRAS, 399, 1839          %  NGC 524 and 2549

\nhi Krajnovi\'c, D., Sharp, R., \& Thatte, N. 2007, MNRAS, 374, 385  %  Cen A

\nhi Krolik, J. H. 2001, ApJ, 551, 72

\nhi Kuo, C. Y., Braatz, J. A., Condon, J. J., \etal 2011, ApJ, 727, 20 

%\nhi Kurk, J. D., Walter, F., Fan, X., et al. 2007, ApJ, 669, 32  %  z ~ 6 quasars

\nhi Lauer, T. R., Ajhar, E. A., Byun, Y.-I., \etal 1995, AJ,   110, 2622        %  Nuker 1

%\nhi Lauer, T. R., Faber, S. M., Currie, D. G., \etal 1992, ApJ,  104, 552      %  PC observations of M32

\nhi Lauer, T. R., Faber, S. M., Gebhardt, K., \etal 2005, AJ,   129, 2138       %  Paper V = nuclear mags

\nhi Lauer, T. R., Faber, S. M., Lynds, C. R., \etal 1992, AJ, 103, 703          %  PC observations of M87

\nhi Lauer, T. R., Gebhardt, K., Faber, S. M., \etal 2007, ApJ, 664, 226         %  Paper 6 = Bimodal 

\nhi Lauer, T. R., Gebhardt, K., Richstone, D., \etal 2002,  AJ,  124, 1975      %  Hollow cores

\nhi Lodato, G., \&  Bertin, G. 2003, A\&A, 398, 517

\nhi Lyutyi, V. M., \& Cherepashchuk, A. M. 1972, Astron. Tsirk., 688, 1

\nhi Macchetto, F., Marconi, A., Axon, D. J., \etal 1997, ApJ, 489, 579

\nhi Maciejewski, W., \& Binney, J. 2001, MNRAS, 323, 831   %   NGC 4374

\nhi Magorrian, J., Tremaine, S., Richstone, D., \etal 1998, AJ, 115, 2285

\nhi Maiolino, R., Krabbe, A., Thatte, N., \& Genzel, R. 1998, ApJ, 493, 650

\nhi Mamyoda, K., Nakai, N., Yamauchi, A., Diamond, P., \& Hur\'e, J.-M. 2009, PASJ, 61, 1143  %  IC 1481

\nhi Marconi, A., Capetti, A., Axon, D. J., \etal 2001, ApJ,    549, 915  %  Cen A

\nhi Marconi, A., Pastorini, G., Pacini, F., \etal 2006, A\&A,  448, 921  %  Cen A

\nhi Marziani, P., \& Sulentic, J. W. 2012, NewAR, 56, 49

%\nhi Marziani, P., Sulentic, J. W., Stirpe, G. M., Zamfir, S., \& Calvani, M. 2009, A\&A, 495, 83  %  z = 2 quasars

\nhi McConnell, N. J., \& Ma, C.-P. 2013, ApJ, 764, 184

\nhi McConnell, N. J., Ma, C.-P., Gebhardt, K., \etal 2011a, Nature, 480, 215

\nhi McConnell, N. J., Ma, C.-P., Graham, J. R., \etal 2011b, ApJ, 728, 100

\nhi McConnell, N. J., Ma, C.-P., Murphy, J. D., \etal 2012, ApJ, 756, 179

\nhi McDermid, R. M., Emsellem, E., Shapiro, K. L. \etal 2006, MNRAS, 373, 906

\nhi McGill, K. L., Woo, J.-H., Treu, T., \& Malkan, M. A. 2008, ApJ, 673, 703

\nhi McLure, R.~J., \& Dunlop, J.~S. 2004, MNRAS, 352, 1390

\nhi McLure, R.~J., \& Jarvis, M.~J. 2002, MNRAS, 337, 109

\nhi Mei, S., Blakeslee, J. P., C\^ot\'e, P., \etal 2007, ApJ, 655, 144

\nhi M\'endez-Abreu, J., Aguerri, J. A. L., Corsini, E. M., \& Simonneau, E. 2008, A\&A, 478, 353  %  B/T = 0.64 for NGC 6861

\nhi Merloni, A., Heinz, S., \& Di Matteo, T. 2003, MNRAS, 345, 1057 

\nhi Miyoshi, M., Moran, J., Herrnstein, J., \etal 1995, Nature, 373, 127  %  N4258 maser BH (checked)

\nhi M\"ollenhoff, C., Matthias, M., \& Gerhard, O. E. 1995, A\&A, 301, 359

\nhi Monachesi, A., Trager, S. C., Lauer, T. R., \etal 2011, ApJ, 727, 55

\nhi Mouhcine, M., Ferguson, H. C., Rich, R. M., Brown, T. M., \& Smith, T. E. 2005, ApJ, 633, 810

%\nhi Mouhcine, M., \& Lan\c on, A. 2003, A\&A, 402, 425

\nhi Mould, J., \& Sakai, S. 2008, ApJ, 686, L75

\nhi Mueller S\'anchez, F., Davies, R. I., Eisenhauer, F., \etal 2006, A\&A, 454, 481

\nhi Mundell, C. G., Pedlar, A., Axon, D. J., Meaburn, J., \& Unger, S. W. 1995, MNRAS, 277, 641

\nhi Nagino, R., \& Matsushita, K. 2009, A\&A, 501, 157

\nhi Nelson, C. H., Green, R. F., Bower, G., Gebhardt, K., \& Weistrop, D. 2004, ApJ, 615, 652

\nhi Netzer, H. 1990, in Active Galactic Nuclei, ed.{\ts}R.{\ts}D.{\ts}Blandford, H.{\ts}Netzer, \& L.{\ts}Woltjer 
     (Berlin:{\ts}Springer),~57

\nhi Netzer, H., Lira, P., Trakhtenbrot, B., Shemmer, O., \& Cury, I. 2007, ApJ, 671, 1256

\nhi Netzer, H, Maoz, D., Laor, A., \etal 1990, ApJ, 353, 108  %  reverb on N5548 ==> BLR smaller than given by ionizn theory

\nhi Netzer, H., \& Marziani, P. 2010, ApJ, 724, 318

%\nhi Netzer, H., \& Peterson, B. M. 1997, in Astronomical Time Series, ed. D. Maoz, A. Sternberg, \& E. M. Leibowitz 
%     (Dordrecht: Kluwer), 85

\nhi Neumayer, N., Cappellari, M., Reunanen, J., \etal 2007, ApJ, 671, 1329  %  Cen A

%\nhi Noordermeer, E., van der Hulst, J.~M., Sancisi, R., Swaters, R.~S., \& van Albada, T.~S.~2007, MNRAS, 276, 1513

\nhi Nowak, N., Saglia, R. P., Thomas, J., \etal 2007, MNRAS, 379, 909     %   NGC 4486A

\nhi Nowak, N., Saglia, R. P., Thomas, J., \etal 2008, MNRAS, 391, 1629    %   Fornax A

\nhi Nowak, N., Thomas, J., Erwin, P., \etal 2010, MNRAS, 403, 646 %  NGC 3368 & NGC 3489

\nhi Oliva, E., Origlia, L., Kotilainen, J. K., \& Moorwood, A. F. M. 1995, A\&A, 301, 55  %  sigmas, starbursts

\nhi Onken, C. A., Ferrarese, L., Merritt, D., \etal 2004, ApJ, 615, 645  %  Calibration of reverberation-mapped MBH

%\nhi Onken, C. A., \& Peterson, B. M. 2002, ApJ, 572, 746  %  One of two sources in 3.5.1 Figure 0

%\nhi Onken, C. A., Valluri, M., Peterson, B. M., et al. 2007, ApJ, 670, 105

\nhi Osterbrock, D. E., \& Mathews, W. G. 1986, ARA\&A, 24, 171

\nhi Pancoast, A., Brewer, B. J., \& Treu, T. 2011, ApJ, 730, 139

\nhi Pancoast, A., Brewer, B. J., Treu, T., et al. 2012, ApJ, 754, 49

\nhi Park, D., Woo, J.-H., Treu, T., et al. 2012, ApJ, 747, 30

\nhi Paturel, G., Fang, Y., Petit, C., Garnier, R., \& Rousseau, J. 2000, A\&AS, 146, 19

\nhi Paturel, G., Petit, C., Prugniel, Ph., \etal 2003, A\&A, 412, 45

\nhi Pellegrini, S., Held, E. V., \& Ciotti, L. 1997, MNRAS, 288, 1

%\nhi Peterson, B. M. 1993, PASP, 105, 247

%\nhi Peterson, B. M. 2001, Advanced Lectures on the Starburst-AGN Connection,
       % , Proceedings of a conference held in Tonantzintla, Puebla, Mexico, 26-30 June, 2000. 
%       ed. I. Aretxaga, D. Kunth, \& R. M\'ujica (Singapore: World Scientific), 3

%\nhi Peterson, B. M. 2008, NewAR, 52, 240

\nhi Peterson, B. M. 2011, in Narrow-Line Seyfert 1 Galaxies and Their Place in the Universe,
                  ed. L. Foschini, et al., 
                  {\bf http://pos.sissa.it/cgi-bin/reader/conf.cgi?confid=126}, id. 32

%\nhi Peterson, B. M., Bentz, M. C., Desroches, L.-B., et al. 2005, ApJ, 632, 799;
%                      Erratum. 2005, ApJ, 641, 638 %  Reverb MBH=3.6e5 for NGC 4395

%\nhi Peterson,{\ts}B.{\ts}M., Berlind,{\ts}P., Bertram,{\ts}R., et al. 2002, ApJ, 581, 197  %  tau propto F_UV**1/2

\nhi Peterson, B. M., Ferrarese, L., Gilbert, K. M., et al. 2004, ApJ, 613, 682  %  Reverb database for 35 gal: 481 cites Oct 14

%\nhi Peterson, B. M., \& Wandel, A. 1999, ApJ, 521, L95

%\nhi Peterson, B. M., \& Wandel, A. 2000, ApJ, 540, L13

\nhi Pinkney, J., Gebhardt, K., Bender, R., \etal 2003, ApJ, 596, 903

\nhi Rix, H.-W., Kennicutt, R. C., Braun, R., \& Walterbos, R. A. M. 1995, ApJ, 438, 155

\nhi Rusli, S. P. 2012, PhD Thesis, Ludwig-Maximilians-University, Munich, Germany;
     {\bf http://edoc.ub.uni-muenchen.de/14493}

\nhi Rusli, S. P., Thomas, J., Erwin, P., \etal 2011, MNRAS, 410, 1223   %  NGC 1332

\nhi Rusli, S. P., Thomas, J., Saglia, R. P., \etal 2013, AJ, 146, 45

\nhi Saglia, R. P., Fabricius, M., Bender, R., \etal 2010, A\&A, 509, A61 

\nhi Samurovi\'c, S., \& Danziger, I. J. 2005, MNRAS, 363, 769

\nhi S\'anchez-Portal, M., D\'\i az, A. I., Terlevich, E., \& Terlevich, R. 2004, MNRAS, 350, 1087

\nhi Sandage, A. 1961, The Hubble Atlas of Galaxies (Washington, DC: Carnegie Institution of Washington)

\nhi Sandage, A., \& Bedke, J. 1994, The Carnegie Atlas of Galaxies (Washington, DC: Carnegie Institution of Washington)

\nhi Sandage, A., \& Tammann, G. A. 1981, A Revised Shapley-Ames Catalog of Bright Galaxies 
     (Washington, DC: Carnegie Institution of Washington)

\nhi Sarzi, M., Rix, H.-W., Shields, J. C., \etal 2001, ApJ, 550, 65  %  NGC 2787 BH

\nhi Sarzi, M., Rix, H.-W., Shields, J. C., \etal 2002, ApJ, 567, 237

\nhi Schlafly, E. F., \& Finkbeiner, D. P. 2011, ApJ, 737, 103

\nhi Schlegel, D. J., Finkbeiner, D. P., \& Davis, M. 1998, ApJ, 500, 525

\nhi Schmidt, M., \& Green, R.~F. 1983, ApJ, 269, 352

\nhi Schombert, J. 2011, arXiv:1107.1728

\nhi Schulze, A., \& Gebhardt, K. 2011, ApJ, 729, 21

\nhi Scorza, C., \& Bender, R. 1995, A\&A, 293, 20

\nhi Seigar, M. S., Barth, A. J., \& Bullock, J. S. 2008, MNRAS, 389, 1911

\nhi S\'ersic, J. L. 1968, Atlas de Galaxias Australes (C\'ordoba: Obs. Astron. Univ. Nacional de C\'ordoba)

\nhi Shen, J., \& Gebhardt, K. 2010, ApJ, 711, 484

\nhi Shen, J., Rich, R. M., Kormendy, J., \etal 2010, ApJ, 720, L72

\nhi Shen, Y. 2013, Bull. Astron. Soc. India, 41, 61 (arXiv:1302.2643)

\nhi Shen, Y., Greene, J. E., Strauss, M. A., Richards, G. T., \& Schneider, D. P. 2008, ApJ, 680, 169

\nhi Shen, Y., \& Liu, X. 2012, ApJ, 753, 125

\nhi Shen, Y., Richards, G. T., Strauss, M. A., et al. 2011, ApJS, 194, 45

\nhi Shostak, G. S. 1987, A\&A, 175, 4

\nhi Silge, J. D., Gebhardt, K., Bergmann, M., \& Richstone, D. 2005, AJ, 130, 406  %  Cen A

\nhi Simien, F., \& Prugniel, Ph. 2000, A\&AS, 145, 263

\nhi Simien, F., \& Prugniel, Ph. 2002, A\&A, 384, 371

\nhi Skrutskie, M. F., Cutri, R. M., Stiening, R., \etal 2006, AJ, 131, 1163

\nhi Tadhunter, C., Marconi, A., Axon, D., \etal 2003, MNRAS, 342, 861                             %  BH in Cygnus A

\nhi Tempel, E., Tamm, A., \& Tenjes, P. 2010, A\&A, 509, A91

\nhi Thomsen, B., Baum, W. A., Hammergren, M., \& Worthey, G. 1997, ApJ, 483, L37  %  Distance to Coma Cluster = 102 Mpc

\nhi Thornton, R. J., Stockton, A., \& Ridgway, S. E. 1999, AJ, 118, 1461

\nhi Tonry, J. L., Dressler, A., Blakeslee, J. P., \etal 2001, ApJ, 546, 681

\nhi Tremaine, S., Gebhardt, K., Bender, R., \etal 2002, ApJ, 574, 740

\nhi Trevese, D., Paris, D., Stirpe, G. M., Vagnetti, F., \& Zitelli, V. 2007, A\&A, 470, 491

\nhi Tully, R. B. 1988, Nearby Galaxies Catalog (Cambridge: Cambridge University Press)

\nhi van den Bosch, R. C. E., \& de Zeeuw, P. T. 2010, MNRAS, 401, 1770

\nhi van den Bosch, R. C. E., Gebhardt, K., G\"ultekin, K., \etal 2012, Nature, 491, 729

\nhi van der Marel, R. P. 1994a, MNRAS, 270, 271

\nhi van der Marel, R. P. 1994b, ApJ, 432, L91  %  HST-expected M87 LOSVDs

\nhi van der Marel, R. P., \& van den Bosch, F. C. 1998, AJ,     116, 2220  %  BH in NGC 7052

\nhi van Driel, W., \& van Woerden, H. 1991, A\&A, 243, 71

\nhi Verdoes Kleijn, G. A., van der Marel, R. P., Carollo, C. M., \& de Zeeuw, P. T. 2000, AJ, 120, 1221

\nhi Vestergaard, M. 2002, ApJ, 571, 733

\nhi Vestergaard, M., \& Peterson, B. M. 2006, ApJ, 641, 689  %  Most used calibration of photoionization models --> MBH

\nhi Walsh, J. L., Barth, A. J., \& Sarzi, M. 2010, ApJ, 721, 762  %  NGC 4374 

\nhi Walsh, J. L., van den Bosch, R. C. E., Barth, A. J., \& Sarzi, M. 2012, ApJ, 753, 79

\nhi Walterbos, R. A. M., \& Kennicutt, R. C. 1987, A\&AS, 69, 311

%\nhi Wandel, A., Peterson, B. M., \& Malkan, M. A. 1999, ApJ, 526, 579

\nhi Wandel, A., \& Yahil, A. 1985, ApJ, 295, L1

\nhi Wang, J.-G., Dong, X.-B., Wang, T.-G., \etal 2009, ApJ, 707, 1334  %  Recalibration of virial masses with variable f

\nhi Wegg, C., \& Gerhard, O. 2013, MNRAS, in press (arXiv:1308.0593)

\nhi Weiland, J. L., Arendt, R. G., Berriman, G. B., \etal 1994, ApJ, 425, L81

\nhi Weinzirl, T., Jogee, S., Khochfar, S., Burkert, A., \& Kormendy, J. 2009, ApJ, 696, 411

%\nhi Willott, C. J., Albert, L., Arzoumanian, D., et al. 2010, ApJ, 140, 546  % z ~ 6 quasars

\nhi Wold, M., Lacy, M., K\"aufl, H. U., \& Siebenmorgen, R. 2006, A\&A, 460, 449

\nhi Woo, J.-H. 2008, AJ, 135, 1849

\nhi Woo, J.-H., Treu, T., Barth, A. J., et al. 2010, ApJ, 716, 269

\nhi Xiao, T., Barth, A. J., Greene, J. E., \etal 2011, ApJ, 739, 28

\nhi Yamauchi, A., Nakai, N., Ishihara, Y., Diamond, P., \& Sato, N. 2012, PASJ, 64, 103

\nhi Yuan, F., Yu, Z., \& Ho, L. C. 2009, ApJ, 703, 1034

\vfill\eject

\end